\RequirePackage{booktabs}\let\LaTeXcline\cline
\documentclass[pdflatex,sn-mathphys-num]{sn-jnl}\let\cline\LaTeXcline
\usepackage{amsthm}
\usepackage{amsmath}
\usepackage{amssymb}
\usepackage{amsfonts}
\usepackage{bm}
\usepackage{bbm}
\usepackage{enumerate}
\usepackage{graphicx}
\usepackage{hyperref}
\usepackage{import}
\usepackage{multirow}
\usepackage{multicol}
\usepackage{psfrag}
\usepackage[dvipsnames]{xcolor}
\usepackage{nicefrac}

\usepackage{tabularx}

\newcommand{\change}[1]{{{#1}}}

\DeclareFontFamily{U}{FdSymbolA}{}
\DeclareFontShape{U}{FdSymbolA}{m}{n}{<-> s * [1] FdSymbolA-Book}{}
\DeclareFontShape{U}{FdSymbolA}{m}{b}{<-> s * [1] FdSymbolA-Medium}{}
\DeclareSymbolFont{fdsymbols}{U}{FdSymbolA}{m}{n}
\SetSymbolFont{fdsymbols}{bold}{U}{FdSymbolA}{m}{b}
\DeclareMathSymbol{\bdiamond}{\mathbin}{fdsymbols}{130}

\usepackage{pmboxdraw}
\usepackage{newunicodechar}
\newunicodechar{└}{\textSFii}
\newunicodechar{├}{\textSFviii}
\newunicodechar{─}{\textSFx}

\DeclareMathOperator*{\conv}{{\bf conv}}
\DeclareMathOperator*{\dist}{{\bf dist}}

\DeclareMathOperator*{\argmin}{argmin}

\newcommand{\Oplus}{\ensuremath{\vcenter{\hbox{\scalebox{1.2}{$\oplus$}}}}}
\newcommand{\medcap}{\ensuremath{\vcenter{\hbox{\scalebox{1.2}{$\cap$}}}}}
\DeclareMathOperator*{\cproduct}{\operatorname*{\Oplus}}

\newcommand{\spe}{{\sf spe}}
\newcommand{\eps}{\varepsilon}

\newcommand{\Tr}{\mathbf{tr}}

\newcommand{\mm}[1]{{\boldsymbol{#1}}}
\newcommand{\bb}[1]{\mathbbm{#1}}
\newcommand{\cc}[1]{\mathcal{#1}}
\newcommand{\xt}[2]{{#1}^{(#2)}}

\newcommand{\inprod}[2]{{\langle{#1}, {#2}\rangle}}
\newcommand{\mmexp}{\mathbf{exp}}

\newcommand{\bJ}{\bb{J}}

\newcommand{\bR}{\bb{R}}

\newcommand{\bU}{\bb{U}}
\newcommand{\bV}{\bb{V}}

\newcommand{\bZ}{\bb{Z}}

\newcommand{\mma}{\mm{a}}
\newcommand{\mmb}{\mm{b}}
\newcommand{\mmc}{\mm{c}}

\newcommand{\mme}{\mm{e}}

\newcommand{\mmh}{\mm{h}}

\newcommand{\mmp}{\mm{p}}
\newcommand{\mmq}{\mm{q}}

\newcommand{\mmu}{\mm{u}}
\newcommand{\mmv}{\mm{v}}
\newcommand{\mmw}{\mm{w}}
\newcommand{\mmx}{\mm{x}}
\newcommand{\mmy}{\mm{y}}
\newcommand{\mmz}{\mm{z}}

\newcommand{\mmV}{\mm{V}}

\newcommand{\mmxi}{{\boldsymbol{\xi}}}

\newcommand{\mmone}{\mm{1}}
\newcommand{\mmzero}{\mm{0}}
\newcommand{\brmf}{{\rm\bf f}}
\newcommand{\brmg}{{\rm\bf g}}
\newcommand{\rmh}{{\rm h}}

\newcommand{\la}{\langle}
\newcommand{\ra}{\rangle}

\def\bpm{\begin{pmatrix}}
\def\epm{\end{pmatrix}}
\def\bal{\begin{aligned}}
\def\eal{\end{aligned}}

\newcommand{\bi}{\begin{itemize}}
\newcommand{\ei}{\end{itemize}}
\newcommand{\beq}{\begin{equation}}
\newcommand{\eeq}{\end{equation}}

\newcommand{\ben}{\begin{enumerate}}
\newcommand{\een}{\end{enumerate}}

\usepackage{graphicx}%
\usepackage{multirow}%
\usepackage{amsmath,amssymb,amsfonts}%
\usepackage{amsthm}%
\usepackage{mathrsfs}%
\usepackage[title]{appendix}%
\usepackage{xcolor}%
\usepackage{textcomp}%
\usepackage{manyfoot}%
\usepackage{booktabs}%
\usepackage{algorithm}%
\usepackage{adjustbox}%
\usepackage{algorithmic}%
\usepackage{diagbox}
\usepackage{listings}%

\newcolumntype{C}[1]{>{\centering\arraybackslash}p{#1}}

\theoremstyle{thmstyleone}%
\newtheorem{theorem}{Theorem}%
\newtheorem{lemma}{Lemma}%
\newtheorem{corollary}{Corollary}%
\newtheorem{proposition}[theorem]{Proposition}%

\theoremstyle{thmstyletwo}%
\newtheorem{remark}{Remark}%
\newtheorem{problem}{Problem}%

\theoremstyle{thmstylethree}%

\begin{document}

\title[Approximation Algorithms for Smallest Intersecting Balls]{Approximation Algorithms for Smallest Intersecting Balls}

\author*[1]{\fnm{Jiaqi} \sur{Zheng}}\email{jiaqi@u.nus.edu}

\author[1]{\fnm{Tiow-Seng} \sur{Tan}}\email{tants@comp.nus.edu.sg}

\affil[1]{\orgdiv{Department of Computer Science}, \orgname{National University of Singapore}, \orgaddress{\street{13 Computing Drive}, \postcode{117417}, \country{Singapore}}}

\abstract{    
We study a general smallest intersecting ball problem and its soft-margin variant in high-dimensional Euclidean spaces for input objects that are compact and convex.
These two problems link and unify a series of fundamental problems in computational geometry and machine learning, including smallest enclosing ball, polytope distance, intersection radius, $\ell_1$-loss support vector machine, $\ell_1$-loss support vector data description, and so on.
Leveraging our novel framework for solving zero-sum games over symmetric cones, we propose general approximation algorithms for the two problems, where implementation details are presented for specific inputs of convex polytopes, reduced polytopes, axis-aligned bounding boxes, balls, and ellipsoids.
For most of these inputs, our algorithms are the first results in high-dimensional spaces, and also the first approximation methods.
Experimental results show that our algorithms can solve large-scale input instances efficiently.
}


\maketitle

\section{Introduction}

Given a set of convex objects, the {\em smallest intersecting ball} (SIB) problem is to find a ball with the smallest radius that intersects all these objects.
It is often considered a variant of the smallest enclosing ball (SEB) problem to find the smallest ball that encloses all the input objects.
Nevertheless, the general understanding of the SIB problem lags behind that of the SEB problem in several aspects.
The first is diversity: SEB has been studied for various types of inputs, such as points, balls and ellipsoids, whereas SIB has only been explored for several specific input objects.
The second is application: SEB has found applications in the domains of computer graphics, machine learning, metrology, and so on, while SIB has attracted much less attention and the studies mainly remain in theory.
The third is algorithm design: numerous algorithms have been proposed for solving the SEB problems, including exact and approximation algorithms, using optimization or coreset techniques, and in parallel or streaming settings, but for SIB, most algorithms are merely designed for solving the problems in fixed dimensions.

In this paper, we study two general problems, one is the original SIB problem where the input convex objects are {\em compact} (i.e.~closed and bounded), and the other is its soft-margin variant designed to handle the presence of outliers.
Both problems are in arbitrary-dimensional Euclidean spaces.
Throughout this work, we use SIB to refer to the original problem that finds the smallest ball intersecting every input object, and term the soft-margin variant as soft-margin SIB or Soft-SIB.
The general setup allows various types of input objects including points, line segments, convex polytopes, reduced polytopes, axis-aligned bounding boxes (AABBs), balls, ellipsoids, and so on.
It links and unifies many important problems in the literature.
For example,
when the input objects are just points (singleton sets), SIB is equivalent to SEB and Soft-SIB is a variant of the $\ell_1$-loss support vector data description ($\ell_1$-SVDD) problem~\cite{svdd:CLL2013};
when line segments are used to simulate network links, the result is the smallest regional disaster that hits all links, which has found applications in backbone network planning~\cite{netplan:TRVG2020};
when AABBs and balls are used to represent imprecise input points in geometric measurement tasks, SIB provides the best lower bound on the radius of SEB~\cite{siblowdim:LK2010}.
\change{
Moreover, elliptical inputs to the problems can guarantee robustness in uncertain scenarios,
since they can model confidence regions for a family of probability distributions~\cite{multichebyshev:AI1960}.
}

As a special case, when the input contains only two convex objects, SIB corresponds to the shortest connector (SC) problem, which is to find the shortest line segment connecting these two objects (assuming they do not intersect).
The dual problem of the SC problem is to find the optimal hyperplane that separates two convex sets~\cite{distance:Achiya2006}, which is a fundamental idea of many classification models in machine learning.
Specifically,
when the objects are convex polytopes generated from finite point sets, the SC problem is often referred to as the polytope distance (PD) problem, and the separation problem corresponds to the hard-margin support vector machine (SVM) problem~\cite{pd:GJ2009};
when the inputs are reduced polytopes, the separation problem is equivalent to $\ell_1$-loss SVM~\cite{svm:BB2000,svm:CB1999};
when they are ellipsoids, the separating hyperplane corresponds to a classifier that minimizes the maximum probability of misclassification of data points sampled from distributions with known means and covariances~\cite{minimaxsvm:GLCM2002}.

\begin{table}[t]
\caption{Summary of the problems and the main results}
\setlength\extrarowheight{2pt}
\begin{tabular}{|c|l|cc|c|c|}
\hline
\multicolumn{2}{|c|}{Problem} & {\bf $n$} & {\bf $m_i$} & Previous Work & This Paper \\
\hline\hline
\parbox[t]{2mm}{\multirow{9}{*}{\rotatebox[origin=c]{90}{SIB of}}} 
& Convex Polytopes \S\ref{sec:sib_cvx} & Any & Any & \scalebox{.85}{$O(M)^\ddag$~\cite{siblowdim:JMB1996}} & \scalebox{.85}{$\widetilde{O}({R^2(N + nd) \over \eps^2})$}\\
& ├ Polytope Distance (Hard-SVM) & 2 & Any & \scalebox{.85}{$\widetilde{O}({R^2 (M + d)\over \eps^2})$~\cite{sublinearopt:CHW2012}} & \scalebox{.85}{$O({R^2(N+d) \over \eps^2})$}\\
& ├ Smallest Enclosing Ball (Hard-SVDD) & Any & 1 & \scalebox{.85}{$\widetilde{O}({n\over \eps^2} + {d \over \eps})$~\cite{sublinearopt:CHW2012}} & \scalebox{.85}{$\widetilde{O}({nd \over \eps^2})$}\\
& └ Intersecting Radius of Line Segments & Any & 2 & \scalebox{.85}{$O(n)^\ddag$~\cite{siblowdim:BJMR1991}} & \scalebox{.85}{$\widetilde{O}({R^2 nd \over \eps^2})$} \\\cline{2-6}
& Reduced Polytopes \S\ref{sec:sib_rcvx} & Any & Any & - & \scalebox{.85}{$\widetilde{O}({R^2(N + nd) \over \eps^2})$} \\
& └ Soft-SVM ($C$-SVM, $\nu$-SVM) & 2 & Any & \scalebox{.85}{$\widetilde{O}({R^2 (M + d)\over \eps^2})$~\cite{sublinearsvm:HKS2011}} & \scalebox{.85}{$O({R^2(N+d) \over \eps^2})$} \\\cline{2-6}
& AABBs (Imprecise Points) \S\ref{sec:sib_box} & Any & - & \scalebox{.85}{$O(n)^\ddag$~\cite{siblowdim:LK2010}} & \scalebox{.85}{$\widetilde{O}({R^2 nd \over \eps^2})$} \\\cline{2-6}
& Balls (Imprecise Points) \S\ref{sec:sib_ball} & Any & - &  \scalebox{.85}{$O({n \over \eps^{(d-1)/2}})$~\cite{sibhighdim:SA2015}} & \scalebox{.85}{$\widetilde{O}({R^2 nd \over \eps^2})$} \\\cline{2-6}
& Ellipsoids (Distributions) \S\ref{sec:sib_ellipsoid} & Any & - &  - & \scalebox{.85}{$\widetilde{O}(nd^\omega + {R^2 nd^2 \over \eps^2})$} \\
\hline
\multicolumn{2}{|c|}{Soft-SIB of Points (Soft-SVDD)$^\dag$ \S\ref{sec:svdd}} & Any & 1 & - & \scalebox{.85}{$\widetilde{O}({R^2 nd \over \eps^2})$} \\
\hline
\end{tabular}%
\vspace{.5em}
\scalebox{0.9}{
\parbox{1.11\linewidth}{
\rule{10em}{.5pt}

Note: $d$ is the dimensionality. $n$ is the number of objects (polytopes, balls,  etc.). $m_i$ is the number of points in each set. $M$ is the total number of points. $N$ is the number of nonzeros in the input. $R$ is an instance-dependent parameter. $\omega$ is the matrix multiplication exponent. We use the notation $\widetilde{O}(\cdot)$ to hide~logarithmic~factors.

$^\dag$ Results for Soft-SIB with other inputs are omitted here since they are similar to the hard-margin counterparts.

$^\ddag$ Running time of exact algorithms for problems in fixed dimensions.
}
}
\label{table:summary}
\end{table}

\medskip
{\bf\em Contributions.}
We model both problems (SIB and Soft-SIB) as convex optimization problems, and propose two deterministic approximation algorithms to solve them respectively.
Both algorithms compute approximate solutions in general dimensions with an approximation ratio of $(1+\eps)$.
Implementation details are given for specific types of inputs such as convex polytopes, reduced polytopes, AABBs, balls, and ellipsoids (points and line segments are considered special cases of polytopes).
For most of these problems, our algorithms are the first results in high-dimensional spaces, and the first fully polynomial-time approximation schemes (FPTAS).
Moreover, the main procedure of the algorithms (except for the pre-processing of ellipsoids) can be directly implemented in parallel settings (for example, in the work-depth or PRAM model).
See Table~\ref{table:summary} for a summary of the main results of this paper.

The theoretical foundations of our algorithms are from Euclidean Jordan algebras (EJA).
Inspired by the seminal work of Arora et al.~\cite{mwu:AHK2012} and Clarkson et al.~\cite{sublinearopt:CHW2012}, we fit the geometric optimization problems into a general framework of zero-sum games over symmetric cones, which we called {\em symmetric cone games} (SCG).
We propose an algorithm to compute nearly optimal strategies for both players in the SCG, which is based on a recent extension of the multiplicative weights update method to symmetric cones~\cite{scmwu:CLPV2023}, and our analysis uses the Golden-Thompson inequality for EJA exponentials~\cite{gtineq:TWK2022}.
The algorithmic framework of SCG can be directly extended to other EJA spaces and symmetric cones, which may have broader applications in algorithm design.

\begin{figure}[t]
\begin{center}
\includegraphics[width=.95\linewidth]{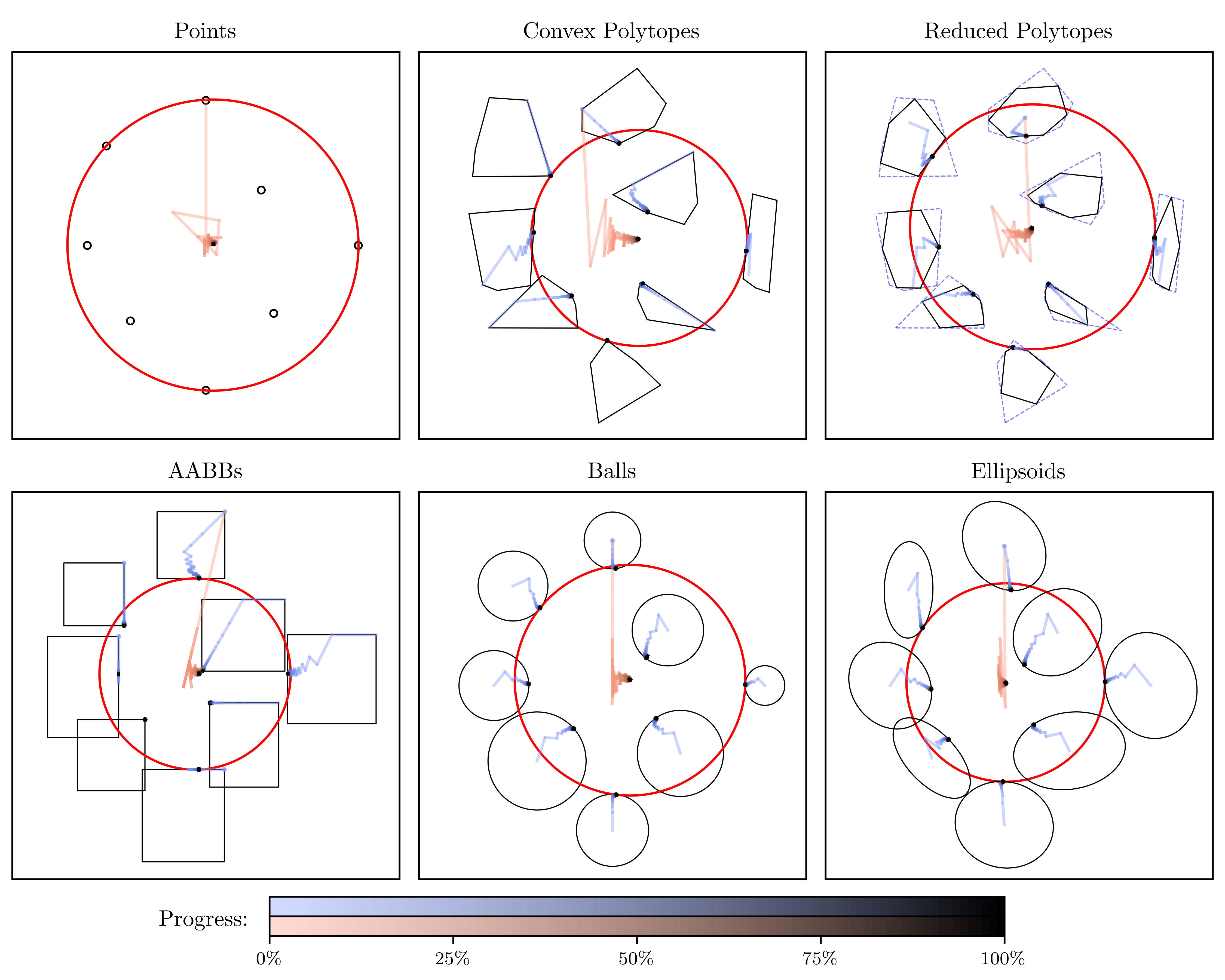}
\end{center}
\caption{Visualization of the 2D results \change{(red circles)} computed by our algorithms. The brown and blue lines illustrate the trajectories of the centers and the point in each object respectively.}
\label{fig:visualization}
\end{figure}

The practical performance of our algorithms is validated through extensive experiments. 
Code that implements the algorithms is available at \cite{libsib}. 
To our knowledge, this is the first software capable of solving SIBs \change{with non-singleton input objects, even in two-dimensional spaces.}
Figure~\ref{fig:visualization} shows some 2D examples.

\medskip
{\bf\em Related Work.}
The SIB problem is also referred in the literature as intersection radius, smallest ball stabber, smallest possible smallest enclosing ball, etc. Bhattacharya et al.~\cite{siblowdim:BJMR1991} proposed the first algorithm for the SIB problem, where the input is a set of line segments on the plane (in which case the intersecting ball is a disk). Their algorithm uses the prune-and-search strategy of Dyer and Megiddo~\cite{Dyer1984,Megiddo1983}, and finds an exact solution in linear time.
The technique is later extended by Jadhav et al.~\cite{siblowdim:JMB1996} to devise a linear-time algorithm for input sets of two-dimensional convex polygons.
Löffler and van Kreveld~\cite{siblowdim:LK2010} investigated the scenario when the input objects are squares and discs.
Nguyen et al.~\cite{sibtheory:NNS2012} formulated the SIB problem as an unconstrained optimization problem in reflective Banach spaces and analyzed its properties. They also showed that a subgradient algorithm will converge to the optimal set, which is the first algorithm for SIB in high dimensions. However, the convergence rate of the subgradient method has not been proved, and computing a subgradient itself could be a difficult problem (for example, when the input objects are convex polytopes).
Son and Afshani~\cite{sibhighdim:SA2015} proposed the first streaming and approximation algorithms for SIB when the input is a set of {\em disjoint} Euclidean balls in $\bR^d$, but the running time has a dependency of $O\big({1\over \eps^{(d-1)/2}}\big)$ on the error $\eps$ if a $(1+\eps)$-approximation is required.

SEB (a.k.a.\ SVDD) and PD (a.k.a.\ SVM) are two important problems that are closely related to SIB, there is a long history of research on their approximation algorithms.
Bădoiu and Clarkson \cite{sebcoreset:BC2008} showed that it is possible to find a small {\em coreset} (a subset of the input points) for SEB that yields a $(1+\eps)$-approximate solution.
Coresets has been used in the development of many approximation algorithms for SEB~\cite{sescoreset:badoiu2003smaller,sebcoreset:KMY2003,seb:yildirim2008,sebcoreset:Clarkson2010}.
Among these works, Yildirim's algorithm~\cite{seb:yildirim2008} is probably the most representative one, which returns a coreset of size $O(1/\eps)$ and solves the problem approximately in linear time.
There are also algorithms that focus on solving the SEB problem using pure optimization techniques, which do not rely on the coresets~\cite{sebopt:BS2000,sebball:ZTS2005,sebopt:ALY2016}.
Gärtner and Jaggi~\cite{pd:GJ2009} gave the first result on coreset for the PD problem and summarized the relationship between PD and the most used SVM variants.
They presented a geometric algorithm that outputs a $O(1/\eps)$ coreset and an approximate solution in linear time.
Clarkson et al.~\cite{sublinearopt:CHW2012} proposed an elegant framework for solving several machine learning problems in sublinear running time, including SEB and hard-margin SVM.
They also gave lower bounds on these two problems and showed that their algorithms are nearly the best possible in the unit-cost RAM model.
This is later extended by Hazan et al.~\cite{sublinearsvm:HKS2011} to develop the first sublinear time algorithm for $\ell_1$-loss SVM.

The framework of Clarkson et al.~\cite{sublinearopt:CHW2012} is a zero-sum game over a standard simplex and a unit ball, which is also known as a $\ell_1$-$\ell_2$ game. Their sublinear approximation algorithm is based on stochastic mirror descent and stochastic gradient descent.
Li et al.~\cite{l1lpgame:LWCW2021} generalized \cite{sublinearopt:CHW2012} to a $\ell_1$-$\ell_p$ game by extending the $\ell_2$-norm unit ball to a $\ell_p$-norm unit ball.
They also introduced sublinear algorithms in both classical and quantum settings for approximating this game. 
Our SCG framework generalizes both \cite{sublinearopt:CHW2012} and \cite{l1lpgame:LWCW2021}, where the strategy sets are a spectraplex over a symmetric cone (which covers the $\ell_1$- and $\ell_2$-ball) and a closed convex set (which covers the $\ell_p$-ball).
Nevertheless, none of the previous sublinear methods can be directly applied to our general framework. 
Recently, Canyakmaz et al.~\cite{scmwu:CLPV2023} proposed a new approach for solving online optimization problems over spectraplexes of symmetric cones, which is later employed by Zheng et al.~\cite{pdscp:ZVTL2024} for solving symmetric cone programs.
Our algorithm for SCG also builds on \cite{scmwu:CLPV2023}, but in a \change{solely} offline setting. 
The choice stems from the fact that their regret bound already matches the lower bound for standard simplices \cite{mwu:AHK2012}. 
The online setting used by \cite{scmwu:CLPV2023} and \cite{pdscp:ZVTL2024} prevents further improvements on the algorithms; \change{on the other hand, under the offline setting,} a better algorithm \change{than ours in this paper} for SCG remains~open.

Comparisons 
of our algorithms and previous methods are also presented in Table~\ref{table:summary}.

\medskip
{\bf\em Outline.}
A brief introduction to EJA and symmetric cones is provided in Section~\ref{sec:pre}.
Section~\ref{sec:game} introduces SCG and an algorithm for finding approximate equilibria.
Section~\ref{sec:sib} presents the approximation algorithm for the SIB problem and states the detailed running time for inputs of convex polytopes, reduced polytopes, AABBs, balls, and ellipsoids.
Section~\ref{sec:soft} discusses the approximation algorithm for the soft-margin SIB problem.
Section~\ref{sec:experiment} reports the experimental results.
Section~\ref{sec:disc} concludes the paper.

\section{Euclidean Jordan algebras and symmetric cones}\label{sec:pre}
In this section, we introduce the necessary tools and concepts of Euclidean Jordan algebras and symmetric cones that will be used in our algorithm design.
Readers who are interested in EJA and symmetric cones can find more in~\cite{symmetriccone:FK1994,eja:Orlitzky2022,symmetriccone:Vandenberghe2016}.
All the vector spaces and algebras in this work are over the real field.

\subsection{Euclidean Jordan algebras}

Let $\bJ$ be a finite-dimensional vector space, and let $\circ: \bJ\times \bJ \rightarrow \bJ$ be a bilinear binary operation on $\bJ$ such that for all $\mmx, \mmy \in \bJ$, we have $\mmx\circ \mmy \in \bJ$. We call any bilinear binary operation $\circ$ a {\em Jordan product} on $\bJ$ if for all $\mmx, \mmy \in \bJ$, it satisfies the following two properties:
\begin{enumerate}
    \item (Commutativity) $\mmx \circ\mmy = \mmy \circ \mmx$,
    \item (Jordan identity) $\mmx^2 \circ (\mmx \circ \mmy) = \mmx \circ (\mmx^2 \circ \mmy)$.
\end{enumerate}
A finite-dimensional vector space $\bJ$ equipped with a Jordan product is a (finite-dimensional) {\em Jordan algebra}, and is denoted as $(\bJ, \circ)$. For convenience, we will sometimes use $\bJ$ to denote $(\bJ, \circ)$ if the Jordan product $\circ$ is clear from the context. An element $\mme\in \bJ$ that satisfies $\mmx \circ \mme = \mme \circ \mmx = \mmx$ for all $\mmx\in \bJ$ is the {\em identity element} of $\bJ$. The power of an element $\mmx \in \bJ$ is defined as
\[
    \mmx^0 := \mme, \quad \mmx^k := \mmx\circ \mmx^{k-1},\ \forall k \ge 1.
\]
The degree of $\mmx\in \bJ$ is defined as the smallest $k\in \mathbbm{Z}_+$ such that the vectors $\mme, \mmx, \dots, \mmx^k$ are linearly dependent. The {\em rank} of a Jordan algebra $\bJ$ is then defined as the largest degree of all elements in $\bJ$.

We call a Jordan algebra $(\bJ, \circ)$ Euclidean if there exists a inner product $\bullet: \bJ\times \bJ \rightarrow \bR$ on $\bJ$ that satisfies
\beq\label{eq:EJA_inner_product_property}
\mmx \bullet (\mmy \circ \mmz) = (\mmx \circ \mmy) \bullet \mmz, \quad \forall \mmx, \mmy, \mmz\in \bJ.
\eeq
Let $(\bJ, \circ, \bullet)$ (or $\bJ$ when the Jordan product $\circ$ and the inner product $\bullet$ are clear from the context) be an Euclidean Jordan algebra (EJA), and let $r$ be its rank. Any $\mmx\in \bJ$ admits a spectral decomposition, that is $\mmx = \sum_{k=1}^r \lambda_k(\mmx)\mmq_k$, where $\lambda_1(\mmx), \dots, \lambda_r(\mmx)\in \bR$ are the eigenvalues of $\mmx$, and $\{\mmq_1,\dots, \mmq_r\} \subseteq \bJ$ is a complete system of primitive orthogonal {\em idempotents}, which satisfies the following three properties:
\begin{enumerate}
    \item (Idempotents and primitiveness) $\mmq_k^2 = \mmq_k$ and $\mmq_k\bullet \mmq_k = 1$, $\forall k \in [r]$,
    \item (Orthogonality) $\mmq_i \circ \mmq_j = 0,\ \forall i \neq j,\ i,j\in [r]$,
    \item (Completeness) $\sum_{k=1}^r \mmq_k = \mme$.
\end{enumerate}
The system $\{\mmq_1,\dots, \mmq_r\}$ is called a {\em Jordan frame} in $\bJ$. Based on the spectral decomposition, the trace, determinant and exponential of $\mmx\in \bJ$ are defined as:
\[
    \Tr(\mmx) := \sum_{k=1}^r \lambda_k(\mmx), \quad {\bf det}(\mmx) := \prod_{k=1}^r \lambda_k(\mmx) \quad \text{and} \quad
    \mmexp(\mmx) := \sum_{k=1}^r \exp(\lambda_k(\mmx))\mmq_k.
\]
Moreover, we have $\sum_{k=1}^r \lambda_k^2(\mmx)\mmq_k = \mmx \circ \mmx = \mmx^2$.
We define the {\em spectral norm} of $\mmx\in\bJ$ as the maximum magnitude of its eigenvalues:
\beq\label{eq:spectral_norm}
    \|\mmx\|_\spe = \max_{k\in [r]} |\lambda_k(\mmx)|.
\eeq

A Jordan algebra $\bJ$ is Euclidean if and only if the symmetric bilinear form $(\mmx, \mmy)\mapsto \Tr(\mmx \circ \mmy)$ is an inner product on $\bJ$ that satisfies~\eqref{eq:EJA_inner_product_property}. As such, throughout this work we fix the inner product of an EJA as:
\[
    \mmx\bullet \mmy = \Tr(\mmx \circ \mmy), \quad \forall \mmx, \mmy \in \bJ.
\]
This is called the {\em canonical trace inner product}. Based on this inner product, the Golden-Thompson inequality can be generalized to the EJA systems~\cite{gtineq:TWK2022}. For all $\mmx,\mmy \in \bJ$, the following inequality holds:
\beq\label{eq:golden-thompson}
\Tr(\mmexp(\mmx + \mmy)) \le \Tr(\mmexp(\mmx)\circ\mmexp(\mmy)) = \mmexp(\mmx)\bullet \mmexp(\mmy).
\eeq

\subsection{Symmetric cones}\label{sec:symmetric_cone}

A {\em symmetric cone} $\cc{K}$ can be characterized as the {\em cone of squares} of an EJA $\bJ$, namely $\cc{K} = \{\mmx^2 : \mmx\in \bJ\}$. This is called the {\em Jordan algebraic characterization} of symmetric cones, which provides a convenient way to study symmetric cones.

Let $\cc{K}$ be the cone of squares of an EJA $\bJ$ and let $r$ be its rank. It is easily seen that
\[
    \cc{K} = \big\{\mmx\in \bJ : \lambda_k(\mmx) \geq 0,\ \forall k\in[r]\big\},
\]
namely $\cc{K}$ consists of all the elements in $\bJ$ with nonnegative eigenvalues.
Therefore, for all $\mmx\in \bJ$, its exponential $\mmexp(\mmx)$ is in the cone of squares $\cc{K}$.
The cone $\cc{K}$ induces partial orders in $\bJ$. For all elements $\mmx, \mmy \in \bJ$, we say $\mmx \succeq_\cc{K} \mmy$ (resp.~$\mmx \preceq_{\cc{K}} \mmy$) if and only if $\mmx - \mmy \in \cc{K}$ (resp.~$\mmy - \mmx \in \cc{K}$).
A useful property of symmetric cones is that they are {\em self-dual}. For any symmetric cone $\cc{K}$, we have $\cc{K} = \cc{K}^*$, where $\cc{K}^*$ is the dual cone of $\cc{K}$ defined as:
\[
    \cc{K}^* = \big\{\mmy \in \bJ : \forall \mmx \in \cc{K},\ \mmx \bullet \mmy\ge 0 \big\}.
\]
Based on this property, for any $\mmx, \mmy \in \cc{K}$, we have $\mmx\bullet \mmy \ge 0$. Particularly, for any $\mmx, \mmy \in \bJ$, if $\mmx\succeq_\cc{K}\mmy$ (resp. $\mmx \preceq_\cc{K} \mmy$), then $\mmx\bullet \mmz \ge \mmy \bullet \mmz$ (resp.~$\mmx\bullet \mmz \le \mmy \bullet \mmz$), $\forall \mmz \in \cc{K}$.
Another important property of symmetric cones is {\em homogeneity}. That is, for any $\mmx, \mmy$ in the interior of $\cc{K}$, there exists a linear map that maps $\mmx$ to $\mmy$ and preserves the cone.

Famous examples of symmetric cones include the nonnegative orthant $\bR^n_+$, the cone of $n\times n$ Hermitian positive-semidefinite (PSD) matrices $\cc{S}^n_+$, and the second-order cone $\cc{Q}^{d+1}$.
Let $\cc{K}_1, \dots, \cc{K}_n, n \ge 1$, be $n$ symmetric cones, where $\cc{K}_i \subseteq \bJ_i,\ \forall i\in [n]$. The Cartesian product $\cc{K}_1 \times \dots\times \cc{K}_n$ is a symmetric cone in the vector space $\bJ_1 \times \dots \times \bJ_n$.
In this work, we focus on a symmetric cone that is the Cartesian product of $n$ second-order cones with equal dimensionalities, namely
\beq\label{eq:product_cone_defn}
\cc{C} := \underbrace{\cc{Q}^{d+1}\times \dots \times \cc{Q}^{d+1} }_{n}.
\eeq
In what follows, we provide some algebraic facts on $\cc{Q}^{d+1}$ and the product cone $\cc{C}$ for later reference.

\medskip
{\bf\em Second-order cone.} The second-order cone $\cc{Q}^{d+1}$ is a symmetric cone in the vector space $\bR^d\times \bR$. For any $\mmx = (\bar{\mmx}, x_0)\in \bR^d \times \bR$ and $\mmy = (\bar{\mmy}, y_0) \in \bR^d \times \bR$, the Jordan product $\circ$ is defined as:
\[
    \mmx \circ \mmy := {1\over \sqrt{2}}(x_0\bar{\mmy} + y_0\bar{\mmx},\ \mmx^T \mmy),
\]
and the identity element $\mme = (\mmzero, \sqrt{2}) \in \bR^d \times \bR$. The rank of $\cc{Q}^{d+1}$ is 2, and for an element $\mmx = (\bar{\mmx}, x_0)\in \bR^d \times \bR$, the eigenvalues and the corresponding idempotents are:
\beq\label{eq:soc_decomposition}
\begin{aligned}
     & \lambda_1(\mmx) = {1\over \sqrt{2}}(x_0 + \|\bar{\mmx}\|), \quad \mmq_1 = {1\over \sqrt{2}}(\mmu, 1),   \\
     & \lambda_2(\mmx) = {1\over \sqrt{2}}(x_0 - \|\bar{\mmx}\|), \quad \mmq_2 = {1\over \sqrt{2}}(- \mmu, 1),
\end{aligned}
\eeq
where $\mmu = \frac{\bar{\mmx}}{\|\bar{\mmx}\|}$ if $\bar{\mmx}\neq \mmzero$, otherwise $\mmu$ can be any unit $\ell_2$-norm vector.
From the spectral decomposition, it is easily seen that the second-order cone $\cc{Q}^{d+1}$ can also be characterized as:
\beq\label{eq:soc_characterization}
\cc{Q}^{d+1} := \big\{ \mmx = (\bar{\mmx}, x_0)\in \bR^d\times \bR : \|\bar{\mmx}\| \le x_0 \big\}.
\eeq
The inner product of $\mmx = (\bar{\mmx}, x_0)\in \bR^d \times \bR$ and $\mmy = (\bar{\mmy}, y_0) \in \bR^d \times \bR$ is then given by:
\[
    \mmx \bullet \mmy = \Tr(\mmx \circ \mmy) = \mmx^T \mmy.
\]

{\bf\em Product cone.}
Let $(\bV, \diamond, \bdiamond)$ be the underlying EJA of the product cone $\cc{C}$, where
\[
    \bV := \underbrace{(\bR^d \times \bR) \times \dots \times (\bR^d \times \bR)}_{n}.
\]
Let $\mmx_1 = (\bar{\mmx}_1, x_{1,0}),\ \dots,\ \mmx_n = (\bar{\mmx}_n, x_{n,0})$ be vectors in $\bR^d \times \bR$. We denote the concatenation of $\mmx_1, \dots, \mmx_n$ as
\[
    \cproduct_{i=1}^n \mmx_i := \cproduct_{i=1}^n (\bar{\mmx}_i, x_{i,0}) = (\bar{\mmx}_1, x_{1,0},\ \dots,\ \bar{\mmx}_n, x_{n,0}).
\]
For any $\cproduct_{i=1}^n \mmx_i = \cproduct_{i=1}^n (\bar{\mmx}_i, x_{i,0}) \in \bV$ and $\cproduct_{i=1}^n \mmy_i = \cproduct_{i=1}^n (\bar{\mmy}_i, y_{i,0})\in \bV$, the Jordan product $\diamond$ is given by:
\[
    \big(\cproduct_{i=1}^n \mmx_i\big) \diamond \big(\cproduct_{i=1}^n \mmy_i\big) := \cproduct_{i=1}^n (\mmx_i \circ \mmy_i),
\]
and the identity element $\mme = \cproduct_{i=1}^n (\mmzero, \sqrt{2}) \in \bV$. The rank of $\cc{C}$ is $2n$, and for any $\cproduct_{i=1}^n \mmx_i = \cproduct_{i=1}^n (\bar{\mmx}_i, x_{i,0})\in\bV$, the eigenvalues and the corresponding idempotents are:
\beq\label{eq:product_cone_decomposition}
\begin{aligned}
    \lambda_{2j-1}\big(\cproduct_{i=1}^n \mmx_i\big) & = \lambda_1(\mmx_j), \quad \mmq_{2j-1} = {1\over \sqrt{2}}\big(\underbrace{0, \dots, 0,}_{(j-1)(d+1)} \mmu_j, 1, \underbrace{0, \dots, 0}_{(n-j)(d+1)} \big), \\
    \lambda_{2j}\big(\cproduct_{i=1}^n \mmx_i\big)   & = \lambda_2(\mmx_j), \quad \mmq_{2j} = {1\over \sqrt{2}}\big(\underbrace{0, \dots, 0,}_{(j-1)(d+1)} -\mmu_j, 1, \underbrace{0, \dots, 0}_{(n-j)(d+1)} \big),
\end{aligned}
\eeq
where $\mmu_j = \frac{\bar{\mmx}_j}{\|\bar{\mmx}_j\|}$ if $\bar{\mmx}_j\neq \mmzero$, otherwise $\mmu_j$ can be any unit $\ell_2$-norm vector.
Based on the spectral decomposition, the trace of $\cproduct_{i=1}^n \mmx_i = \cproduct_{i=1}^n (\bar{\mmx}_i, x_{i,0})\in\bV$ equals the sum of the traces of its components, namely, $\Tr\big(\cproduct_{i=1}^n \mmx_i\big) = \sum_{i=1}^n \Tr(\mmx_i)$, and its inner product with another element $\cproduct_{i=1}^n \mmy_i = \cproduct_{i=1}^n (\bar{\mmy}_i, y_{i,0})\in \bV$ is given by:
\[
    \big(\cproduct_{i=1}^n \mmx_i\big)\bdiamond \big(\cproduct_{i=1}^n \mmy_i\big) = \sum_{i=1}^n (\mmx_i \bullet \mmy_i) = \sum_{i=1}^n \mmx_i^T \mmy_i.
\]
The exponential of $\cproduct_{i=1}^n \mmx_i = \cproduct_{i=1}^n (\bar{\mmx}_i, x_{i,0})\in \bV$ is the concatenation of the exponentials of its components:
\beq\label{eq:product_cone_exp}
\mmexp\big(\cproduct_{i=1}^n \mmx_i\big) = \cproduct_{i=1}^n \mmexp(\mmx_i).
\eeq

\section{Symmetric cone games}\label{sec:game}

In this section, we introduce SCG and present a method to solve them approximately, which form the basis for designing our algorithms for the SIB problems.

Let $(\bJ, \circ, \bullet)$ be a EJA and let $\cc{K}$ be the associated symmetric cone.
Let $\bU$ be a finite-dimensional vector space and let $\brmf: \bU \rightarrow \bJ$ be an affine transformation that maps a vector $\mmx \in \bU$ to $\bJ$.
Let $\cc{A}$ be a compact convex subset of $\bU$ and let $\cc{B}$ be the {\em spectraplex} of $\cc{K}$, namely
$\cc{B} := \big\{\mmy\in \cc{K} : \Tr(\mmy) = 1 \big\}$.
Consider a zero-sum game between two players Alice and Bob as follows: Alice plays a point $\mmx\in \cc{A}$, and Bob plays a point $\mmy\in \cc{B}$. For a pair of points $(\mmx, \mmy)\in \cc{A}\times \cc{B}$ played by Alice and Bob, the payoff gained by Bob (and the loss Alice incurs) is given by $\brmf(\mmx)\bullet \mmy$. It can be formulated as the following saddle point problem:
\beq\label{eq:scg}
    \min_{\mmx \in \cc{A}}\ \max_{\mmy \in \cc{B}}\ \brmf(\mmx) \bullet \mmy.
\eeq
Since the function $\brmf(\mmx)\bullet \mmy$ is bilinear, and $\cc{A}$ and $\cc{B}$ are compact convex, a generalized version of von Neumann's minimax theorem~\cite{minimax:Sion1958} implies:
\[
    \min_{\mmx\in \cc{A}} \max_{\mmy\in \cc{B}}\ \brmf(\mmx) \bullet \mmy = \max_{\mmy\in \cc{B}} \min_{\mmx\in \cc{A}}\ \brmf(\mmx) \bullet \mmy = \lambda^*,
\]
where the common value $\lambda^*$ is called the {\em value} of this game.
Let $\eps > 0$ be an error parameter. 
We say $(\tilde{\mmx}, \tilde{\mmy}) \in \cc{A} \times \cc{B}$ is a $\eps$-Nash equilibrium of the SCG if
\beq\label{eq:game_condition}
\begin{aligned}
    \max_{\mmy\in \cc{B}}\ \brmf(\tilde{\mmx})\bullet \mmy \le \lambda^* + \eps
    \hspace{1em} \text{and} \hspace{1em}
    \min_{\mmx\in \cc{A}}\ \brmf(\mmx)\bullet \tilde{\mmy} \ge \lambda^* - \eps.
\end{aligned}
\eeq

Next, we provide an algorithm for computing such a $\eps$-Nash equilibirum.
We assume that for any given $\mmy\in \cc{B}$, there is an efficient algorithm which provides the best response for Alice, i.e.~the algorithm outputs a point $\mmx\in \cc{A}$ that minimizes $\brmf(\mmx)\bullet\mmy$. We call this algorithm the {\sc Oracle}. Moreover, we assume that for all $\mmy \in \cc{B}$, the {\sc Oracle} response $\mmx\in \cc{A}$ always satisfies $\|\brmf(\mmx)\|_\spe\le \rho$, \change{where $\|\cdot\|_\spe$ is the spectral norm (recall \eqref{eq:spectral_norm}).} The quantity $\rho$ is designated as the {\em width} of the {\sc Oracle}.
Let $\mme$ be the identity and $r$ be the rank of the EJA. Let $T \in \bZ_+$ be an iteration count, and $\eta > 0$ be a step size. Consider the following iterative algorithm.

\begin{algorithm}
\caption{Approximation algorithm for SCG}\label{algo:scg}
\begin{algorithmic}[1]
\STATE Let $\xt{\mmy}{1} = {\mme \over r}$
\FOR{$t=1,\dots, T$}
\STATE Find $\xt{\mmx}{t} \in \displaystyle\argmin_{\mmx\in \cc{A}} \ \brmf(\mmx) \bullet \xt{\mmy}{t}$ using the {\sc Oracle} \label{algo:scg_oracle}
\STATE Let $\xt{\mmw}{t+1} = \mmexp(\frac{\eta}{\rho}\sum_{\tau = 1}^{t} \brmf(\xt{\mmx}{\tau}))$ \label{algo:scg_wt}
\STATE Let $\xt{\mmy}{t+1} = {\xt{\mmw}{t+1} / \Tr(\xt{\mmw}{t+1})}$ \label{algo:scg_yt}
\ENDFOR
\RETURN $\tilde{\mmx} = {1\over T} \sum_{t=1}^T \xt{\mmx}{t}$ and $\tilde{\mmy} = {1\over T}\sum_{t=1}^T \xt{\mmy}{t}$
\end{algorithmic}
\end{algorithm}

\begin{theorem}\label{thm:zero-sum_game}
Let $\eps\in (0, 2\rho]$ be the error, and set the parameters $T = \lceil\frac{4\rho^2 \ln r}{\eps^2}\rceil$ and $\eta = \sqrt{\ln r \over T}$. Then the output $(\tilde{\mmx}, \tilde{\mmy})$ of Algorithm~\ref{algo:scg} is a $\eps$-Nash equilibrium of the SCG.
\end{theorem}

\begin{proof}
It is easily seen that every $\xt{\mmy}{t}$ (including $\xt{\mmy}{1}$ and those computed in Line~\ref{algo:scg_yt} of the algorithm) is in $\cc{B}$ because its eigenvalues are nonnegative (which means it is in the cone $\cc{K}$) and its trace is 1.
Let $\tilde{\mmx} = \frac{1}{T}\sum_{t=1}^T \xt{\mmx}{t}$ and $\tilde{\mmy} = {1\over T} \sum_{t=1}^T \xt{\mmy}{t}$.
By convexity of the sets $\cc{A}$ and $\cc{B}$, we know that $\tilde{\mmx}\in\cc{A}$ and $\tilde{\mmy}\in\cc{B}$.
Furthermore, we claim that:
\beq\label{eq:game_theorem}
\lambda^* - \eps\
\overset{(i)}{\le}\
\max_{\mmy\in\cc{B}}\ \brmf(\tilde{\mmx})\bullet \mmy - \eps\
\overset{(ii)}{\le}\
{1\over T}\sum_{t=1}^T \brmf(\xt{\mmx}{t}) \bullet \xt{\mmy}{t} \
\overset{(iii)}{\le}\
\min_{\mmx\in\cc{A}}\ \brmf(\mmx)\bullet\tilde{\mmy}\
\overset{(iv)}{\le}\
\lambda^*.
\eeq
Thus $(\tilde{\mmx}, \tilde{\mmy})$ is a pair of points that satisfies~\eqref{eq:game_condition}.
The inequalities $(i)$ and $(iv)$ are because
\[\begin{aligned}        
    \max_{\mmy\in\cc{B}}\ \brmf(\tilde{\mmx}) \bullet \mmy
    \ge \min_{\mmx\in\cc{A}} \max_{\mmy\in\cc{B}}\ \brmf(\mmx)\bullet \mmy
    = \lambda^*
    \quad\text{and}\quad
    \min_{\mmx\in\cc{A}}\ \brmf(\mmx)\bullet \tilde{\mmy}
    \le \max_{\mmy\in\cc{B}} \min_{\mmx\in\cc{A}}\ \brmf(\mmx)\bullet \mmy
    = \lambda^*.
\end{aligned}
\]
Let $\mmx^*$ be any point in $\cc{A}$. Since $\xt{\mmx}{t}$ is the best response in each round, we know that
\[
    \brmf(\mmx^*)\bullet \tilde{\mmy} = {1\over T}\sum_{t=1}^T \brmf(\mmx^*) \bullet \xt{\mmy}{t} \ge {1\over T}\sum_{t=1}^T \brmf(\xt{\mmx}{t})\bullet \xt{\mmy}{t}.
\]
Let $\mmx^* \in \cc{A}$ be a minimizer of $\brmf(\mmx)\bullet\tilde{\mmy}$ in the above, we obtain inequality $(iii)$. In what follows, we justify the correctness of inequality $(ii)$ thus the proof of~\eqref{eq:game_theorem} is completed.

Let $\xt{\mmw}{1} = \mme$ and $\xt{\mmw}{t}$ as in Line~\ref{algo:scg_wt} for $t = 2, \dots, T + 1$.
By the generalized Golden-Thompson inequality~\eqref{eq:golden-thompson}, we have:
\[\begin{aligned}
        \Tr(\xt{\mmw}{T+1}) & = \Tr\Big(\mmexp\big({\eta\over \rho}\sum_{t=1}^{T}\brmf(\xt{\mmx}{t})\big)\Big) \\
                            & \le
        \Tr\Big(\mmexp\big({\eta\over \rho}\sum_{t=1}^{T-1}\brmf(\xt{\mmx}{t})\big)
        \circ \mmexp\big({\eta\over \rho}\brmf(\xt{\mmx}{T})\big)\Big)                                      \\
        & = \xt{\mmw}{T}\bullet \mmexp\big({\eta\over \rho}\brmf(\xt{\mmx}{T})\big)      \\
    \end{aligned}\]
Let $\sum_{k=1}^{r} \lambda_k\big({\eta \over \rho}\brmf(\xt{\mmx}{T})\big)\mmq_k$ be the spectral decomposition of ${\eta\over \rho}\brmf(\xt{\mmx}{T})$.
Since $\eta \le 1$, and by our assumption $\|\brmf(\xt{\mmx}{T})\|_\spe\le \rho$,
we have $\lambda_k\big({\eta \over \rho}\brmf(\xt{\mmx}{T})\big) \in [-1, 1],\ \forall k \in [r]$. Then,
\beq\label{eq:game_exponential_upperbound}
\begin{aligned}
    \mmexp\big({\eta \over \rho}\brmf(\xt{\mmx}{T})\big)
     & = \sum_{k=1}^{r} \exp\Big(\lambda_k\big({\eta \over \rho}\brmf(\xt{\mmx}{T})\big)\Big)\mmq_k \\
     & \overset{(a)}{\preceq_{\cc{K}} }\ \sum_{k=1}^{r} \Big(1 + \lambda_k\big({\eta \over \rho}\brmf(\xt{\mmx}{T})\big) + \lambda_k^2\big({\eta \over \rho}\brmf(\xt{\mmx}{T})\big)\Big)\mmq_k \\
     & = \sum_{k=1}^{r} \mmq_k + {\eta \over \rho} \sum_{k=1}^{r} \lambda_k\big(\brmf(\xt{\mmx}{T})\big) \mmq_k + {\eta^2 \over \rho^2} \sum_{k=1}^{r} \lambda^2_k\big(\brmf(\xt{\mmx}{T})\big)\mmq_k \\
     & = \mme + {\eta\over \rho} \brmf(\xt{\mmx}{T}) + {\eta^2\over \rho^2} \big(\brmf(\xt{\mmx}{T})\big)^2,
\end{aligned}
\eeq
where the generalized inequality $(a)$ is because $e^x \le 1 + x + x^2,\ \forall x \in [-1,1]$. Consequently,
\[\begin{aligned}
        \Tr(\xt{\mmw}{T+1})
         & \le \xt{\mmw}{T}\bullet \mmexp\big({\eta\over \rho}\brmf(\xt{\mmx}{T})\big)                                                                                                               \\
         & \overset{(a)}{\le} \xt{\mmw}{T}\bullet \Big(\mme + {\eta\over \rho}\brmf(\xt{\mmx}{T}) + {\eta^2\over \rho^2}\big(\brmf(\xt{\mmx}{T})\big)^2\Big)                                           \\
         & = \Tr(\xt{\mmw}{T}) + {\eta\over \rho}\brmf(\xt{\mmx}{T})\bullet\xt{\mmw}{T} + {\eta^2\over \rho^2}\big(\brmf(\xt{\mmx}{T})\big)^2\bullet\xt{\mmw}{T}                                    \\
         & \overset{(b)}{=} \Tr(\xt{\mmw}{T})\cdot\Big(1 + {\eta\over \rho}\brmf(\xt{\mmx}{T})\bullet\xt{\mmy}{T} + {\eta^2\over \rho^2}\big(\brmf(\xt{\mmx}{T})\big)^2\bullet\xt{\mmy}{T}\Big)     \\
         & \overset{(c)}{\le} \Tr(\xt{\mmw}{T})\cdot\exp\Big({\eta\over \rho}\brmf(\xt{\mmx}{T})\bullet\xt{\mmy}{T} + {\eta^2\over \rho^2}\big(\brmf(\xt{\mmx}{T})\big)^2\bullet\xt{\mmy}{T}\Big)
    \end{aligned}\]
Here $(a)$ follows from~\eqref{eq:game_exponential_upperbound} and the self-duality of $\cc{K}$; $(b)$ uses the fact that $\xt{\mmy}{T} = \frac{\xt{\mmw}{T}}{\Tr(\xt{\mmw}{T})}$; and $(c)$ is because $1+x\le e^{x},\ \forall x\in \bR$.
Since $\Tr(\xt{\mmw}{1}) = r$, by induction, we have
\[
    \Tr(\xt{\mmw}{T+1}) \le r\cdot\exp\Big(\sum_{t=1}^T {\eta\over \rho} \brmf(\xt{\mmx}{t})\bullet \xt{\mmy}{t} + \sum_{t=1}^T {\eta^2\over \rho^2}\big(\brmf(\xt{\mmx}{t})\big)^2\bullet\xt{\mmy}{t}\Big),
\]
Moreover, since $\|{1\over \rho}\brmf(\xt{\mmx}{t})\|_\spe\le 1$, we also have $\|{1\over \rho^2}\big(\brmf(\xt{\mmx}{t})\big)^2\|_\spe\le 1$. Together with the fact that $\|\xt{\mmy}{t}\|_\spe \le 1$, we obtain an upperbound on $\Tr(\xt{\mmw}{T+1})$:
\beq\label{eq:potential_upperbound}
\Tr(\xt{\mmw}{T+1}) \le r\cdot\exp\Big({\eta\over \rho} \sum_{t=1}^T \brmf(\xt{\mmx}{t}) \bullet \xt{\mmy}{t} + \eta^2 T\Big).
\eeq

On the other hand, we can also establish a lowerbound on $\Tr(\xt{\mmw}{T+1})$. Let $\lambda_{\max}(\cdot)$ be the largest eigenvalue. Then:
\beq\label{eq:potential_lowerbound}
\begin{aligned}
    \Tr(\xt{\mmw}{T+1}) & = \sum_{k=1}^{r} \lambda_k\Big(\mmexp\big({\eta\over \rho}\sum_{t=1}^{T}\brmf(\xt{\mmx}{t})\big)\Big) \ge \lambda_{\max}\Big(\mmexp\big({\eta\over \rho}\sum_{t=1}^{T}\brmf(\xt{\mmx}{t})\big)\Big) \\
    & = \exp\Big(\lambda_{\max}\big({\eta\over \rho}\sum_{t=1}^{T}\brmf(\xt{\mmx}{t})\big)\Big)
    = \exp\Big({\eta\over \rho}\lambda_{\max}\big(\sum_{t=1}^{T}\brmf(\xt{\mmx}{t})\big)\Big).
\end{aligned}
\eeq

Combine the upperbound~\eqref{eq:potential_upperbound} and the lowerbound~\eqref{eq:potential_lowerbound}, we have:
\[
    r\cdot\exp\Big({\eta\over \rho} \sum_{t=1}^T \brmf(\xt{\mmx}{t}) \bullet \xt{\mmy}{t} + \eta^2 T\Big) \ge \exp\Big({\eta\over \rho}\lambda_{\max}\big(\sum_{t=1}^{T}\brmf(\xt{\mmx}{t})\big)\Big).
\]
Taking logarithmic on both sides and scaling by $\rho\over \eta T$ gives:
\[
    {\rho\ln(r)\over \eta T} + {1\over T} \sum_{t=1}^T \brmf(\xt{\mmx}{t}) \bullet \xt{\mmy}{t} + \eta \rho \ge {1\over T}\lambda_{\max}\big(\sum_{t=1}^{T}\brmf(\xt{\mmx}{t})\big) = \lambda_{\max}\big(\brmf(\tilde{\mmx})\big).
\]
Substitute $T=\lceil{4\rho^2\ln(r)\over \eps^2}\rceil$ and $\eta = \sqrt{\ln(r)\over T}$ in the above, and rearrange the terms, we have:
\beq\label{eq:zero-sum_game_eigenvalue}
{1\over T} \sum_{t=1}^T \brmf(\xt{\mmx}{t})\bullet \xt{\mmy}{t} \ge \lambda_{\max}\big(\brmf(\tilde{\mmx})\big) - \eps.
\eeq

To show that inequality $(ii)$ in~\eqref{eq:game_theorem} is true, it remains to prove that~$\lambda_{\max}\big(\brmf(\tilde{\mmx})\big) \ge \max_{\mmy\in\cc{B}}\ \brmf(\tilde{\mmx})\bullet\mmy$.
Let $\mmy^*$ be any point in the set $\cc{B}$, and let $\sum_{k=1}^{r} \lambda_k\big(\brmf(\tilde{\mmx})\big) \mmq_k$ be the spectral decomposition of $\brmf(\tilde{\mmx})$. Then,
\[\begin{aligned}
        \brmf(\tilde{\mmx})\bullet \mmy^*
         & = \Tr\big(\brmf(\tilde{\mmx})\circ \mmy^*\big)
        = \Tr\Big(\sum_{k=1}^{r}\lambda_{k}\big(\brmf(\tilde{\mmx})\big)\mmq_k \circ \mmy^* \Big)
        = \sum_{k=1}^{r}\lambda_{k}\big(\brmf(\tilde{\mmx})\big)\cdot \Tr(\mmq_k\circ \mmy^*)     \\
        & \le \lambda_{\max}\big(\brmf(\tilde{\mmx})\big) \sum_{k=1}^{r} \Tr(\mmq_k\circ \mmy^*)
        = \lambda_{\max}\big(\brmf(\tilde{\mmx})\big) \cdot \Tr(\mme\circ \mmy^*)
        = \lambda_{\max}\big(\brmf(\tilde{\mmx})\big).
    \end{aligned}
\]
Replacing $\mmy^*$ by the maximizer of $\brmf(\tilde{\mmx})\bullet\mmy$ in the above completes the proof.
\end{proof}

\subsection{SCG over the product cone}

We now analyze the details of the algorithm for SCG over the specific cone $\cc{C}$ associated with the EJA $(\bV, \diamond, \bdiamond)$ (recall the definitions in Section~\ref{sec:symmetric_cone}).
Let $\brmg_1, \dots, \brmg_n$ be affine transformations mapping vector from $\bU$ to $\bR^d$, and let ${\rm h}_1, \dots, {\rm h}_n$ be real-valued affine functions mapping $\bU$ to $\bR$. Then any affine map $\brmf : \bU \rightarrow \bV$ can be written as $\brmf(\mmx) = \cproduct_{i=1}^n (\brmg_i(\mmx), \rmh_i(\mmx))$. 
Consequently, letting $\mmy = \cproduct_{i=1}^n (\bar{\mmy}_i, y_{i,0})$, the SCG over $\cc{C}$ can be re-formulated as:
\beq\label{eq:scg_product}
\begin{aligned}
    &\min_{\mmx \in \cc{A}}\ \max_{\mmy \in \cc{B}}\ 
    \Big(\underbrace{
    \cproduct_{i=1}^n
    \begin{pmatrix}
        \brmg_i(\mmx) \\
        \rmh_i(\mmx)        
    \end{pmatrix}}_{\brmf(\mmx)}
    \Big) 
    \bdiamond 
    \Big(\underbrace{
    \cproduct_{i=1}^n \begin{pmatrix}
        \bar{\mmy}_i \\
        y_{i,0}
    \end{pmatrix}}_{\mmy}
    \Big) \\
    =\
    &\min_{\mmx \in \cc{A}} \ \max_{\mmy \in \cc{B}}\ 
    \sum_{i=1}^n \big(
        \la \brmg_i(\mmx), \bar{\mmy}_i \ra + \rmh_i(\mmx)\cdot y_{i,0}
    \big).
\end{aligned}
\eeq

To facilitate both the running time analysis and the accessibility of the algorithm for the specific SCG
we express the computational procedures as Algorithm~\ref{algo:scg_product} using elementary arithmetic operations rather than the EJA notation.

\begin{algorithm}
\caption{Approximation algorithm for SCG over $\cc{C}$}\label{algo:scg_product}
\begin{algorithmic}[1]
\STATE Let $T = \lceil {4\rho^2 \ln (2n) \over \eps^2} \rceil$ and $\eta = \sqrt{\ln(2n) \over T}$
\STATE Let $\xt{\mm{\alpha}}{0}_i = \mmzero$ and $\xt{\beta}{0}_i = 0$ for all $i \in [n]$
\STATE Let $\xt{\bar{\mmy}}{1}_i = \mmzero$ and $\xt{y}{1}_{i,0} = {1\over \sqrt{2} n}$ for all $i \in [n]$
\FOR{$t = 1, \dots, T$}
\STATE Find $\xt{\mmx}{t}\in \displaystyle\argmin_{\mmx \in \cc{A}}\ \sum_{i=1}^n \big( \la \brmg_i(\mmx), \xt{\bar{\mmy}}{t}_i \ra + \rmh_i(\mmx)\cdot \xt{y}{t}_{i,0} \big)$ using the {\sc Oracle} \label{algo:scg_product_oracle}
\FOR{$i = 1, \dots, n$}
\STATE Let $\xt{\mm{\alpha}}{t}_i = \xt{\mm{\alpha}}{t-1}_i + \brmg_i(\xt{\mmx}{t})$ and $\xt{\beta}{t}_i = \xt{\beta}{t-1}_i + \rmh_i(\xt{\mmx}{t})$ \label{algo:scg_product_map}
\STATE Let $\xt{\mmu}{t}_i = \xt{\mm{\alpha}}{t}_i / \|\xt{\mm{\alpha}}{t}_i\|$ {\bf if} $\|\xt{\mm{\alpha}}{t}_i\| > 0$, {\bf else} let $\xt{\mmu}{t}_i$ be any unit vector
\STATE Let $\xt{\mu}{t}_{i} = \exp\big({\eta \over \sqrt{2} \rho} (\xt{\beta}{t}_i + \|\xt{\mm{\alpha}}{t}_i\|)\big) + \exp\big({\eta \over \sqrt{2} \rho} (\xt{\beta}{t}_i - \|\xt{\mm{\alpha}}{t}_i\|)\big)$
\STATE Let $\xt{\lambda}{t}_{i} = \exp\big({\eta \over \sqrt{2} \rho} (\xt{\beta}{t}_i + \|\xt{\mm{\alpha}}{t}_i\|)\big) - \exp\big({\eta \over \sqrt{2} \rho} (\xt{\beta}{t}_i - \|\xt{\mm{\alpha}}{t}_i\|)\big)$
\ENDFOR
\FOR{$i = 1,\dots, n$}
\STATE Let $\xt{\bar{\mmy}}{t+1}_i = {\xt{\lambda}{t}_i \over \sqrt{2}\sum_i \xt{\mu}{t}_i}\cdot\xt{\mmu}{t}_i$ and $\xt{y}{t+1}_{i,0} = {\xt{\mu}{t}_i \over \sqrt{2} \sum_i \xt{\mu}{t}_i}$
\ENDFOR
\ENDFOR
\RETURN $\tilde{\mmx} = {1\over T} \sum_{t=1}^T \xt{\mmx}{t}$ and $\tilde{\mmy} = {1\over T} \sum_{t=1}^T \big( \cproduct_{i=1}^n (\xt{\bar{\mmy}}{t}_i, \xt{y}{t}_{i,0})\big)$
\end{algorithmic}
\end{algorithm}

\begin{theorem}\label{thm:scg_product}
Suppose the affine map $\brmf : \bU \rightarrow \bV$ can be computed in $O(F)$ time.
Given an error parameter $\eps \in (0, 2\rho]$, Algorithm~\ref{algo:scg_product} computes a $\eps$-Nash equilibrium of the SCG~\eqref{eq:scg_product} using $O({\rho^2\log n \over \eps^2})$ calls to the {\sc Oracle}, with an additional processing time of $O(F + nd)$ per call.
\end{theorem}

\begin{proof}
The number of iterations (also number of calls to the {\sc Oracle}) $T$ is due to Theorem~\ref{thm:zero-sum_game} and the fact that the rank of the underlying EJA of $\cc{C}$ is $2n$. In each iteration, besides the procedure of the {\sc Oracle} (in Line~\ref{algo:scg_product_oracle} of the Algorithm) and the computation of the affine map $\brmf$ (which is decomposed into $\brmg_i$ and $\rmh_i$ in Line~\ref{algo:scg_product_map}), it only remains doing some simple arithmetic operations such as vector addition and scaling, which can be finished in $O(nd)$ time. Therefore, the processing time between two consective {\sc Oracle} calls is $O(F + nd)$.
\end{proof}

\subsection{Approximate {\sc Oracle}s}

The algorithms discussed above also work with approximate {\sc Oracle}s. We discuss this for the algorithm in the general setting (Algorithm~\ref{algo:scg}); the algorithm for the specific product cone $\cc{C}$ (Algorithm~\ref{algo:scg_product}) is analogous.

Define a $\delta$-approximate {\sc Oracle} to be an algorithm that solves the linear optimization problem $\min_{\mmx\in\cc{A}} \brmf(\mmx)\bullet \mmy$ up to an additive error $\delta$. That is, for any given vector $\mmy \in \cc{B}$, it finds $\bar{\mmx} \in \cc{A}$ such that $\brmf(\bar{\mmx})\bullet \mmy \le \min_{\mmx\in \cc{A}} \brmf(\mmx) \bullet \mmy + \delta$. Then we can replace Line~\ref{algo:scg_oracle} of Algorithm~\ref{algo:scg} with a $\delta$-approximate {\sc Oracle}.
We have the following theorem.

\begin{theorem}
Suppose Algorithm~\ref{algo:scg} is run with a $\delta$-approximate {\sc Oracle} for $T = \lceil {4\rho^2 \ln r \over \eps^2} \rceil$ iterations with $\eta = \sqrt{\ln r \over T}$. Then the output $(\tilde{\mmx}, \tilde{\mmy})$ is a $(\eps + \delta)$-Nash equilibrium of the SCG.
\end{theorem}

\begin{proof}
We show that running Algorithm~\ref{algo:scg} with a $\delta$-approximate {\sc Oracle} only introduces an additive error $\delta$ to the right-hand side of inequality $(iii)$ in \eqref{eq:game_theorem}, and everything else remains the same as in Theorem~\ref{thm:zero-sum_game}. That is, \eqref{eq:game_theorem} now becomes
\[
\lambda^* - \eps\
\overset{(i)}{\le}\
\max_{\mmy\in\cc{B}}\ \brmf(\tilde{\mmx})\bullet \mmy - \eps\
\overset{(ii)}{\le}\
{1\over T}\sum_{t=1}^T \brmf(\xt{\mmx}{t}) \bullet \xt{\mmy}{t} \
\overset{(iii)}{\le}\
\min_{\mmx\in\cc{A}}\ \brmf(\mmx)\bullet\tilde{\mmy} + \delta\
\overset{(iv)}{\le}\
\lambda^* + \delta,
\]
and rearranging the terms shows that $(\tilde{\mmx}, \tilde{\mmy})$ is a $(\eps +\delta)$-Nash equilibirum.

Let $\mmx^*$ be any point in $\cc{A}$, and let $\xt{\mmx}{t}_{\rm best}\in \cc{A}$ be a minimizer of $\brmf(\mmx) \bullet \xt{\mmy}{t}$ for $t = 1, \dots, T$. 
Then we have
\[
    \brmf(\mmx^*)\bullet \tilde{\mmy} = {1\over T}\sum_{t=1}^T \brmf(\mmx^*) \bullet \xt{\mmy}{t} \ge {1\over T}\sum_{t=1}^T \brmf(\xt{\mmx}{t}_{\rm best})\bullet \xt{\mmy}{t}.
\]
Substituting the above $\mmx^* \in \cc{A}$ by a minimizer of $\brmf(\mmx)\bullet\tilde{\mmy}$, we have
\beq\label{eq:approx_oracle_1}
    \min_{\mmx \in \cc{A}} \brmf(\mmx)\bullet \tilde{\mmy} \ge {1\over T} \sum_{t=1}^T \brmf(\xt{\mmx}{t}_{\rm best}) \bullet \xt{\mmy}{t}.
\eeq
By definition of $\delta$-approximate {\sc Oracle}, we know that every $\xt{\mmx}{t}$ it produces satisfies 
\[
\brmf(\xt{\mmx}{t}) \bullet \xt{\mmy}{t} - \delta \le \brmf(\xt{\mmx}{t}_{\rm best}) \bullet \xt{\mmy}{t}.
\]
Therefore, summing up the above over $t= 1, \dots, T$ and taking the average give
\beq\label{eq:approx_oracle_2}
{1\over T} \sum_{t=1}^T \brmf(\xt{\mmx}{t}) \bullet \xt{\mmy}{t} - \delta \le {1\over T} \sum_{t=1}^T \brmf(\xt{\mmx}{t}_{\rm best}) \bullet \xt{\mmy}{t}.
\eeq
Combining the inequalities \eqref{eq:approx_oracle_1} and \eqref{eq:approx_oracle_2} gives the desired result.
\end{proof}

\section{Smallest intersecting balls}\label{sec:sib}

In this section, we introduce the approximation algorithm for the SIB problem. The section is organized as follows: We first introduce the formulation of the SIB problem and analyse its properties. Then we show that the optimization problem can be recasted as a symmetric cone game. 
Next, we analyse the algorithm for computing an approximate Nash equilibrium of the SCG (which corresponds to an approximate solution of SIB).
At the end of the section, we provide details of the algorithm's implementation for specific examples of input objects.

Let $\Omega_i, i = 1,\dots, n, n > 1,$ be nonempty compact convex sets in the Euclidean space $\bR^d$. The SIB problem is to find a ball $B(\mmz, r)$ with the smallest radius $r\ge 0$ such that $B(\mmz, r)\cap \Omega_i \ne \varnothing, \ \forall i \in [n]$.
It can be modeled as the following optimization~problem.

\begin{problem}[Smallest intersecting ball]
\label{prob:sib_general_problem}
\[
\begin{aligned}
    \underset{\mmz, \mmv_1, \dots, \mmv_n, r}{\rm minimize} \quad& r\\
    {\rm subject\ to} \quad& \|\mmz - \mmv_i\| \le r, \ \forall i \in [n],\\
    & \mmv_i \in \Omega_i,\ \forall i \in [n].
\end{aligned}
\]
\end{problem}

\begin{proposition}\label{prop:sib_general}
Let $\Omega_i, i = 1,\dots,n, n > 1,$ be compact convex sets in $\bR^d$. Then,
\begin{enumerate}
\item[(i)] $(\mmz, \mmv_1, \dots, \mmv_n, r)$ is an optimal solution of Problem~\ref{prob:sib_general_problem} if and only if $B(\mmz, r)$ is a smallest intersecting ball satisfies $B(\mmz, r) \cap \Omega_i \ne \varnothing,\ \forall i \in [n]$.
\item[(ii)] $(\mmz, \mmv_1, \dots, \mmv_n, r)$ is an optimal solution of Problem~\ref{prob:sib_general_problem} only if $\mmz \in \conv(\{\Omega_i\}_{i=1}^n)$.
\item[(iii)] Problem~\ref{prob:sib_general_problem} is a convex program and a solution always exists.
\item[(iv)] In general, the solution of Problem~\ref{prob:sib_general_problem} may not be unique.
\end{enumerate}
\end{proposition}

\begin{proof}
$(i)$
Let $(\mmz, \mmv_1, \dots, \mmv_n, r)$ be an optimal solution of Problem~\ref{prob:sib_general_problem}, so $r$ is the optimal value. Then the ball $B(\mmz, r)$ intersects every $\Omega_i$ with radius $r$.
Assume there exists a ball $B(\mmz', r')$ with a smaller radius $r' < r$ that also intersects every $\Omega_i$.
For $i = 1,\dots, n$, let $\mmv_i'$ be a point in $B(\mmz', r') \cap \Omega_i$, so $\|\mmz' - \mmv_i'\| \le r'$ and $\mmv_i' \in \Omega_i$.
Then $(\mmz', \mmv_1', \dots, \mmv_n', r')$ is a feasible solution of the Problem~\ref{prob:sib_general_problem} and its objective value $r'$ is smaller than $r$, which is a contradictory result.

We now justify the converse. 
Let $B(\mmz, r)$ be a smallest intersecting ball, so $B(\mmz, r)\cap \Omega_i \ne \varnothing,\forall i\in [n]$, and $r$ is the smallest radius.
For $i=1,\dots, n$, let $\mmv_i$ be a point in $B(\mmz, r)\cap \Omega_i$. Then $(\mmz, \mmv_1, \dots, \mmv_n, r)$ is a feasible solution of Problem~\ref{prob:sib_general_problem} and its objective value is $r$. 
Assume that Problem~\ref{prob:sib_general_problem} has a solution $(\mmz', \mmq_1', \dots, \mmq_n', r')$ with smaller objective value $r' < r$. Then $B(\mmz', r')$ intersects every $\Omega_i$ and its radius is smaller than $B(\mmz, r)$, which is a contradiction.

\medskip
$(ii)$
Let $\cc{D}$ be the convex hull $\conv(\{\Omega_i\}_{i=1}^n)$.
Assume there exists a solution $(\mmz, \mmv_1, \dots, \mmv_n, r)$ such that $\mmz\notin \cc{D}$. Let $\mmz' = \Pi(\mmz, \cc{D})$ be the Euclidean projection from $\mmz$ to $\cc{D}$. We have $\|\mmz - \mmz'\| > 0$. For every $i\in [n]$,
\[
    r^2 \ge \|\mmv_i - \mmz\|^2 = \|\mmv_i - \mmz'\|^2 + \|\mmz' - \mmz\|^2 + 2\inprod{\mmv_i - \mmz'}{\mmz' - \mmz}.
\]
From the projection theorem~\cite[Theorem 9.8]{projtheory:Beck2014}, we know $\inprod{\mmv_i - \mmz'}{\mmz' - \mmz} \ge 0,\ \forall i \in [n]$. Therefore, $\|\mmv_i - \mmz\|^2 > \|\mmv_i - \mmz'\|^2,\ \forall i \in [n]$. Let $r' = \max_{i\in [n]} \|\mmv_i - \mmz'\|$. Then $r' < r$, and $(\mmz', \mmv_1, \dots, \mmv_n, r')$ is a better solution.

\medskip
$(iii)$
Since the objective function is linear and $\Omega_i$ are closed convex sets,
to see that the problem is a convex problem, it suffices to show the constraint functions $\|\mmz - \mmv_i\| - r$ are convex.
Let $g(\mmz, \mmv, r) = \|\mmz - \mmv\| - r$. Then $g$ is linear w.r.t.~$r$. Since the $\ell_2$ distance function $\|\mmz - \mmv\|$ is convex, $g$ is a convex function. Consequently, every constraint $g(\mmz, \mmv_i, r) \le 0$ generates a convex region.

From $(ii)$ we know that introducing an additional constraint $\mmz \in \conv(\{\Omega_i\}_{i=1}^n)$ will not change the optimal set. 
Moreover, let $\bar{\mmz}$ be any point in the convex hull $\conv(\{\Omega_i\}_{i=1}^n)$ and $D$ be the diameter of the convex hull, then the ball $B(\bar{\mmz}, D)$ intersects every input object. Therefore, the optimal radius $r^*$ satisfies $0 \le r^* \le D$. Consequently, 
Problem~\ref{prob:sib_general_problem} is equivalent to:
\[\begin{aligned}
    \underset{\mmz, \mmv_1, \dots, \mmv_n, r}{\rm minimize} \quad& r\\
    {\rm subject\ to} \quad& \|\mmz - \mmv_i\| \le r, \ \forall i \in [n],\\
    & \mmv_i \in \Omega_i,\ \forall i \in [n],\\
    & \mmz \in \conv(\{\Omega_i\}_{i=1}^n),\\
    & 0 \le r \le D.
\end{aligned}\]
Since the objective function of the above problem is continuous and the feasible region is closed and bounded, by the Weierstrass theorem~\cite[Theorem 2.30]{projtheory:Beck2014}, an optimal solution must exist.

\medskip
$(iv)$
Let $n = 2$ and $d = 2$. Consider the problem    generated by the convex sets
\[\begin{aligned}
    \Omega_1 &= \{(x,y)\in \bR^2 : 0 \le x \le 1,\ y = 0 \},\\
    \Omega_2 &= \{(x,y)\in \bR^2 : 0 \le x \le 1,\ y = 1 \}.
\end{aligned}\]
Then for any $x \in [0,1]$, $\mmz = (x, {1\over 2}), \mmv_1 = (x, 0), \mmv_2 = (x, 1), r = {1\over 2}$ is a solution of this problem.
\end{proof}

Throughout this work, we assume $\medcap_{i=1}^n \Omega_i = \varnothing$. As a consequence, the optimal value $r^*$ of Problem~\ref{prob:sib_general_problem} satisfies $r^* > 0$.
Let $\eps > 0$ be an error parameter. We say $(\mmz, \mmv_1, \dots, \mmv_n, r)$ is a $(1+\eps)$-approximate solution of Problem~\ref{prob:sib_general_problem} 
(and accordingly $B(\mmz, r)$ is a $(1+\eps)$-approximate SIB of the input objects)
if it is feasible and satisfies $r \le (1+\eps) r^*$.
To find a $(1+\eps)$-approximate soluton, we show that Problem~\ref{prob:sib_general_problem} can be modeled as a SCG over the product cone $\cc{C}$, and the approximation algorithm for SCG presented in Theorem~\ref{thm:scg_product} can be applied.

\begin{lemma}\label{lem:sib_game}
Problem~\ref{prob:sib_general_problem} can be modeled as the following symmetric cone game:
\beq\label{eq:sib_game}
    \min_{\underbrace{\scalebox{.85}{$(\mmz, \mmv_1, \dots, \mmv_n)$}}_{\scalebox{.88}{$\hspace{1.3em}\mmx$}} \in \cc{A}}\ \max_{\mmy \in \cc{B}}\ 
    \Big(
        \underbrace{\cproduct_{i=1}^n \begin{pmatrix}
            \mmv_i - \mmz \\
            0
        \end{pmatrix}}_{\brmf(\mmx)}
    \Big) \bdiamond \mmy,
\eeq
where the convex set $\cc{A}$ is defined as:
\[
    \cc{A} = \big\{ (\mmz, \mmv_1, \dots, \mmv_n) \in \bR^d\times \dots \times \bR^d : \mmz \in \conv(\{\Omega_i\}_{i=1}^n) \text{ and } \mmv_i \in \Omega_i \ \forall i \in [n] \big\},
\]
and $\cc{B}$ is the spectraplex of the product cone $\cc{C}$.
Solving the game up to an additive error of ${\eps r^* \over \sqrt{2}}$ gives a $(1+\eps)$-approximate solution of Problem~\ref{prob:sib_general_problem}.
\end{lemma}

\begin{proof}
We first show that the value of the SCG \eqref{eq:sib_game} is ${r^* \over \sqrt{2}}$, then an approximate solution of the SCG is an approximate solution of SIB.

\medskip
$(i)$ {\em Value of the SCG.}
Let $\hat{r}$ be a valid intersecting radius for the input objects, i.e. $\hat{r} \ge r^*$. Then there exist a point $(\hat{\mmz}, \hat{\mmv}_1, \dots, \hat{\mmv}_n) \in \cc{A}$ such that $(\hat{\mmz}, \hat{\mmv}_1, \dots, \hat{\mmv}_n, \hat{r})$ is a feasible solution of Problem \ref{prob:sib_general_problem}.
That is, the point satisfies
\[
\begin{aligned}
    & \hat{r} \ge 0, \\
    & \|\hat{\mmz} - \hat{\mmv}_i\| \le \hat{r}, \ \forall i \in [n],\\
    & (\hat{\mmz}, \hat{\mmv}_1, \dots, \hat{\mmv}_n) \in \cc{A}.
\end{aligned}
\]
By the characterization of the product cone $\cc{C}$ (recall \eqref{eq:product_cone_defn}), the second condition above can be rewrite as the following conic constraint:
\[
    \cproduct_{i=1}^n \begin{pmatrix}
        \hat{\mmz} - \hat{\mmv}_i \\
        \hat{r}
    \end{pmatrix} \in \cc{C}.
\]
Consequently, by the self-duality and homogeneity of $\cc{C}$, for any point $\mmy \in \cc{B}$, we have:
\beq\label{eq:sib_game_self_dual}
    \Big(
    \cproduct_{i=1}^n \begin{pmatrix}
        \hat{\mmz} - \hat{\mmv}_i \\
        \hat{r}
    \end{pmatrix}
    \Big) \bdiamond \mmy \ge 0.
\eeq
By definition of the identity $\mme$ in $\cc{C}$, the left-hand side can be expanded as follows:
\[
    \Big(
    \cproduct_{i=1}^n \begin{pmatrix}
        \hat{\mmz} - \hat{\mmv}_i \\
        0
    \end{pmatrix} + \hat{r} {\mme \over \sqrt{2}} \Big) \bdiamond \mmy 
    = \Big(
    \cproduct_{i=1}^n \begin{pmatrix}
        \hat{\mmz} - \hat{\mmv}_i \\
        0
    \end{pmatrix} \Big) \bdiamond \mmy  +  {\hat{r} \over \sqrt{2}}(\mme \bdiamond \mmy)
    = \Big(
    \cproduct_{i=1}^n \begin{pmatrix}
        \hat{\mmz} - \hat{\mmv}_i \\
        0
    \end{pmatrix} \Big) \bdiamond \mmy  +  {\hat{r} \over \sqrt{2}},
\]
where the last equality is due to the fact that $\Tr(\mmy) = \mme \bdiamond \mmy = 1$. Substituting the above into \eqref{eq:sib_game_self_dual} and rearranging the terms gives
\[
    \Big(
    \cproduct_{i=1}^n \begin{pmatrix}
        \hat{\mmv}_i - \hat{\mmz} \\
        0
    \end{pmatrix} \Big) \bdiamond \mmy \le {\hat{r} \over \sqrt{2}}.
\]
Since the above is true for all $\mmy\in \cc{B}$ and at least one $(\hat{\mmz}, \hat{\mmv}_1, \dots, \hat{\mmv}_n) \in \cc{A}$, we have
\beq\label{eq:sib_game_ub}
    \min_{(\mmz, \mmv_1, \dots, \mmv_n)\in \cc{A}}\ \max_{\mmy \in \cc{B}}\ 
    \Big(
        \cproduct_{i=1}^n \begin{pmatrix}
            \mmv_i - \mmz \\
            0
        \end{pmatrix}
    \Big) \bdiamond \mmy \le {\hat{r} \over \sqrt{2}}.
\eeq

On the other hand, let $\check{r}$ be an invalid intersecting radius, i.e. $\check{r} < r^*$. Then for all the points $(\mmz, \mmv_1, \dots, \mmv_n) \in \cc{A}$, we have:
\[
    \cproduct_{i=1}^n \begin{pmatrix}
        \mmz - \mmv_i \\
        \check{r}
    \end{pmatrix} \notin \cc{C}.
\]
By self-duality and homogeneity of $\cc{C}$, there exist a point $\check{\mmy} \in \cc{B}$ such that
\[
    \Big(\cproduct_{i=1}^n \begin{pmatrix}
        \mmz - \mmv_i \\
        \check{r}
    \end{pmatrix}\Big) \bdiamond \check{\mmy} < 0.
\]
Expanding the left-hand side and rearranging the terms gives
\[
    \Big(\cproduct_{i=1}^n \begin{pmatrix}
        \mmv_i - \mmz \\
        0
    \end{pmatrix}\Big) \bdiamond \check{\mmy} > {\check{r} \over \sqrt{2}}.
\]
Since the above is true for at least one $\check{\mmy} \in \cc{B}$ and all $(\mmz, \mmv_1, \dots, \mmv_n) \in \cc{A}$, we have
\beq\label{eq:sib_game_lb}
    \max_{\mmy \in \cc{B}}\ \min_{(\mmz, \mmv_1, \dots, \mmv_n)\in \cc{A}}\ 
    \Big(
        \cproduct_{i=1}^n \begin{pmatrix}
            \mmv_i - \mmz \\
            0
        \end{pmatrix}
    \Big) \bdiamond \mmy > {\check{r} \over \sqrt{2}}.
\eeq

Combining the inequalities \eqref{eq:sib_game_ub} and \eqref{eq:sib_game_lb}, and by the generalized minimax theorem, we conclude that
\beq
    \min_{(\mmz, \mmv_1, \dots, \mmv_n)\in \cc{A}}\ \max_{\mmy \in \cc{B}}\ 
    \Big(
        \cproduct_{i=1}^n \begin{pmatrix}
            \mmv_i - \mmz \\
            0
        \end{pmatrix}
    \Big) \bdiamond \mmy = {r^* \over \sqrt{2}}.
\eeq

\medskip
$(ii)$ {\em From SCG to SIB.}
Let $(\tilde{\mmz}, \tilde{\mmv}_1, \dots, \tilde{\mmv}_n) \cproduct \tilde{\mmy} \in \cc{A}\times \cc{B}$ be a ${\eps r^* \over \sqrt{2}}$-Nash equilibrium of the SCG, which satisfies 
\begin{align}
&\Big(
    \cproduct_{i=1}^n \begin{pmatrix}
        \tilde{\mmv}_i - \tilde{\mmz} \\
        0
    \end{pmatrix}
\Big) \bdiamond \mmy \le {r^* \over \sqrt{2}} + {\eps r^* \over \sqrt{2}}
\quad \text{ for all }
\mmy \in \cc{B},
\label{eq:sib_game_equiv} \\
\text{and}\quad  &\Big(
    \cproduct_{i=1}^n \begin{pmatrix}
        {\mmv}_i - {\mmz} \\
        0
    \end{pmatrix}
\Big) \bdiamond \tilde{\mmy} \ge {r^* \over \sqrt{2}} - {\eps r^* \over \sqrt{2}}
\quad \text{ for all }
(\mmz, \mmv_1, \dots, \mmv_n)\in \cc{A}.\notag
\end{align}
Rearranging the terms in \eqref{eq:sib_game_equiv} and using the fact that $\mme \bullet \mmy = 1\ \forall \mmy \in \cc{B}$, we have
\[
\Big(
    \cproduct_{i=1}^n \begin{pmatrix}
        \tilde{\mmz} - \tilde{\mmv}_i \\
        (1 + \eps) r^*
    \end{pmatrix}
\Big) \bdiamond \mmy \ge 0 \quad \text{for all } \mmy \in \cc{B},
\]
which implies $\displaystyle\cproduct_{i=1}^n \begin{pmatrix}
    \tilde{\mmz} - \tilde{\mmv}_i \\
    (1 + \eps) r^*
\end{pmatrix} \in \cc{C}$, or equivalently:
\[
    \|\tilde{\mmz} - \tilde{\mmv}_i\| \le (1 + \eps) r^*, \ \forall i \in [n].
\]
Let $\tilde r = \max_{i\in [n]} \|\tilde{z} - \tilde{v}_i\|$. From the above we have $\tilde r \le (1 + \eps) r^*$. Therefore, $(\tilde{\mmz}, \tilde{\mmv}_1, \dots,\tilde{\mmv}_n, \tilde r)$ is a $(1+\eps)$-approximate solution of Problem~\ref{prob:sib_general_problem}.
\end{proof}

Before presenting our main result for SIB, we introduce a sub-problem that recurs throughout the execution of our algorithm.
\begin{problem}\label{prob:sib_subproblem}
Given vectors $\mmh_1, \dots, \mmh_n\in \bR^d$,
find $\mmu_1, \dots, \mmu_n \in \bR^d$ such that 
\[
    \mmu_i \in \argmin_{\mmu \in \Omega_i}\ \mmh_i^T \mmu \quad \text{for } i = 1, \dots, n.
\]
\end{problem}

Problem~\ref{prob:sib_subproblem} consists of $n$ linear optimization problems over compact regions. The existence of an optimal solution is evident. For $\Omega_i$ with favorable structures, a solution can be computed efficiently, as will be discussed later in this section.
Now we are ready to present our main result for SIB.

\begin{theorem}\label{thm:sib_general}
Let $D$ be the diameter of the input and let $R = {D \over r^*}$. Assume a subroutine that solves Problem~\ref{prob:sib_subproblem} with running time $O(S)$. Then there is an iterative algorithm that computes a $(1 + \eps)$-approximate solution of Problem~\ref{prob:sib_general_problem} in $O({R^2 (S + nd) \log n \over \eps^2})$ time.
\end{theorem}

\begin{proof}
From Lemma~\ref{lem:sib_game}, we know that Problem~\ref{prob:sib_general_problem} can be modeled as the SCG \eqref{eq:sib_game}, and solving the SCG up to an additive error ${\eps r^*\over \sqrt{2}}$ gives a $(1+\eps)$-approximate solution of Problem~\ref{prob:sib_general_problem}. 
Since it conforms to the structure of \eqref{eq:scg_product}, we can employ Algorithm~\ref{algo:scg_product} to solve the problem.
The implementation of this algorithm necessitates clarification of several details: What is the process of the {\sc Oracle} that solves the optimization problem in Line~\ref{algo:scg_product_oracle}? What is the width of the {\sc Oracle}? How does the algorithm work with the unkown parameters $D$ and $r^*$?
We proceed by answering these questions in sequence, and conclude with an analysis of the total running time of the algorithm.

\medskip
$(i)$ {\em The {\sc Oracle} process.}
For any given $\mmy\in\cc{B}$, the {\sc Oracle} finds $\mmx \in \argmin_{\mmx\in\cc{A}} \brmf(\mmx)\bdiamond\mmy$.
Let $\mmx = (\mmz, \mmv_1, \dots, \mmv_n)\in \bR^d \times \dots \times \bR^d$ and $\mmy = \cproduct_{i=1}^n (\bar{\mmy}_i, y_{i,0})\in\bV$. Then the problem can be formulated as:
\beq\label{eq:sib_general_oracle}
\begin{aligned}
    \underset{\mmz, \mmv_1, \dots, \mmv_n}{\rm minimize}\quad & \sum_{i=1}^n \bar{\mmy}_i^T \mmv_i - \big(\sum_{i=1}^n \bar{\mmy}_i \big)^T \mmz\\
    {\rm subject\ to} \quad
    & \mmz \in \conv(\{\Omega_i\}_{i=1}^n),\\
    & \mmv_i \in \Omega_i, \ \forall i \in [n].
\end{aligned}
\eeq
Observe that in the above problem the constraints on $\mmz$ and each $\mmv_i$ are independent. The problem can be decomposed into smaller sub-problems and be solved separately. 
For each $\mmv_i$, the corresponding sub-problem is
$\min\big\{ \bar{\mmy}_i^T \mmv_i : \mmv_i \in \Omega_i \big\}$. Therefore, finding solutions for all $\mmv_i$ conforms to the structure of Problem~\ref{prob:sib_subproblem} and can be finished in $O(S)$ time as assumed.
Let $\mmh = \sum_{i=1}^n \bar{\mmy}_i$. Then the sub-problem for $\mmz$ can be simplified to $\max\big\{\mmh^T \mmz : \mmz \in \conv(\{\Omega_i\}_{i=1}^n)\big\}$. 
Since the sets $\Omega_i$ are compact, we have 
\[    
\max\Big\{\mmh^T \mmz : \mmz \in \conv(\{\Omega_i\}_{i=1}^n)\Big\} = \max_{i\in [n]} \Big(\max\big\{\mmh^T \mmz : \mmz \in \Omega_i \big\}\Big),
\]
which also reduces to solving Problem~\ref{prob:sib_subproblem}. With an additional processing time of $O(nd)$ for computing the vector $\mmh$, the problem can also be solved in $O(S)$ time.

\medskip
$(ii)$ {\em Width of {\sc Oracle}.}
Let $\mmx = (\mmz, \mmv_1, \dots, \mmv_n)$ be the output of the {\sc Oracle}.
The width $\rho$ is an upperbound on the spectral norm of $\brmf(\mmx) = \cproduct_{i=1}^n (\mmv_i - \mmz, 0)$ over all possible inputs $\mmy\in\cc{B}$. Since $\mmz \in \conv(\{\Omega_i\}_{i=1}^n)$ and $\mmv_i\in \Omega_i$, we have $\|\mmv_i - \mmz\| \le D,\ \forall i \in [n]$. Consequently,
\[\begin{aligned}
    \Big\|\cproduct_{i=1}^n (\mmv_i - \mmz, 0)\Big\|_\spe = \max_{k\in [2n]}\ \Big|\lambda_{k}\big(\cproduct_{i=1}^n (\mmv_i - \mmz, 0)\big) \Big| \le \max_{i\in [n]}\ {1\over \sqrt{2}} \|\mmv_i - \mmz\| \le {D\over \sqrt{2}},
\end{aligned}\]
where the second last inequality is due to the definition of eigenvalues in the algebra associated with $\cc{C}$.
Therefore, the {\sc Oracle} has a width of $D \over \sqrt{2}$.

\medskip
$(iii)$ {\em Working with unkown $D$.}
From $(ii)$ we know that the width of the {\sc Oracle} is proportional to the diameter $D$, which serves as a parameter in Algorithm~\ref{algo:scg_product}. However, computing the diameter itself is a difficult problem. Here we introduce a doubling trick that enables the algorithm to work without determining the exact value of $D$.
Recall that the width $\rho$ is an upperbound on $\|\brmf(\mmx)\|_\spe$ for all possible output $\mmx$ of the {\sc Oracle}. 
The purpose of introducing $\rho$ is to ensure that the eigenvalues of $\brmf(\mmx) / \rho$ consistently fall within the range $[-1,1]$, thereby all the inequalities in the proof of Theorem~\ref{thm:zero-sum_game} can work appropriately.
Therefore, instead of bounding the spectral norm of $\brmf(\mmx)$ for all possible $\mmx$, it suffices to find an upperbound on $\|\brmf(\xt{\mmx}{t})\|_\spe$ for all $\xt{\mmx}{t}$ that appear in the runtime. 

Pick a point $\hat{\mmv}_i$ in each object $\Omega_i$, and compute $E = \max_{j \ge 2} \|\mmv_1 - \mmv_j\|$. Clearly, we have $E \le D$.
Let $\rho_1 = 2E$ be a foresight estimation on the upperbound. We first run Algorithm~\ref{algo:scg_product} with the estimated width $\rho_1$. 
During the runtime, if there exists a certain iteration $t$ such that the value of $\|\brmf(\xt{\mmx}{t})\|_\spe$ exceeds $\rho_1$, we terminate the algorithm and set $\rho_2 = 2\rho_1$. 
Repeat this process by setting $\rho_\tau = 2\rho_{\tau-1}$ for $\tau\ge 2$, until we reach a sufficiently large value $\rho_{\rm final}$ so that the algorithm can complete its full process using the width $\rho_{\rm final}$. 
If $\rho_{\rm final} \neq \rho_1$, there must exists an {\sc Oracle} output $\xt{\mmx}{t}$ such that $\|\brmf(\xt{\mmx}{t})\|_\spe > \rho_{\rm final} / 2$; together with $\|\brmf(\xt{\mmx}{t})\|_\spe \le \rho$, we see that $\rho_{\rm final} / 2  < \rho$ and consequently $\rho_{\rm final} < 2\rho \le \sqrt{2}D$. 
Otherwise, if $\rho_{\rm final} = \rho_1 = 2E$, we have $\rho_{\rm final} \le 2D$. 

\medskip
$(iv)$ {\em Working with unknown $r^*$.} 
Similar to $(iii)$, we introduce a halving trick which enables the algorithm to work without prior knowledge of $r^*$.
Let $E$ be the quantity defined in $(iii)$.
Then the ball $B(\mmv_1, E)$ intersects every input object, which implies $r^* \le E$. 
Setting $r_1 = {E \over 2}$, we first run Algorithm~\ref{algo:scg_product} with error parameter $\eps r_1 \over \sqrt{2}$. 
Let $\tilde{\mmx} = (\tilde{\mmz}, \tilde{\mmv}_1, \dots, \tilde{\mmv}_n) \in \cc{A}$ and $\tilde{\mmy} \in \cc{B}$ be the output of the algorithm, and compute
\[
\begin{aligned}
    &\nu_{\tilde{\mmx}} = \max_{\mmy \in \cc{B}}\ 
    \Big(
    \cproduct_{i=1}^n \begin{pmatrix}
        \tilde{\mmv}_i - \tilde{\mmz} \\
        0
    \end{pmatrix}
    \Big) \bdiamond \mmy, \\
    &\nu_{\tilde \mmy} = \min_{(\mmz, \mmv_i, \dots, \mmv_n) \in \cc{A}} \Big(
        \cproduct_{i=1}^n \begin{pmatrix}
            {\mmv}_i - {\mmz} \\
            0
        \end{pmatrix}
    \Big) \bdiamond \tilde{\mmy}.
\end{aligned}
\]
These two quantities can be efficiently computed. Specifically, computing $\nu_{\tilde{\mmy}}$ is equivalent to solving the {\sc Oracle} problem, and computing $\nu_{\tilde{\mmx}}$ is equivalent to computing $\max_{i\in [n]} \|\tilde{\mmv}_i - \tilde{\mmz}\|$.
Theorem~\ref{thm:scg_product} guarantees that $\nu_{\tilde \mmx} \le {r^* \over \sqrt{2}} + {\eps r_1 \over \sqrt{2}}$ and $\nu_{\tilde \mmy} \ge {r^* \over \sqrt{2}} - {\eps r_1 \over \sqrt{2}}$. 
Then we check whether $\nu_{\tilde \mmy} > 0$ and ${\nu_{\tilde \mmx} - \nu_{\tilde \mmy} \over \nu_{\tilde \mmy}} \le \eps$.
If both are true,  we can conclude that
\[
    \nu_{\tilde \mmx} \le (1 + \eps) \nu_{\tilde \mmy} \le (1 + \eps) {r^* \over \sqrt{2}},
\]
and a $(1 + \eps)$-approximate solution of Problem~\ref{prob:sib_general_problem} can be easily constructed from $(\tilde \mmz, \tilde \mmv_1, \dots, \tilde \mmv_n)$.
Otherwise, we apply the halving trick, reducing the parameter to $r_2 = {r_1 \over 2}$ and executing the algorithm again. We repeat the process by setting $r_\tau = {r_{\tau-1}\over 2}$ for $\tau \ge 2$, until reaching a sufficiently small value $r_{\rm final}$ such that both conditions $\nu_{\tilde \mmy} > 0$ and ${\nu_{\tilde \mmx} - \nu_{\tilde \mmy} \over \nu_{\tilde \mmy}} \le \eps$ are satisfied.
For the first condition, we have 
\[
    {r^* \over \sqrt{2}} - {\eps r_{\rm final} \over \sqrt{2}} > 0 \quad \Rightarrow \quad r_{\rm final} < {r^* \over \eps}. 
\]
Suppose $\eps$ is small, say $\eps < 1$, then $r_{\rm final} \le {r^*} < {r^* \over \eps}$ is sufficient for the first condition.
As for the second condition, we have
\[
    {\nu_{\tilde \mmx} - \nu_{\tilde \mmy} \over \nu_{\tilde \mmy}} \le \eps 
    \quad \Rightarrow \quad 
    {2 \eps r_{\rm final}\over r^* - \eps r_{\rm final}} \le \eps 
    \quad \Rightarrow \quad 
    r_{\rm final} \le {r^* \over 2 + \eps}.
\]
Therefore, $r_{\rm final} \le {r^* \over 3} < {r^* \over 2 + \eps}$ is sufficient for the second condition. Combined, we know that once $r_{\rm final} \le {r^* \over 3}$ we can find a $(1+\eps)$-approximate solution.
Additionally, if $r_{\rm final} \ne r_1$, then we have $2 r_{\rm final} > {r^* \over 3}$ and consequently $r_{\rm final} > {r^* \over 6}$; otherwise, if $r_{\rm final} = r_1 = {E\over 2}$, then due to the fact that $r^* \le E$ we have $r_{\rm final} \ge {r^* \over 2}$.

\medskip
$(v)$ {\em Running time analysis.}
We first analyse the time complexity under the assumption that both $D$ and $r^*$ are predetermined. From Theorem~\ref{thm:scg_product} we know that Algorithm~\ref{algo:scg_product} can find a ${\eps r^* \over \sqrt{2}}$-Nash equilibrium of the SCG using $O\big({\rho^2 \log n \over (\eps r^*)^2}\big)$ calls to the {\sc Oracle}, with an additional processing time of $O(F + nd)$ per call.
Substituting the width $\rho = {D \over \sqrt{2}}$ into $T$, then the iteration complexity becomes $O\big({D^2 \log n \over (\eps r^*)^2} \big)$. 
For any given $(\mmz, \mmv_1, \dots, \mmv_n)$, the affine map $\brmf$ can be computed in $O(nd)$ time, so the processing time between two {\sc Oracle} calls is $O(nd)$. From $(i)$ we know that the {\sc Oracle} problem can be solved in $O(S + nd)$ time. 
Adding all together, the total running time of the algorithm becomes $O\big({R^2 (S + nd) \log n \over \eps^2 } \big)$.

We now demonstrate that the adaptive approaches introduced in $(iii)$ and $(iv)$ preserve the algorithm's computational efficiency.
In $(iii)$, we know that the parameters $\rho_1, \dots, \rho_{\rm final}$ form an increasing geometric series. Since the running time scales quadratically with $\rho$, the total running time is dominated by the last step with the parameter $\rho_{\rm final} \le 2D$, which is of the same asymptotic order as before. Similarly, in $(iv)$, the parameters $r_1, \dots, r_{\rm final}$ form a decreasing geometric series. The total running time is determined by the last step with the parameter $r_{\rm final} > {r^* \over 6}$, and preserves the same asymptotic order.
\end{proof}

In the remainder of this section, we present details of the {\sc Oracle} for specific examples of SIB and state the corresponding \change{running time of our algorithm.}

\subsection{Convex polytopes}\label{sec:sib_cvx}
Let $\cc{P}_i, i = 1, \dots, n, n > 1$, be non-empty finite point sets in the Euclidean space $\bR^d$, where $\cc{P}_i$ consists of $m_i$ points.
For each $i\in [n]$, the convex hull $\conv(\cc{P}_i)$ is a convex polytope. 
Let $\Omega_i = \conv(\cc{P}_i)$ for $i = 1, \dots, n$.
Then Problem~\ref{prob:sib_general_problem} becomes finding a smallest ball that intersects every polytope $\Omega_i$.
SIB of convex polytopes covers many important problems in the literature. 
For example, when each point set $\cc{P}_i$ contains only one point (in which $\Omega_i$ are singletons), the problem becomes the smallest enclosing ball problem;
when there are only two polytopes (i.e.~$n=2$), it becomes the polytope distance problem, which is also known to be the dual problem of hard-margin SVM if the polytopes are linearly separable~\cite{pd:GJ2009};
when the polytopes are line segments (i.e.~$m_i = 2, \forall i\in [n]$), the problem is to find the intersection radius for a set of line~segments~\cite{siblowdim:BJMR1991}.

Employing the algorithm described in Theorem~\ref{thm:sib_general} to find a $(1+\eps)$-approximate SIB for the set of convex polytopes, the {\sc Oracle} problem~\eqref{eq:sib_general_oracle} 
can be reduced to a sequence of linear optimization problems in the form of Problem~\ref{prob:sib_subproblem},
which in the current context becomes:
\beq\label{eq:sib_cvx_oracle_sequence}
    \min\Big\{\mmh_i^T\mmu_i : \mmu_i \in \conv(\cc{P}_i) \Big\}
    \quad \text{for $i = 1,\dots, n$}.
\eeq
For each $i\in [n]$, the above can be solved by computing the function value $\mmh_i^T \mmu_i$ for every point $\mmu_i\in \cc{P}_i$ because $\min\big\{\mmh_i^T \mmu_i : \mmu_i\in \conv(\cc{P}_i) \big\} = \min\big\{\mmh_i^T \mmu_i : \mmu_i\in \cc{P}_i\big\}$.
Therefore, let $N$ be the number of non-zero coordinates in the input sets $\cc{P}_1, \dots, \cc{P}_n$, Problem~\ref{prob:sib_subproblem} can be solved in $O(N)$ time.

\begin{corollary}
Let $D$ be the diameter of the input and let $R = {D \over r^*}$.
Given an error parameter $\eps>0$, there is an algorithm that computes a $(1+\eps)$-approximate SIB for a set of convex polytopes with a running time of $O({R^2 (N + nd) \log n \over \eps^2})$.
\end{corollary}

\begin{remark}
Here we present the running time of our algorithm for several important cases:
\begin{itemize}
    \item {\em Smallest enclosing ball.} When $\Omega_i$ are singletons, the SIB problem is identical to SEB. The optimal radius $r^*$ satisfies $r^* \ge {D\over 2}$, and consequently the ratio $R = {D \over r^*} \le 2$.  Moreover, we have $N \le nd$. 
    The running time can be simplified to $O({nd \log n \over \eps^2})$.
    \item {\em Polytope distance.} In this case, we have $n=2$ and the running time is $O({R^2 (N + d) \over \eps^2})$.
    \item {\em Intersection radius of line segments.} In this case, $m_i = 2$ for all $i \in [n]$, and consequently $O(N)$ is at most $O(nd)$. The running time can be simplified to $O({R^2 nd \log n \over \eps^2})$.
\end{itemize}
\end{remark}

\subsection{Reduced polytopes}\label{sec:sib_rcvx}

Given a point set $\cc{P} = \{\mmp_1, \dots, \mmp_m\}$ in $\bR^d$, every point in $\conv(\cc{P})$ can be represented by a convex combination of the points in $\cc{P}$.
In other words, $\conv(\cc{P}) = \big\{\sum_{j=1}^m b_j \mmp_j : \mmb \in \Delta^{m-1}\big\}$,
where $\Delta^{m-1} = \big\{\mmb \in \bR^m_+ : \mmone^T \mmb = 1 \big\}$ is the $(m-1)$-dimensional {\em simplex}.
For a given point $\mmv\in\conv(\cc{P})$, we say $\mmb$ is a {\em barycentric coordinate} of $\mmv$ if $\mmb \in \Delta^{m-1}$ and $\sum_j b_j \mmp_j = \mmv$. 

\begin{figure}[t]
\begin{center}
\includegraphics[width=\linewidth]{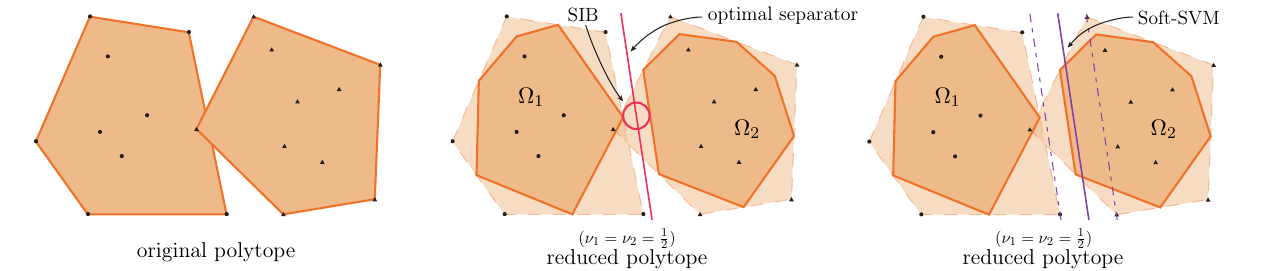}
\end{center}
\vspace{-1.5em}
\caption{{\em Left:} the original convex polytope. 
{\em Middle:} the reduced polytopes (with $\nu_1 = \nu_2 = {1\over 2}$), the SIB and the optimal separating hyperplane thereof.
{\em Right:} the corresponding soft-margin SVM.
}\label{fig:cvx_vx_rcvx}
\end{figure}

Let $\cc{P}_1, \dots, \cc{P}_n,\ n > 1$, be non-empty finite point sets in the Euclidean space $\bR^d$, where $\cc{P}_i$ consists of $m_i$ points.
For each $i\in [n]$, let $\mmV_i$ be a ${d\times m_i}$ matrix such that the entires in its $j$-th column corresponds to the $j$-th point in $\cc{P}_i$.
Then the convex polytopes generated by $\cc{P}_i$ can be characterized as
$\conv(\cc{P}_i) = \big\{ \mmV_i \mmb_i : \mmb  _i \in \Delta^{m_i-1} \big\}$.
Let $\nu_1, \dots, \nu_n$ be parameters satisfing $\nu_i \in [{1\over m_i}, 1]$. We say $\Omega_i$ is a {\em reduced polytope} generated by $\cc{P}_i$ and $\nu_i$ if
\[
    \Omega_i = \Big\{ \mmV_i \mmb_i : \mmb_i \in \Delta^{m_i-1}, \ \mmb_i \le \nu_i \Big\}.
\]
Let $\Omega_1, \dots, \Omega_n$ be convex sets defined as above. Then Problem~\ref{prob:sib_general_problem} becomes finding the smallest ball that intersects the reduced polytopes. Note that if $\nu_i = 1$ for all $i\in [n]$, the problem is identical to the SIB problem for convex polytopes in Section~\ref{sec:sib_cvx}. 
In general, the radius $r^*$ of an SIB that intersects $\Omega_1, \dots, \Omega_n$ may be larger than the one that intersects $\conv(\cc{P}_1), \dots, \conv(\cc{P}_n)$. On the other hand, the diameter $D$ of $\conv(\{\Omega_i\}_{i=1}^n)$ may be smaller than the diameter of $\conv(\{\cc{P}_i\}_{i=1}^n)$.
As a special case, when there are only two reduced polytopes (i.e.~$n=2$), the problem becomes finding the shortest line segment connecting them, which is known to be the dual problem of the famous $\ell_1$-loss $C$-SVM and $\nu$-SVM problems~\cite{svm:BB2000, svm:CB1999}.
See Figure~\ref{fig:cvx_vx_rcvx} for an illustration of their relationship.

Employing the algorithm described in Theorem~\ref{thm:sib_general} to find a $(1+\eps)$-approximate SIB for the set of reduced polytopes, then the subproblem Problem~\ref{prob:sib_subproblem} becomes:
\beq\label{eq:sib_rcvx_oracle_sequence}
\begin{aligned}
    \min \Big\{\mmh_i^T \mmu_i : \mmu_i = \mmV_i \mmb_i, \mmb_i \in \Delta^{m_i - 1}, \mmb_i \le \nu_i \Big\}, \quad \text{for $i = 1, \dots, n$}.
\end{aligned}
\eeq
For each $i \in [n]$, if we substitute the variable $\mmu_i$ by $\mmV_i \mmb_i$ and let $\bar{\mmh}_i = \mmV_i^T \mmh_i$, then the above problem becomes $\min\big\{\bar{\mmh}_i^T \mmb_i : \mmb_i \in \Delta^{m_i -1}, \mmb_i \le \nu_i \big\}$, which is optimizing a linear function w.r.t.\ the barycentric coordinate over the ``reduced'' simplex.
Let $k = \lceil {1\over \nu_i} \rceil$. The problem can be solved by identifying the top-$k$ smallest entries in $\bar{\mmh}_i$. Let $j_1, \dots, j_k$ be the indices of these entries, where $\bar{h}_{i,j_1} \le \bar{h}_{i,j_2} \le \dots \le \bar{h}_{i,j_k}$. The optimum is given by:
\[
    b_{i,j}  =
    \begin{cases}
        \nu_i, &\text{for $j = j_1, \dots, j_{k-1}$},\\
        1 - \nu_i(k-1), &\text{for $j = j_k$},\\
        0, &\text{otherwise}.
    \end{cases}
\]
Employing the median-of-medians algorithm~\cite[Sec.~9.3]{intro2algo:CLRS2022} to find the $k$-th smallest entry in $\mmh_i$, the above solution can be computed in $O(m_i)$ time. 
As a consequence,
let $M = \sum_i m_i$ and let $N$ be the number of non-zeros in the input, 
Problem~\ref{prob:sib_subproblem} can be solved in $O(M)$ time, with an additional time of $O(N)$ to compute $\bar{\mmh}_i = \mmV_i^T \mmh_i$.
Assuming each point set $\cc{P}_i$ contains at most one origin point (otherwise we remove duplications), then the time is dominated by $O(N)$.

\begin{corollary}
Let $D$ be the diameter of the input and let $R = {D \over r^*}$.
Given an error parameter $\eps>0$, there is an algorithm that computes a $(1+\eps)$-approximate SIB for a set of reduced polytopes with a running time of $O({R^2 (N + nd) \log n \over \eps^2})$.
\end{corollary}

\begin{remark}
The running time of our algorithm for a set of reduced polytopes is no worse than the one for the original convex polytopes because the ratio $R = {D\over r^*}$ can only decrease. For the special case of soft-margin SVM where there are only two reduced polytopes (i.e.~$n=2$), the running time of our algorithm is $O({R^2 (N + d) \over \eps^2})$.
\end{remark}

\subsection{Axis-aligned bounding boxes (AABBs)}\label{sec:sib_box}

AABBs are frequently used as approximations of point sets, and are also being used to model imprecise input points~\cite{siblowdim:LK2010}.
Let $\Omega_1, \dots, \Omega_n$, be non-empty AABBs in the Euclidean space $\bR^d$. 
Then Problem~\ref{prob:sib_general_problem} becomes finding an SIB for a set of AABBs.
To solve this problem, we can represent each AABB as a convex polytope and apply the algorithm described in Section~\ref{sec:sib_cvx}.
The same idea is used by Löffler and van Kreveld~\cite{siblowdim:LK2010} to solve the problem in two dimensions, where they applied the algorithm of Jadhav et al.~\cite{siblowdim:JMB1996}.
Nevertheless, it is inefficient in high dimensions because the number of vertices of an AABB grows exponentially with the dimensionality.
Therefore, instead of viewing each AABB as a convex polytope, we represent it by the minimal and maximal value of its coordinates in each dimension. 
That is,
\[
    \Omega_i = \Big\{ \mmv_i : \mma_i \le \mmv_i \le \mmb_i \Big\},
    \quad i = 1, \dots, n,
\]
where $\mma_1, \dots, \mma_n$ and $\mmb_1,\dots, \mmb_n$ are vectors in $\bR^d$ satisfying $\mma_i \le \mmb_i, \forall i \in [n]$.

Employing the algorithm described in Theorem~\ref{thm:sib_general} to find a $(1+\eps)$-approximate SIB for the set of AABBs, then the subproblem Problem~\ref{prob:sib_subproblem} becomes:
\beq\label{eq:sib_box_oracle_sequence}
\min\Big\{\mmh_i^T \mmu_i : \mma_i \le \mmu_i \le \mmb_i \Big\} \quad \text{for $i = 1, 
\dots, n$}.
\eeq
For each $i \in [n]$, the above can be solved by identifying the sign of $h_{i,j}$ for $j \in [d]$.
That is,
\[
    w_j = 
    \begin{cases}
        a_{i,j}, & \text{if $h_{i,j} > 0$,} \\
        b_{i,j}, & \text{otherwise.}
    \end{cases}
\]
The above solution can be computed in $O(d)$ time, and consequently Problem~\ref{prob:sib_subproblem} can be solved in $O(nd)$ time.

\begin{corollary}
Let $D$ be the diameter of the input and let $R = {D \over r^*}$.
Given an error parameter $\eps>0$, there is an algorithm that computes a $(1+\eps)$-approximate SIB for a set of AABBs with a running time of $O({R^2 nd \log n \over \eps^2})$.
\end{corollary}

\subsection{Balls}\label{sec:sib_ball}
Another important example is to find an SIB for a set of balls.
This problem is considered by Löffler and van Kreveld~\cite{siblowdim:LK2010} when they use balls to model imprecise input points in two-dimensional Euclidean space. \change{Subsequently,} Nguyen et al.~\cite{sibtheory:NNS2012} showed that a subgradient algorithm applied to this problem in a Banach space will converge.
Son and Afshani~\cite{sibhighdim:SA2015} investigated this problem in Euclidean spaces in the streaming setting and proposed the first approximation algorithm.
In this section, we consider Euclidean balls defined as follows:
\[
    \Omega_i = B(\mmc_i, r_i) = \Big\{ \mmv_i : \|\mmv_i - \mmc_i\| \le r_i \Big\},
    \quad i = 1, \dots, n,
\]
where $\mmc_1, \dots, \mmc_n\in \bR^d$ and $r_1, \dots, r_n \in \bR_+$ are the centers and radii of the balls.

Employing the algorithm described in Theorem~\ref{thm:sib_general} to find a $(1+\eps)$-approximate SIB for the set of Euclidean balls, then Problem~\ref{prob:sib_subproblem} becomes:
\beq
    \min\Big\{\mmh_i^T \mmu_i :  \|\mmu_i - \mmc_i\| \le r_i \Big\} \quad \text{for $i = 1, \dots, n$}.
\eeq
For each $i\in [n]$, the above has an analytical solution $\mmu_i = \mmc_i - r_i {\mmh_i \over \|\mmh_i\|}$, which can be computed in $O(d)$ time. Consequently, Problem~\ref{prob:sib_subproblem} can be solved in $O(nd)$ time.

\begin{corollary}
Let $D$ be the diameter of the input and let $R = {D \over r^*}$.
Given an error parameter $\eps>0$, there is an algorithm that computes a $(1+\eps)$-approximate SIB for a set of Euclidean balls with a running time of $O({R^2 nd \log n \over \eps^2})$.
\end{corollary}

\subsection{Ellipsoids}\label{sec:sib_ellipsoid}
A natural extension following balls is ellipsoids. 
They have been frequently used to model confidence regions of multivariate probability distributions~\cite{multichebyshev:AI1960} and provide good approximations for compact convex sets~\cite{ellipsoid:fritz1948}.
The problem of finding the smallest enclosing ball (SEB) for a set of ellipsoids is well-studied in the literature (see, e.g.~\cite{sdp:LS1996,seb:yildirim2008}). Nonetheless, to the best of our knowledge, the SIB problem for a set of ellipsoids has not been explored.
The existing algorithms for SEB usually involve solving semidefinite programs as a subroutine and exhibit running time scales at least cubically w.r.t.\ the dimensionality.
However, as we will see in the following, this can be avoided in our algorithm.
Consider the SIB problem where $\Omega_i$ are $d$-dimensional ellipsoids defined as:
\[
    \Omega_i = \cc{E}(\mmc_i, \mm{\Sigma}_i) = \Big\{ \mmv_i : (\mmv_i - \mmc_i)^T \mm{\Sigma}_i (\mmv_i - \mmc_i) \le 1 \Big\},
    \quad i = 1,\dots, n,
\]
where $\mmc_1, \dots, \mmc_n \in \bR^d$ and $\mm{\Sigma}_1, \dots, \mm{\Sigma}_n$ are $d\times d$ symmetric positive-definite matrices.

Employing the algorithm described in Theorem~\ref{thm:sib_general} to find a $(1+\eps)$-approximate SIB for the set of ellipsoids, then Problem~\ref{prob:sib_subproblem} becomes:
\[
    \min\Big\{ \mmh_i^T \mmu_i : (\mmu_i - \mmc_i)^T \mm{\Sigma}_i (\mmu_i - \mmc_i) \le 1 \Big\}, \quad \text{for $i = 1, \dots, n$}.
\]
For each $i\in [n]$, the above has an analytical solution $\mmu_i = \mmc_i - {\mm{\Sigma}_i^{-1} \mmh_i \over ({\mmh_i^T \mm{\Sigma}_i^{-1} \mmh_i})^{1/2}}$.
Let $\omega$ be the matrix multiplication exponent. The matrix inverses $\mm{\Sigma}_1^{-1}, \dots, \mm{\Sigma}_n^{-1}$ can be computed in $O(nd^\omega)$ time.
Therefore, assuming a pre-processing time of $O(nd^\omega)$ and an auxiliary space of $O(nd^2)$ to compute and store the matrix inverses, then the subroutine that solves Problem~\ref{prob:sib_subproblem} only involves matrix-vector multiplications and some simple arithmetic operations, which takes $O(nd^2)$ time. 

\begin{corollary}
Let $D$ be the diameter of the input and let $R = {D \over r^*}$.
Given an error parameter $\eps>0$, there is an algorithm that computes a $(1+\eps)$-approximate SIB for a set of ellipsoids with a running time of $O(nd^\omega + {R^2 nd^2 \log n \over \eps^2})$.
\end{corollary}

\subsection{Other input objects}\label{sec:extension}

Besides the above examples, our algorithm can be extended to solve the SIB problem with various types of input objects (e.g.\ half-ellipsoids), as long as they are compact and convex and there is an algorithm to solve linear optimization problems over their regions.
In addition, our algorithm not only works for the {\em pure} settings (where the input objects are of the same type) but is also applicable for {\em mixed} settings (where the input can be a mixture of the primitive objects, e.g.\ polytopes, balls, ellipsoids, etc.).

\begin{figure}[t]
\begin{center}
\includegraphics[width=.8\linewidth]{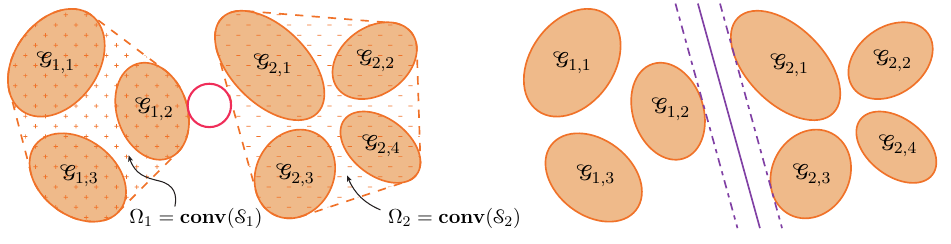}
\end{center}
\vspace{-1.5em}
\caption{{\em Left:} the SIB that intersects the convex hulls of two sets of ellipsoids. 
{\em Right:} the corresponding linear classifier that separates the two sets of ellipsoids with the largest margin.
}\label{fig:ellipsoid_svm}
\end{figure}

It is worth noting that $\Omega_i$ can also be convex hulls of compact (not necessarily convex) objects.
That is, let $\cc{S}_1, \dots, \cc{S}_n, n > 1$, be collections of objects, where $\cc{S}_i = \{\cc{G}_{i, 1}, \dots, \cc{G}_{i, m_i}\}$ and every $\cc{G}_{i, j}$ is compact.
We can set each $\Omega_i$ to be the convex hull of $\cc{S}_i$, namely $\Omega_i = \conv(\cc{S}_i)$, which can be viewed as a generalization of convex polytopes. 
With this setup, for each $i\in [n]$, the subproblem in Problem~\ref{prob:sib_subproblem} becomes $\min\big\{\mmh_i^T \mmu_i : \mmu_i \in \conv(\cc{S}_i) \big\}$. Moreover, because every $\cc{G}_{i,j}$ is compact, we have:
\[
    \min\Big\{\mmh_i^T \mmu_i : \mmu_i \in \conv(\cc{S}_i) \Big\} = \min_{j \in [m_i]} \Big(\min\big\{\mmh_i^T \mmu_i : \mmu_i \in \cc{G}_{i,j} \big\} \Big).
\]
Therefore, if we can solve linear optimization problems over each $\cc{G}_{i,j}~(i\in [n], j \in [m_i])$, we can also solve Problem~\ref{prob:sib_subproblem} and consequently find an approximate SIB.

As a concrete example, let $\cc{S}_1 = \{\cc{G}_{1,1}, \dots, \cc{G}_{1,m_1}\}$ and $\cc{S}_2 = \{\cc{G}_{2,1}, \dots, \cc{G}_{2, m_2}\}$, where each $\cc{G}_{i, j}~(i\in [2], j\in [m_i])$ is an ellipsoid denoted as $\cc{E}(\mmc_{i,j}, \mm{\Sigma}_{i,j})$, and let $\Omega_1 = \conv(\cc{S}_1),\ \Omega_2 = \conv(\cc{S}_2)$. 
Assuming $\Omega_1 \cap \Omega_2 = \varnothing$, then Problem~\ref{prob:sib_general_problem} becomes finding an SIB that intersects the convex hulls of two sets of ellipsoids.
Accordingly, the {\sc Oracle} problem \eqref{eq:sib_general_oracle} can be reduced to a sequence of problems in the form of $\min \big\{\mmh_i^T \mmu_i : \mmu_i \in \cc{E}(\mmc_{i,j}, \mm{\Sigma}_{i,j}) \big\}$, which has an analytical solution. 
In Section~\ref{sec:sib_cvx}, we saw that an SIB that intersects two convex polytopes implies a hard-margin SVM that separates the two corresponding point sets. 
Similarly, here an SIB that intersects $\Omega_1$ and $\Omega_2$ also corresponds to a linear classifier that separates two sets of ellipsoids with the largest margin. 
The latter coincides with Shivaswamy et al.~\cite[Sec.~3.3]{ellipsoidsvm:PCA2006}, where they used ellipsoids to model uncertain data points in classification problems. 
A simplified model using balls is proposed by Trafalis and Alwazzi~\cite{ballsvm:TS2010}, which is also covered by this example when $\cc{G}_{i,j}$ are just balls.
Figure~\ref{fig:ellipsoid_svm} illustrates the SIB and linear classification problems for two sets of ellipsoids.

\section{Soft-margin SIB}\label{sec:soft}

\begin{figure}[h]
\begin{center}
\includegraphics[width=.6\linewidth]{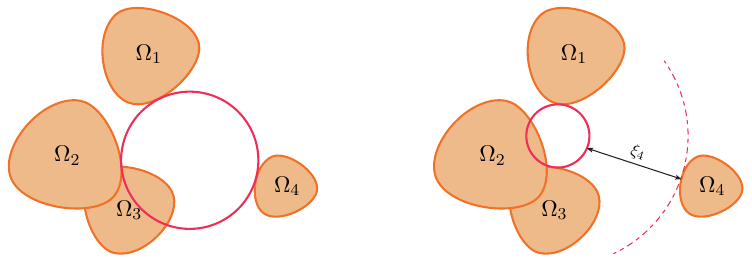}
\end{center}
\vspace{-1.5em}
\caption{{\em Left:} the original SIB. 
{\em Right:} the corresponding soft-margin SIB\change{, where $\xi_1 = \xi_2 = \xi_3 = 0$.}
}\label{fig:soft_sib}
\end{figure}

Real-world datasets are often noisy and contain outliers. For the SIB problem, it is easily seen that even one outlier object can lead to completely different results. We are therefore motivated to investigate a more robust version of the SIB problem that handles the presence of outliers.
One possible approach is to choose an integer parameter $k\in [n]$ and find the smallest ball that intersects at least $k$ objects in the input, in which case the objects that are not intersected by the ball can be considered outliers. 
Nevertheless, this problem is shown to be strongly NP-hard even when the input objects are just points, and there is no fully polynomial-time approximation scheme (FPTAS) for this problem unless P=NP~\cite{kSEBhardness:Vladimir2015}.
Therefore, in this section, we propose a non-combinatorial variant to handle the outliers, which is inspired by $\ell_1$-loss SVDD. Formally, we consider the following optimization problem.
\begin{problem}[Soft-margin SIB]\label{prob:soft-sib}
\[
\begin{aligned}
    \underset{\mmz, \mmv_1, \dots, \mmv_n, \mmxi, r}{\rm minimize} \quad
    & r + C \sum_{i=1}^n \xi_i\\
    {\rm subject\ to} \quad
    & \|\mmz - \mmv_i\| \le r + \xi_i,\ \forall i \in [n],\\
    & \mmv_i \in \Omega_i,\ \forall i \in [n],\\
    & \mmxi\ge 0,\ r \ge 0.
\end{aligned}
\]    
\end{problem}
In the above problem, each $\xi_i$ is a slack variable (or a ``penalty'' for each input object), and $C>0$ is a user-specified parameter that controls the total penalty.
We call this problem the soft-margin SIB problem or Soft-SIB.
Figure~\ref{fig:soft_sib} illustrates an instance of the problem with an appropriate choice of $C$.

\begin{proposition}\label{prop:soft_sib}
Let $\Omega_1, \dots, \Omega_n$ be compact convex sets in $\bR^d$, and let $D$ be the diameter of $\conv(\{\Omega_i\}_{i=1}^n)$. Consider Problem~\ref{prob:soft-sib}:
\begin{enumerate}
    \item[$(i)$] A point $(\mmz, \mmv_1, \dots, \mmv_n, \mmxi, r)$ is an optimal solution only if $\mmz \in \conv(\{\Omega_i\}_{i=1}^n),\ \mmxi \le D$, and $r \le D$.
    \item[$(ii)$] It is a convex program and a solution always exists. 
    \item[$(iii)$] If $C > 1$, then $(\mmz, \mmv_1, \dots, \mmv_n, \mmxi, r)$ is an optimal solution only if $\mmxi = 0$. 
    \item[$(iv)$] If $C < {1\over n}$, then $(\mmz, \mmv_1, \dots, \mmv_n, \mmxi, r)$ is an optimal solution only if $r = 0$.
\end{enumerate}
\end{proposition}

\begin{proof}
$(i)$
Let $\cc{D}$ be the convex hull $\conv(\{\Omega_i\}_{i=1}^n)$.
Assume there exists a solution $(\mmz, \mmv_1, \dots, \mmv_n, \mmxi, r)$ such that $\mmz\notin \cc{D}$. Let $\mmz' = \Pi(\mmz, \cc{D})$ be the Euclidean projection from $\mmz$ to $\cc{D}$. Similar to $(ii)$ in Proposition~\ref{prop:sib_general}, we have
\[
    \|\mmv_i - \mmz' \| < \|\mmv_i - \mmz\| \le r + \xi_i,\ \forall i \in [n].
\]
If $r = 0$, for $i = 1,\dots,n$, we let $\xi_i' = \|\mmv_i - \mmz'\|$. Then $\xi_i' < \xi_i, \forall i \in [n]$, and $(\mmz', \mmv_1, \dots, \mmv_n, \mmxi', r)$ is a better solution. 
Otherwise if $r > 0$, let $r' = \max_{i\in [n]} \|\mmv_i - \mmz'\| - \xi_i$ and let $r'' = \max\{r', 0\}$. Then we have $0\le r'' < r$, and $(\mmz', \mmv_1, \dots, \mmv_n, \mmxi, r'')$ is a better solution.

Let $(\mmz, \mmv_1, \dots, \mmv_n, \mmxi, r)$ be an optimal solution of Problem~\ref{prob:soft-sib}.
Because $\mmz \in \cc{D}$ and $\mmv_i \in \cc{D}, \forall i \in [n]$, we have $\|\mmz - \mmv_i\| \le D, \forall i \in [n]$. As a consequence, $r + \xi_i$ is at most $D$ and the optimum will have $\mmxi \le D$ and $r\le D$.

\medskip
$(ii)$
The proof of convexity is almost identical to $(iii)$ in Proposition~\ref{prop:sib_general}, thus is omitted here.
From the above $(i)$ we know that introducing the additional constraints $\mmz \in \conv(\{\Omega\}_{i=1}^n)$, $\mmxi \le D$ and $r\le D$ will not change the optimal set of Problem~\ref{prob:soft-sib}. The problem is equivalent to:
\[\begin{aligned}
    \underset{\mmz, \mmv_1, \dots, \mmv_n, \mmxi, r}{\rm maximize} \quad
    & r + C \sum_{i=1}^n \xi_i \\
    {\rm subject\ to} \quad 
    & \|\mmz - \mmv_i\| \le r + \xi_i,\ \forall i \in [n],\\
    & \mmv_i \in \Omega_i,\ \forall i \in [n],\\
    & \mmz \in \conv(\{\Omega_i\}_{i=1}^n), \\
    & 0\le \mmxi \le D,\\
    & 0\le r \le D.
\end{aligned}\]
Since the objective function of the above problem is continuous and the feasible region is closed and bounded, by the Weierstrass theorem~\cite[Theorem 2.30]{projtheory:Beck2014}, an optimal solution must exist.

\medskip
$(iii)$
Suppose $C > 1$.
Assume by contradiction that $(\mmz, \mmv_1, \dots, \mmv_n, \mmxi, r)$ is a solution with $\mmxi \ne 0$. 
Let $\xi_{\max} = \max_{i\in [n]} \xi_i$ and $r' = r + \xi_{\max}$.
We have:
\[\begin{aligned}
    & \|\mmz - \mmv_i\| \le r + \xi_i \le r + \xi_{\max} = r' + 0, \ \forall i \in [n],\\
    & r' + C\cdot 0 = r + \xi_{\max} < r + C\xi_{\max} \le r + C \sum_{i = 1}^n \xi_i,
\end{aligned}\]
which implies $(\mmz, \mmv_1, \dots, \mmv_n, \mmzero, r')$ is a better solution.

\medskip
$(iv)$
Suppose $C < {1\over n}$.
Assume by contradition that $(\mmz, \mmv_1, \dots, \mmv_n, \mmxi, r)$ is a solution with $r > 0$.
Let $\xi_i' = \xi_i + r$.
We have:
\[\begin{aligned}
    & \|\mmz - \mmv_i \| \le  r + \xi = 0 + \xi_i' , \ \forall i \in [n],\\
    & 0 + C \sum_{i=1}^n \xi_i' = C \sum_{i=1}^n (\xi_i + r) = C n r + C\sum_{i=1}^n \xi_i < r + C\sum_{i=1}^n \xi_i,
\end{aligned}\]
which implies $(\mmz, \mmv_1, \dots, \mmv_1, \mmxi', 0)$ is a better solution.
\end{proof}

\begin{remark}
From Proposition~\ref{prop:soft_sib}, we know that if the parameter $C > 1$, an optimal solution will have $\mmxi = 0$, in which case Problem~\ref{prob:soft-sib} is equivalent to the hard-margin problem (Problem~\ref{prob:sib_general_problem}). 
Let $\dist(\mmz, \Omega_i)$ be the minimum distance from $\mmz$ to any point in $\Omega_i$.
If $C < {1 \over n}$, an optimal solution must have $r = 0$, in which case Problem~\ref{prob:soft-sib} is to find a point $\mmz$ such that the sum of $\dist(\mmz, \Omega_i)$ is minimized. When the input sets $\Omega_i$ are singletons, the latter becomes the Geometric Median (a.k.a.\ Fermat-Weber) problem~\cite{geometricmedian:CLMPS2016}.
\end{remark}

In what follows, without loss of generality, we assume that the parameter $C$ satisfies ${1\over n}\le C \le 1$.
Let $\alpha^*$ be the optimal value of Problem~\ref{prob:soft-sib}, and let $\eps > 0$ be an error parameter. We say $(\mmz, \mmv_1, \dots, \mmv_n, \mmxi, r)$ is a $(1+\eps)$-approximate solution of Problem~\ref{prob:soft-sib} if it is feasible and its objective value satisfies $r + C \sum_i \xi_i \le (1 + \eps) \alpha^*$.

To find such an approximate solution, we first consider an approximate decision problem, which we called the {\em feasibility test problem} (FTP): given $\hat{\alpha} > 0$, we want to either determine $\hat{\alpha} < \alpha^*$ or verify that $(1 + \eps)\hat{\alpha} \ge \alpha^*$; and for the latter case, we find a feasible solution of Problem~\ref{prob:soft-sib} with objective value at most $(1 + \eps)\hat{\alpha}$. 
\begin{problem}[Feasibility test for Soft-SIB]\label{prob:soft-sib-ftp}
Given $\hat{\alpha} > 0$ and $\eps > 0$, verify that $\hat{\alpha} < \alpha^*$ or find $\mmz, \mmv_1, \dots, \mmv_n, \mmxi, r$ satisfying the following conditions:
\[
\begin{aligned}
    &r + C \sum_{i=1}^n \xi_i \le (1+\eps)\hat{\alpha},\\
    &\|\mmz - \mmv_i\| \le r + \xi_i, \ \forall i \in [n],\\
    &\mmv_i \in \Omega_i, \ \forall i \in [n],\\
    &\mmz \in \conv(\{\Omega_i\}_{i=1}^n),\\
    &\mmxi\ge 0, \ r \ge 0.
\end{aligned}
\]
\end{problem}

\begin{lemma}\label{lem:soft_sib_scg}
Problem~\ref{prob:soft-sib-ftp} can be modeled as the following symmetric cone game:
\beq\label{eq:soft-sib-game}
    \min_{\underbrace{\scalebox{.85}{$(\mmz, \mmv_1, \dots, \mmv_n, \mmxi, r)$}}_{\scalebox{.88}{$\hspace{1.6em}\mmx$}} \in \cc{A}}\ 
    \max_{\mmy \in \cc{B}}\ 
    \Big(
        \underbrace{\cproduct_{i=1}^n \begin{pmatrix}
            \mmv_i - \mmz \\
            -r -\xi_i
        \end{pmatrix}}_{\brmf(\mmx)}
    \Big) \bdiamond \mmy,
\eeq
where the convex set $\cc{A}$ is defined as:
\[
\begin{aligned}
    \cc{A} = \Big\{(\mmz, \mmv_1,\dots,\mmv_n, \mmxi, r) \in \bR^d \times \dots \times \bR^d \times \bR^n \times \bR : \mmz\in\conv(\{\Omega_i\}_{i=1}^n),&\\ 
    \mmv_i \in \Omega_i, \forall i\in [n],\  r + C\sum_{i=1}^n \xi_i \le \hat{\alpha},\ 0\le \mmxi \le D,\ 0 \le  r \le D &\Big\},
\end{aligned}    
\]
and $\cc{B}$ is the spectraplex of the product cone $\cc{C}$.
If the value of the game is greater than 0, then $\hat{\alpha} < \alpha^*$; otherwise, solving the game up to an additive error of ${\eps\hat{\alpha} \over \sqrt{2}}$ gives a solution satisfying all the conditions in Problem~\ref{prob:soft-sib-ftp}.
\end{lemma}

\begin{proof}

$(i)$ {\em Invalid case.} 
If the value of the game is greater than 0, then we have
\[
    \max_{\mmy \in \cc{B}}\ 
    \Big(
        \cproduct_{i=1}^n \begin{pmatrix}
            \mmv_i - \mmz \\
            -r -\xi_i
        \end{pmatrix}
    \Big) \bdiamond \mmy > 0 \quad 
    \text{for all } (\mmz, \mmv_1, \dots, \mmv_n, \mmxi, r) \in \cc{A}.
\]
Let $\hat{\mmy} \in \cc{B}$ be a maximier of the bilinear objective, then we have
\[
    \Big(
        \cproduct_{i=1}^n \begin{pmatrix}
            \mmz - \mmv_i \\
            r + \xi_i
        \end{pmatrix}
    \Big) \bdiamond \hat{\mmy} < 0
    \quad
    \text{for all } (\mmz, \mmv_1, \dots, \mmv_n, \mmxi, r) \in \cc{A}.
\]
By self-duality of $\cc{C}$, from the above we know that for all $(\mmz, \mmv_1, \dots, \mmv_n, \mmxi, r) \in \cc{A}$ 
\[
    \cproduct_{i=1}^n \begin{pmatrix}
        \mmz - \mmv_i \\
        r + \xi_i
    \end{pmatrix} \notin \cc{C}.
\]
That is, there is no point that can satisfy $\|\mmz - \mmv_i\| \le r + \xi_i \ \forall i \in [n]$ and the conditions in $\cc{A}$ simultaneously. 
However, from Proposition~\ref{prop:soft_sib}$(i)$ we know that if $\hat{\alpha} \ge \alpha^*$, any point in the optimal set of Problem~\ref{prob:soft-sib} satisfies these conditions. 
Therefore, if the value of the game is greater than 0, we can conclude that $\hat{\alpha} < \alpha^*$.

\medskip
$(ii)$ {\em From SCG to FTP.}
Let $(\tilde{\mmz}, \tilde{\mmv}_1, \dots, \tilde{\mmv}_n, \tilde{\mmxi}, \tilde{r}) \cproduct \tilde{\mmy} \in \cc{A} \times \cc{B}$ be a ${\eps\hat{\alpha} \over \sqrt{2}}$-Nash equilibrium of the SCG, which satisfies
\begin{align}
    &\Big(
        \cproduct_{i=1}^n \begin{pmatrix}
            \tilde{\mmz} - \tilde{\mmv}_i \\
            -\tilde{r} - \tilde{\xi}_i
        \end{pmatrix}
    \Big) \bdiamond \mmy \le \lambda^* + {\eps \hat{\alpha} \over \sqrt{2}}
    \quad
    \text{for all } \mmy \in \cc{B},
    \label{eq:soft-sib-game-equiv}
    \\
    \text{and}\quad
    &\Big(
        \cproduct_{i=1}^n \begin{pmatrix}
            \mmz - \mmv_i \\
            -r - \xi_i
        \end{pmatrix}
    \Big) \bdiamond \tilde{\mmy} \ge \lambda^* - {\eps\hat{\alpha} \over \sqrt{2}}
    \quad
    \text{for all } (\mmz, \mmv_1, \dots, \mmv_n, \mmxi, r) \in \cc{A},\notag
\end{align}
where $\lambda^*$ is the value of the game. 
If $\lambda^* \le 0$, then from \eqref{eq:soft-sib-game-equiv} we have
\[
    \Big(
        \cproduct_{i=1}^n \begin{pmatrix}
            \tilde{\mmz} - \tilde{\mmv}_i \\
            -\tilde{r} - \tilde{\xi}_i
        \end{pmatrix}
    \Big) \bdiamond \mmy \le {\eps \hat{\alpha} \over \sqrt{2}}
    \quad \Rightarrow \quad 
    \Big(
        \cproduct_{i=1}^n \begin{pmatrix}
            \tilde{\mmv}_i - \tilde{\mmz} \\
            \tilde{r} + \tilde{\xi}_i
        \end{pmatrix}
    \Big) \bdiamond \mmy + {\eps \hat{\alpha} \over \sqrt{2}} \ge 0
    \quad \text{for all } \mmy \in \cc{B},
\]
which implies $\cproduct_{i=1}^n \begin{pmatrix}
    \tilde{\mmv}_i - \tilde{\mmz} \\
    \tilde{r} + \tilde{\xi}_i 
\end{pmatrix} + {\eps \hat{\alpha} \over \sqrt{2}} \mme \in \cc{C}$, or equivalently:
\beq\label{eq:soft-sib-game-dist-condition}
    \|\tilde{\mmv}_i  - \tilde{\mmz}\| \le \tilde{r} + \tilde{\xi}_i + \eps \hat{\alpha}, \ \forall i \in [n].
\eeq
Let $\tilde{r}' = \tilde{r} + \eps \hat{\alpha}$, then we have
\[
    \tilde{r}' + C \sum_{i=1}^n \tilde{\xi}_i = \tilde{r} + C \sum_{i=1}^n \tilde{\xi}_i + \eps \hat{\alpha} \le (1 + \eps) \hat{\alpha},
\]
where the inequality is due to $\tilde{r} + C \sum_{i} \tilde{\xi}_i \le \hat{\alpha}$.
Together with \eqref{eq:soft-sib-game-dist-condition}, we conclude that the point $(\tilde{\mmz}, \tilde{\mmv}_1, \dots, \tilde{\mmv}_n, \tilde{\mmxi}, \tilde{r}')$ satisfies all the conditions in Problem~\ref{prob:soft-sib-ftp}.
\end{proof}

It is worth noting that there always exists a feasible solution for Problem~\ref{prob:soft-sib} where $\mmxi = 0$ and $r = D$ (which can be directly obtained from a feasible solution of Problem~\ref{prob:sib_general_problem}). 
So $D + C\cdot 0 = D$ is always an upper bound on the optimal value $\alpha^*$. Therefore, without loss of generality, we can assume that the input parameter $\hat{\alpha}$ of Problem~\ref{prob:soft-sib-ftp} always satisfies~$\hat{\alpha} \le D$.

\begin{theorem}\label{thm:soft_sib_ftp}
Let $D$ be the diameter of the input and let $\hat{\alpha} \in (0, D]$. Assume a subroutine that solves Problem~\ref{prob:sib_subproblem} with running time $O(S)$. Then there is an algorithm that solves the feasibility test problem (Problem~\ref{prob:soft-sib-ftp}) in $O({D^2 (S + nd)\log n \over \eps^2 \hat{\alpha}^2})$ time.
\end{theorem}

\begin{proof}
We employ Algorithm~\ref{algo:scg_product} to compute a ${\eps \hat{\alpha} \over \sqrt{2}}$-Nash equilibrium of the SCG~\eqref{eq:soft-sib-game}. 
During the process, if there exists a certain iteration $t$ such that $\displaystyle\min_{\mmx \in \cc{A}}\ \brmf(\mmx) \bdiamond \xt{\mmy}{t} > 0$, then we know that the value of the game $\lambda^* = \displaystyle \max_{\mmy\in\cc{B}} \min_{\mmx\in\cc{A}}\ \brmf(\mmx) \bdiamond \mmy > 0$, and by Lemma~\ref{lem:soft_sib_scg} the value $\hat{\alpha} < \alpha^*$.
Otherwise, the output of the algorithm provides a solution satisfying all the conditions in Problem~\ref{prob:soft-sib-ftp}.
In what follows, we discuss the details of the {\sc Oracle} and analyse the total running time of the algorithm.

\medskip
$(i)$ {\em The {\sc Oracle} process.}
For any given $\mmy\in\cc{B}$, the {\sc Oracle} finds $\mmx = \argmin_{\mmx\in\cc{A}} \brmf(\mmx)\bdiamond\mmy$.
Let $\mmx = (\mmz, \mmv_1, \dots, \mmv_n, \mmxi, r)\in \bR^d \times \dots \times \bR^d \times \bR^n \times \bR$ and $\mmy = \cproduct_{i=1}^n (\bar{\mmy}_i, y_{i,0})\in\bV$. Then the problem can be formulated as:
\beq\label{eq:soft_sib_oracle}
\begin{aligned}
    \underset{\mmz, \mmv_1, \dots, \mmv_n, \mmxi, r}{\rm minimize}\quad 
    & \sum_{i=1}^n \bar{\mmy}_i^T \mmv_i - \big(\sum_{i=1}^n \bar{\mmy}_i \big)^T \mmz - \sum_{i=1}^n y_{i,0}\xi_i - \big(\sum_{i=1}^n y_{i,0}\big)r \\
    {\rm subject\ to}\quad & \mmz\in\conv(\{\Omega_i\}_{i=1}^n),\\  
    & \mmv_i\in \Omega_i,\ \forall i \in [n],\\
    & r + C \sum_{i=1}^n \xi_i \le \hat{\alpha},\\
    & 0 \le \mmxi \le D, \\
    & 0 \le r \le D.
\end{aligned}
\eeq
Observe that the constraints on $\mmz$ and each $\mmv_i$ are independent from those on $\mmxi$ and $r$. Moreover, the sub-problem corresponds to $(\mmz, \mmv_1, \dots, \mmv_n)$ is identical to the {\sc Oracle} problem \eqref{eq:sib_general_oracle} corresponding to the hard-margin SIB problem, which can be solved in $O(S + nd)$ time.
Next, we introduce an approach to address the sub-problem corresponds to $(\mmxi, r)$.
For $\mmxi$ and $r$, the problem can be simplified to:
\beq\label{eq:soft_sib_oracle_simple}
\max \Big\{ 
    \sum_{i=1}^n y_{i,0} \xi_i + {1\over \sqrt{2}} r\  :\
    r + C \sum_{i=1}^n \xi_i \le \hat{\alpha},\
    0\le \mmxi \le D,\
    0\le r \le D
\Big\},
\eeq
which is optimizing a linear function over the intersection of a halfspace and an AABB. 
Observe that the optimum will be an extreme point of the intersected region, and if the solution has nonzero $\xi_i$'s, the values should be preferentially allocated to $\xi_i$'s with larger coefficients $y_{i,0}$ in the objective.
Moreover, on the other hand, the maximum value for any $\xi_i$ is $\min\{D, {\hat{\alpha}\over C}\}$.
Consider a simplex-like process as follows: 
Let $\beta = \min\{D, {\hat{\alpha}\over C}\}$, $k = \lceil {\hat{\alpha}\over C\beta} \rceil$ and let $y_{i_1,0}, \dots, y_{i_k,0}$ be the $k$ largest values in $\{y_{1,0}, \dots, y_{n,0}\}$, where $y_{i_1, 0} \ge \dots \ge y_{i_k, 0}$. Let $\cc{I}_k = \{i_1, \dots, i_k\}$ be the set of indices.
Then the point
\beq\label{eq:soft_sib_slx_init_point}
    r = 0,\quad \xi_i = \begin{cases}
        \beta, & \text{if $i \in \cc{I}_k \backslash \{i_k\} $},\\
        {\hat{\alpha}\over C} - {(k-1)\beta}, & \text{if $i = i_k$},\\
        0, &\text{otherwise},
    \end{cases}
\eeq
is an extreme point of the feasible region of~\eqref{eq:soft_sib_oracle_simple} because it is feasible and is the intersection point of the hyperplanes $r = 0,\ \xi_i = \beta\ \forall i \in \cc{I}_k \backslash \{i_k\},\ \xi_i = 0\ \forall i \in [n]\backslash \cc{I}_k$, and $r + C\sum_i \xi_i = \hat{\alpha}$.
Its objective value is $(\sum_{j=1}^{k-1} y_{i_j, 0})\cdot \beta + y_{i_k, 0}\cdot\big( {\hat{\alpha}\over C} - {(k-1)\beta} \big)$. Consider another extreme point by setting $\xi_{i_k} = 0$ and $r = \hat{\alpha} - (k-1)C\beta$ (which simulates a pivoting step of the simplex method). Its objective value is $(\sum_{j=1}^{k-1} y_{i_j, 0})\cdot \beta + {1\over \sqrt{2}}\big(\hat{\alpha} - (k-1)C\beta\big)$. If ${y_{i_k, 0}\over C} < {1\over \sqrt{2}}$, then the new objective value is larger and we move to the new point. 
Continue to check the value of $y_{i_j,0}$ for $j = k-1, k-2, \dots, 1$, one can find that if ${y_{i_j, 0}\over C} < {1\over \sqrt{2}}$, then setting $\xi_{i_j} = 0$ and increase the value of $r$ by $C\beta$ correspondingly gives a better result. 
Once we reach ${y_{i_j, 0}\over C}\ge {1\over \sqrt{2}}$ or all the $\xi_i$ are set to zero (in which case $r = \hat{\alpha}$), the solution cannot be further improved and is therefore optimal.

Fortunately, due to the nice property of the final solution, it is not necessary to perform the above simplex method in the actual implementation.
Let $\cc{I}_L\subseteq [n]$ be the collection of all indices $i$ satisfying ${y_{i,0}\over C} \ge {1\over \sqrt{2}}$. From the above discussion, we see that if $|\cc{I}_L|\ge k$ (in which case ${y_{i_k,0}\over C} \ge {1\over \sqrt{2}}$), then the solution of~\eqref{eq:soft_sib_oracle_simple} is given by~\eqref{eq:soft_sib_slx_init_point}, which can be computed in $O(n)$ time using the median-of-medians algorithm. Otherwise, if $|\cc{I}_L| < k$, the solution~is:
\beq\label{eq:soft_sib_slx_end_point}
    r = \hat{\alpha} - |\cc{I}_L|\cdot \beta, \quad
    \xi_i = \begin{cases}
        \beta, &\text{if $i\in \cc{I}_L$},\\
        0, &\text{otherwise},
    \end{cases}
\eeq
which can also be computed in $O(n)$ time.

In summary, if there is an {\sc Oracle} that solves problem~\eqref{eq:sib_general_oracle} for the hard-margin SIB problem, then with slight modification it can also solve~\eqref{eq:soft_sib_oracle}. Following $(i)$ of Theorem~\ref{thm:sib_general}, if there is an algorithm that solves Problem~\ref{prob:sib_subproblem} in $O(S)$ time, then there exists an {\sc Oracle} that solves~\eqref{eq:soft_sib_oracle} with a processing time of $O(S + nd + n) = O(S + nd)$.

\medskip
$(ii)$ {\em Width of {\sc Oracle}.}
Let $(\mmz, \mmv_1, \dots, \mmv_n, \mmxi, r)$ be the solution of~\eqref{eq:soft_sib_oracle}.
The width $\rho$ of the {\sc Oracle} is an upperbound on the spectral norm of $\brmf(\mmx) = \cproduct_{i=1}^n (\mmv_i - \mmz, -r - \xi_i)$ over all possible inputs $\mmy\in\cc{B}$. 
Since $\mmz \in \conv(\{\Omega_i\}_{i=1}^n)$ and $\mmv_i \in \Omega_i$, we have $\|\mmz - \mmv_i\| \le D,\ \forall i \in [n]$. Consequently,
\[\begin{aligned}
    \Big\|\cproduct_{i=1}^n (\mmv_i - \mmz, - r - \xi_i)\Big\|_\spe 
    &= \max_{k\in [2n]}\ \Big|\lambda_{k}\big(\cproduct_{i=1}^n (\mmv_i - \mmz, - r - \xi_i)\big) \Big| \\
    &= \max_{i\in [n]}\ {1\over \sqrt{2}} \big( \|\mmv_i - \mmz\| + r + \xi_i \big)\\
    &\le {1\over \sqrt{2}} ( D + r + \beta )\\
    &\le {3D \over \sqrt{2} }.
\end{aligned}\]
The last inequality is because $r\le D$ and $\beta\le D$.
Therefore, the {\sc Oracle} width is ${3D \over \sqrt{2}}$.

\medskip
$(iii)$ {\em Working with unknown $D$.} 
Similar to the hard-margin SIB problem, here the width is also proportional to the diameter $D$ and serves as a parameter in Algorithm~\ref{algo:scg_product}.
We can use the same approach (the doubling trick) discussed in $(iii)$ of Theorem~\ref{thm:sib_general} to address the scenarios where $D$ is unknown.

\medskip
$(iv)$ {\em Running time analysis.} 
We analyse the running time by assuming that the value of $D$ is determined. Running time with the doubling trick in $(iii)$ is the same.
From Theorem~\ref{thm:scg_product} we know that Alogorithm~\ref{algo:scg_product} can find a ${\eps\hat{\alpha} \over \sqrt{2}}$-Nash equilibrium of the SCG using $O({\rho^2 \log n \over (\eps \hat{\alpha})^2})$ calls to the {\sc Oracle}, with an additional processing time of $O(F + nd)$ per call. Substituting the width $\rho = {3 D \over \sqrt{2}}$ into $T$, then the iteration complexity becomes $O({D^2 \log n \over (\eps \hat{\alpha})^2})$. For any given $(\mmz, \mmv_1, \dots, \mmv_n, \mmxi, r)$, the affine map $\brmf$ can be computed in $O(nd)$ time, so the processing time between two {\sc Oracle} calls is $O(nd)$.
From $(i)$ we know that the {\sc Oracle} problem can be solved in $O(S + nd)$ time. Adding all together, we conclude that the total running time of the algorithm is $O({D^2 (S + nd) \log n \over (\eps \hat{\alpha})^2})$.
\end{proof}

Given the algorithm described in Theorem~\ref{thm:soft_sib_ftp} that solves the feasibility test problem (Problem~\ref{prob:soft-sib-ftp}), we can use a searching approach to narrow the range of $\hat{\alpha}$. Once the estimated value $\hat{\alpha}$ is close enough to the optimal value $\alpha^*$, we can obtain an approximation solution to the Soft-SIB problem (Problem~\ref{prob:soft-sib}).

\begin{theorem}\label{thm:soft_sib_opt}
Let $R = {D \over \alpha^*}$. Assume a subroutine that solves Problem~\ref{prob:sib_subproblem} with running time $O(S)$. Then there is an algorithm that computes a $(1+\eps)$-approximate solution of the Soft-SIB problem (Problem~\ref{prob:soft-sib}) in $O({R^2 (S + nd) \log n \over \eps^2 })$ time.
\end{theorem}

\begin{proof}
Pick a point $\hat{\mmv}_i$ in each object $\Omega_i$, and compute $E = \max_{j \ge 2} \|\mmv_1 - \mmv_j\|$. 
It is easy to see that $\alpha^* \le E$.
Define
\[
    L_0 = 0,\ U_0 = E,\ \hat{\alpha}_1 = L_0 + {1\over 3}(U_0 - L_0),\ \eps_1 = {1\over 3\hat{r}_1}(U_0 - L_0).
\]
Employ the algorithm described in Theorem~\ref{thm:soft_sib_ftp} to solve the FTP with the parameters $\hat{\alpha}_1$ and $\eps_1$. Then we can either conclude $\hat{\alpha}_1 < \alpha^*$ or find a feasible solution $(\mmz, \mmv_1, \dots, \mmv_n, \mmxi, r)$ with objective value at most $(1+\eps_1)\hat{\alpha}_1$. For the former case, we set $L_1 = L_0 + {1\over 3}(U_0 - L_0)$ and $U_1 = U_0$; and for the latter case, we set $L_1 = L_0$ and $U_1 = L_0 + {2\over 3}(U_0 - L_0)$.
Then for both cases, we have:
\[
    L_1 < \alpha^* \le U_1,\ \text{where}\ (U_1 - L_1) = {2\over 3}(U_0 - L_0).
\]
Repeat this process by setting $\hat{r}_\tau = L_{\tau-1} + {1\over 3}(U_{\tau-1} - L_{\tau-1})$ and $\eps_\tau = {1\over 3\hat{r}_\tau}(U_{\tau-1} - L_{\tau-1})$ for $\tau\ge 2$, until the resultant $L_\tau$ and $U_\tau$ satisfy $U_\tau \le (1+\eps) L_\tau$. Then the last solution $(\mmz, \mmv_1, \dots, \mmv_n, \mmxi, r)$ we obtained from this process has objective value at most 
\[
U_\tau \le (1+\eps)L_\tau < (1+\eps)\alpha^*,
\] 
which is an $(1 + \eps)$-approximate solution of Problem~\ref{prob:soft-sib}. 
For all the steps $\tau$, the parameters $\hat{\alpha}_\tau$ and $\eps_\tau$ satisfy $\eps_\tau \hat{\alpha}_\tau = {1\over 2}(U_\tau - L_\tau)$. Since the quantity $(U_\tau - L_\tau)$ shrinks by a constant fraction in each step, we know that the value of $\eps_\tau \hat{\alpha}_\tau$ decreases geometrically. 
The total running time of the algorithm will be dominated by the time for solving the last FTP (with the smallest $\eps_\tau \hat{\alpha}_\tau$ satisfying ${1\over 4} \eps \alpha^*\le \eps_\tau \hat{\alpha}_\tau \le {1\over 2} \eps \alpha^*$), which is $O({R^2 (S + nd) \log n \over \eps^2})$.
\end{proof}

In general, given a collection of compact convex objects as the input, if the algorithm introduced in Section~\ref{sec:sib} is capable of solving the hard-margin SIB problem (Problem~\ref{prob:sib_general_problem}), then with slight modification it can also solve the soft-margin SIB problem (Problem~\ref{prob:soft-sib}). Specifically, the algorithms introduced in Section~\ref{sec:sib_cvx} -- \ref{sec:extension} can all be adapted to solve the corresponding soft-margin SIB problems with input objects of convex polytopes, reduced polytopes, AABBs, balls, ellipsoids, etc.
The running time is similar to the corresponding hard-margin problems, except that the ratio $R$ is defined differently here.
In the following subsection, we discuss details of the Soft-SIB algorithm for the most basic case where the input objects are just points.

\subsection{$\ell_1$-loss SVDD}\label{sec:svdd}
When the input objects are simply points, the Soft-SIB problem becomes a variant of the $\ell_1$-loss SVDD problem~\cite{svdd:CLL2013}.
SVDD is a model which aims at finding spherically shaped boundaries around data sets. It is useful for abnormal detection and has been applied to a variety of application domains including biometrics, surveillance, and social network~\cite{lpsvdd:Arashloo2022}.
Let $\cc{P} = \{\mmp_1, \dots, \mmp_n\}$ be a point set in $\bR^d$, and let $\Omega_i = \{\mmp_i\}$ for $i = 1, \dots, n$.
Then the Soft-SIB problem (Problem~\ref{prob:soft-sib}) can be simplified to:
\begin{problem}[$\ell_1$-loss SVDD]\label{prob:l1-svdd}
\[
\begin{aligned}
    \underset{\mmz, \mmxi, r}{\rm maximize} \quad
    & r + C \sum_{i=1}^n \xi_i\\
    {\rm subject\ to} \quad
    & \|\mmz - \mmp_i\| \le r + \xi_i,\ \forall i \in [n],\\
    & \mmxi\ge 0,\ r \ge 0.
\end{aligned}
\]
\end{problem}

\begin{remark}
Note that Problem~\ref{prob:l1-svdd} is different from the commonly used formulation of $\ell_1$-loss SVDD~\cite[Equation~(7)]{svdd:CLL2013}, where the latter uses squared distance in the constraints, i.e.~$\|\mmz - \mmp_i\|^2 \le r + \xi_i,\ \forall i \in [n]$. 
The reason behind the preference for the squared distance in prior work is its differentiability, due to which the Lagrange dual problem can be easily derived and existing optimization tools can be used to solve the problem.
However, square distance provides less geometric intuition and makes it difficult to choose an appropriate hyperparameter $C$.
As a comparision, our formulation (Problem~\ref{prob:l1-svdd}) possesses a clearer geometric interpretation, but cannot be efficiently solved using existing optimization methods.
\end{remark}

We can employ the algorithm stated in Theorem~\ref{thm:soft_sib_opt} to solve Problem~\ref{prob:l1-svdd}. Then the {\sc Oracle} problem~\eqref{eq:soft_sib_oracle} becomes:
\beq\label{eq:svdd_oracle}
\begin{aligned}
    \underset{\mmz, \mmxi, r}{\rm minimize}\quad 
    & \sum_{i=1}^n \bar{\mmy}_i^T \mmp_i - \big(\sum_{i=1}^n \bar{\mmy}_i \big)^T \mmz - \sum_{i=1}^n y_{i,0}\xi_i - \big(\sum_{i=1}^n y_{i,0}\big)r\\
    {\rm subject\ to}\quad & \mmz\in\conv(\{\mmp_i\}_{i=1}^n),\\  
    & r + C \sum_{i=1}^n \xi_i \le \hat{\alpha},\\
    & 0 \le \mmxi \le D,\\
    & 0 \le r \ge D.
\end{aligned}
\eeq
Here the first term in the objective is a constant.
The optimum of $\mmxi$ and $r$ is given by either~\eqref{eq:soft_sib_slx_init_point} or~\eqref{eq:soft_sib_slx_end_point}, which can be computed in $O(n)$ time. Take a processing time of $O(nd)$ to compute $\mmh = \sum_i \bar{\mmy}_i$, the sub-problem for $\mmz$ reduces to $\max_{i \in [n]} \mmh^T \mmp_i$, which can be computed in $O(N)$ time, where $N$ is the number of non-zero coordinates in the input points. Since $N\le nd$, the complexity of the {\sc Oracle} is~$O(N + nd + n) = O(nd)$.

\begin{corollary}
Let $D$ be the diameter of the input and let $R = {D \over \alpha^*}$.
Given an error parameter $\eps > 0$, there is an algorithm that computes a $(1+\eps)$-approximate solution of Problem~\ref{prob:l1-svdd} with a running time of $O({R^2 nd \log n \over \eps^2})$. 
\end{corollary}

\section{Experimental results}\label{sec:experiment}

\begin{figure}[t]
\includegraphics[width=\linewidth]{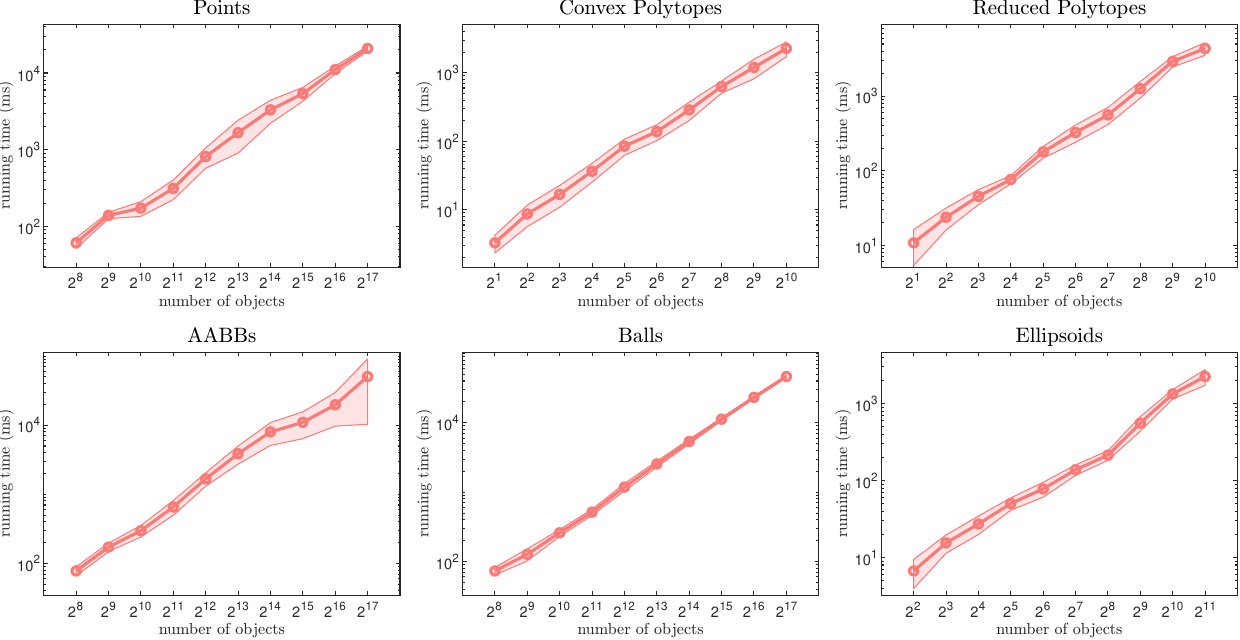}
\vspace{-1em}
\caption{Running time of our algorithm with different input sizes. 
The shaded regions represent the standard deviations.
\change{
The number of objects considered differs across object types due to their varying levels of complexity.
For each convex polytope and reduced polytope, we fix $m_i = 128$.
}
}
\label{fig:varying_n}
\end{figure}

\begin{table}[t]
\centering
\caption{Relative duality gap of the results computed by our algorithm with different input sizes.}\label{table:varying_n}
\setlength\extrarowheight{1pt}
\fontsize{6}{9.6}\selectfont
\begin{tabular}{|@{}C{5.8em}@{}||@{\hskip 4pt}c@{\hskip 4pt}|@{\hskip 4pt}c@{\hskip 4pt}|@{\hskip 4pt}c@{\hskip 4pt}|@{\hskip 4pt}c@{\hskip 4pt}|@{\hskip 4pt}c@{\hskip 4pt}|@{\hskip 4pt}c@{\hskip 4pt}|@{\hskip 4pt}c@{\hskip 4pt}|@{\hskip 4pt}c@{\hskip 4pt}|@{\hskip 4pt}c@{\hskip 4pt}|@{\hskip 4pt}c@{\hskip 4pt}|}
\hline
\diagbox[width=5em,height=2em,innerleftsep=1pt,innerrightsep=2pt]{Type}{$n$} 
& 256 & 512 & 1024 & 2048 & 4096 & 8192 & 16384 & 32768 & 65536 & 131072 \\
\hline\hline
Points & 1.76E-2 & 1.69E-2 & 1.69E-2 & 1.53E-2 & 1.49E-2 & 1.45E-2 & 1.26E-2 & 1.35E-2 & 1.20E-2 & 1.24E-2 \\
AABBs & 3.44E-2 & 3.34E-2 & 3.42E-2 & 3.03E-2 & 2.96E-2 & 2.96E-2 & 2.84E-2 & 3.08E-2 & 2.78E-2 & 2.69E-2 \\
Balls & 2.09E-2 & 1.75E-2 & 1.73E-2 & 1.26E-2 & 1.15E-2 & 1.15E-2 & 1.23E-2 & 1.06E-2 & 1.03E-2 & 9.88E-3 \\
\hline
\end{tabular}
\vspace{.3em}
\begin{tabular}{|@{}C{5.8em}@{}||@{\hskip 4pt}c@{\hskip 4pt}|@{\hskip 4pt}c@{\hskip 4pt}|@{\hskip 4pt}c@{\hskip 4pt}|@{\hskip 4pt}c@{\hskip 4pt}|@{\hskip 4pt}c@{\hskip 4pt}|@{\hskip 4pt}c@{\hskip 4pt}|@{\hskip 4pt}c@{\hskip 4pt}|@{\hskip 4pt}c@{\hskip 4pt}|@{\hskip 4pt}c@{\hskip 4pt}|@{\hskip 4pt}c@{\hskip 4pt}|}
\hline
\diagbox[width=5em,height=2em,innerleftsep=1pt,innerrightsep=2pt]{Type}{$n$} 
& 2 & 4 & 8 & 16 & 32 & 64 & 128 & 256 & 512 & 1024 \\
\hline\hline
Cvx. Poly. & 8.26E-2 & 2.57E-2 & 2.13E-2 & 1.92E-2 & 1.69E-2 & 1.94E-2 & 2.13E-2 & 2.04E-2 & 1.95E-2 & 2.01E-2 \\
Red. Poly. & 2.01E-2 & 1.70E-2 & 2.20E-2 & 2.09E-2 & 1.83E-2 & 1.74E-2 & 1.89E-2 & 1.76E-2 & 1.73E-2 & 1.70E-2 \\
\hline
\end{tabular}
\vspace{.3em}
\begin{tabular}{|@{}C{5.8em}@{}||@{\hskip 4pt}c@{\hskip 4pt}|@{\hskip 4pt}c@{\hskip 4pt}|@{\hskip 4pt}c@{\hskip 4pt}|@{\hskip 4pt}c@{\hskip 4pt}|@{\hskip 4pt}c@{\hskip 4pt}|@{\hskip 4pt}c@{\hskip 4pt}|@{\hskip 4pt}c@{\hskip 4pt}|@{\hskip 4pt}c@{\hskip 4pt}|@{\hskip 4pt}c@{\hskip 4pt}|@{\hskip 4pt}c@{\hskip 4pt}|}
\hline
\diagbox[width=5em,height=2em,innerleftsep=1pt,innerrightsep=2pt]{Type}{$n$} 
& 4 & 8 & 16 & 32 & 64 & 128 & 256 & 512 & 1024 & 2048 \\
\hline\hline
Ellipsoids & 5.20E-2 & 3.21E-2 & 2.51E-2 & 2.26E-2 & 3.01E-2 & 2.70E-2 & 2.58E-2 & 2.67E-2 & 2.54E-2 & 2.12E-2 \\
\hline
\end{tabular}
\end{table}

We implemented the algorithm for SIB introduced in Section~\ref{sec:sib} using C++. Our code is available at \cite{libsib}. To our knowledge, this is the first software capable of computing SIBs for \change{non-singleton objects}, even in two-dimensional spaces.
The algorithm for Soft-SIB described in Section~\ref{sec:soft} should share similar performance to the one for SIB, thus we did not explicitly conduct experiments for Soft-SIB.
We emphasize that all computational operations in our algorithms are inherently parallelizable, rending them well-suited for implementation on modern computing architectures like GPUs to benefit applications.

In practice, our program does not need to finish the full number of iterations as stated in the theoretical analysis. Instead, we employ an early-stopping strategy that is often used in convex optimization software.
Let $(\xt{\tilde{\mmz}}{t}, \xt{\tilde{\mmv}}{t}_1, \dots, \xt{\tilde{\mmv}}{t}_n, \xt{\tilde{\mmy}}{t})$ be the average point of the first $t$ iterates produced by the algorithm.
Then we compute
\[
\begin{aligned}
    &\nu_{\xt{\tilde{\mmx}}{t}} = \max_{\mmy \in \cc{B}}\ 
    \Big(
    \cproduct_{i=1}^n \begin{pmatrix}
        \xt{\tilde{\mmv}}{t}_i - \xt{\tilde{\mmz}}{t} \\
        0
    \end{pmatrix}
    \Big) \bdiamond \mmy, \\
    &\nu_{\xt{\tilde{\mmy}}{t}} = \min_{(\mmz, \mmv_i, \dots, \mmv_n) \in \cc{A}} \Big(
        \cproduct_{i=1}^n \begin{pmatrix}
            {\mmv}_i - {\mmz} \\
            0
        \end{pmatrix}
    \Big) \bdiamond \xt{\tilde{\mmy}}{t}.
\end{aligned}
\]
Note that the ball centered at $\xt{\tilde \mmz}{t}$ with radius $\sqrt{2}\cdot \nu_{\xt{\tilde{\mmx}}{t}}$ intersects every input object.
From Lemma~\ref{lem:sib_game} we know that $\nu_{\xt{\tilde{\mmy}}{t}} \le {r^* \over \sqrt{2}} \le \nu_{\xt{\tilde{\mmx}}{t}}$. Therefore, if $\nu_{\xt{\tilde{\mmy}}{t}} > 0$, then the quantity ${\nu_{\xt{\tilde{\mmx}}{t}} - \nu_{\xt{\tilde{\mmy}}{t}} \over \nu_{\xt{\tilde{\mmy}}{t}}}$ is an upper bound on the relative error incurred by the solution $(\xt{\tilde \mmz}{t}, \xt{\tilde\mmv}{t}_1, \dots, \xt{\tilde\mmv}{t}_n, \sqrt{2}\cdot \nu_{\xt{\tilde{\mmx}}{t}})$. We referred to this quantity as the {\em relative duality gap}.
Our program terminates once there is no significant improvement on the relative duality gap across iterations, and report $(\xt{\tilde \mmz}{t}, \xt{\tilde\mmv}{t}_1, \dots, \xt{\tilde\mmv}{t}_n, \sqrt{2}\cdot \nu_{\xt{\tilde{\mmx}}{t}})$ as the~final~result.
Figure~\ref{fig:visualization} visualizes the trajectories of $\xt{\tilde{\mmz}}{t}$ and $\{\xt{\tilde{\mmv}}{t}_i\}_{i=1}^n$ for some 2D examples. \change{For the case of a point set (the upper left figure), we omit the trajectories of $\{\xt{\tilde{\mmv}}{t}_i\}_{i=1}^n$.}
It is evident that the points converge to their final positions very quickly.
Interestingly, for those input objects with a large overlap with the final SIB, the corresponding $\xt{\tilde{\mmv}}{t}_i$ also strives to converge to the point closest to the center $\xt{\tilde\mmz}{t}$.

We examine the performance of our program in terms of the running time and the quality of the result (measured by the relative duality gap). 
All the experiments are executed on a machine with an Intel Core i7-9700K CPU with 32GB memory.
Each result reported below is averaged over ten independently generated instances.

\begin{figure}[t]
\includegraphics[width=\linewidth]{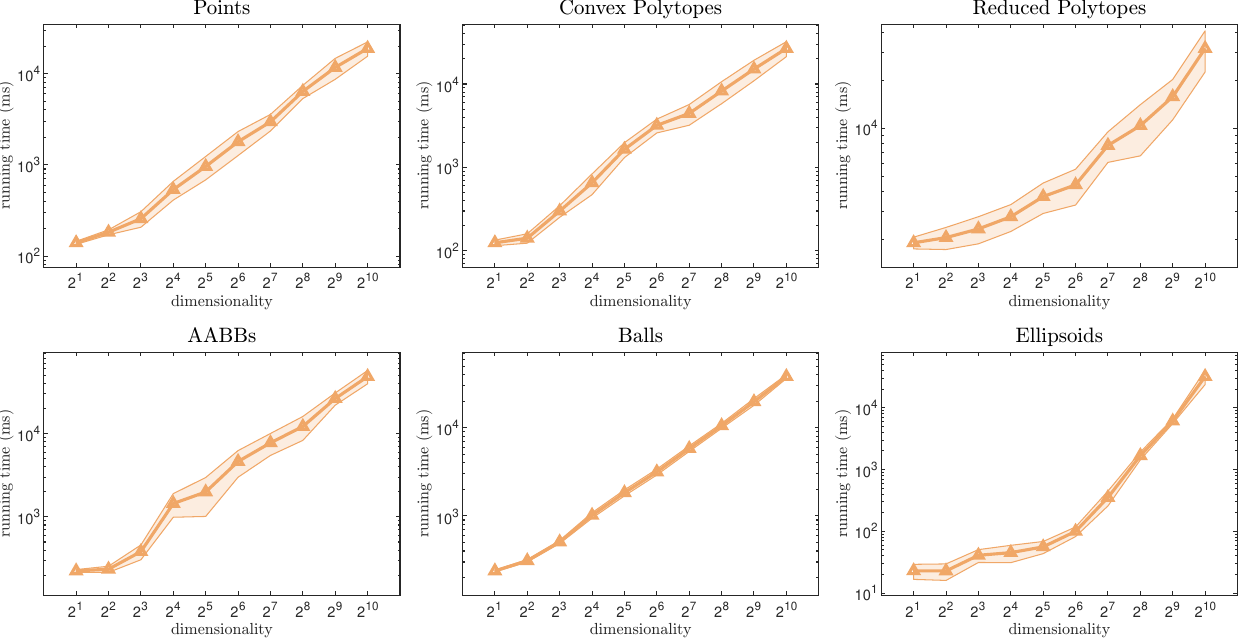}
\vspace{-1em}
\caption{Running time of our algorithm in different dimensions. 
The shaded regions represent the standard deviations.
\change{
For points, AABBs, and balls, we fix $n = 10^4$.
For convex polytopes, reduced polytopes, and ellipsoids, $n = 100$.
For each convex polytope and reduced~polytope,~$m_i=1280$.
}
}\label{fig:varying_d}
\end{figure}

\begin{table}[t]
\centering
\caption{Relative duality gap of the results computed by our algorithm in different dimensions.}\label{table:varying_d}
\setlength\extrarowheight{1pt}
\fontsize{6}{9.6}\selectfont
\begin{tabular}{|@{}C{5.8em}@{}||@{\hskip 4pt}c@{\hskip 4pt}|@{\hskip 4pt}c@{\hskip 4pt}|@{\hskip 4pt}c@{\hskip 4pt}|@{\hskip 4pt}c@{\hskip 4pt}|@{\hskip 4pt}c@{\hskip 4pt}|@{\hskip 4pt}c@{\hskip 4pt}|@{\hskip 4pt}c@{\hskip 4pt}|@{\hskip 4pt}c@{\hskip 4pt}|@{\hskip 4pt}c@{\hskip 4pt}|@{\hskip 4pt}c@{\hskip 4pt}|}
\hline
\diagbox[width=5em,height=2em,innerleftsep=1pt,innerrightsep=2pt]{Type}{$d$} 
& 2 & 4 & 8 & 16 & 32 & 64 & 128 & 256 & 512 & 1024 \\
\hline\hline
Points & 6.46E-2 & 5.30E-2 & 3.45E-2 & 2.31E-2 & 1.85E-2 & 1.52E-2 & 1.24E-2 & 1.18E-2 & 1.01E-2 & 1.03E-2 \\
Cvx. Poly. & 1.02E-1 & 9.80E-2 & 5.82E-2 & 3.49E-2 & 2.56E-2 & 2.06E-2 & 1.50E-2 & 1.40E-2 & 1.23E-2 & 1.16E-2 \\
Red. Poly. & 9.55E-2 & 5.94E-2 & 4.77E-2 & 2.65E-2 & 2.22E-2 & 1.89E-2 & 1.69E-2 & 1.41E-2 & 1.26E-2 & 9.88E-3 \\      
AABBs & 5.80E-2 & 7.99E-2 & 8.37E-2 & 7.25E-2 & 3.44E-2 & 3.06E-2 & 2.30E-2 & 1.98E-2 & 1.55E-2 & 1.40E-2 \\
Balls & 9.58E-2 & 6.21E-2 & 4.34E-2 & 2.55E-2 & 1.60E-2 & 1.05E-2 & 9.87E-3 & 9.95E-3 & 9.87E-3 & 9.95E-3 \\
Ellipsoids & 9.45E-2 & 6.89E-2 & 6.58E-2 & 4.38E-2 & 3.40E-2 & 2.96E-2 & 2.00E-2 & 1.96E-2 & 1.61E-2 & 1.34E-2 \\ 
\hline
\end{tabular}
\end{table}

To understand how our algorithm scales with the input sizes, we generated input instances with varying number of objects $n$ across a large range, while fixing the dimensionality $d$ to a moderate value of 64.
The running time is shown in Figure~\ref{fig:varying_n}. 
It is noticeable that the growth rate of the running time of our algorithm is consistent across the tested ranges. This matches the theoretical analysis which indicates the running time to be nearly linear in $n$.
The relative duality gap incurred by the final results is reported in Table~\ref{table:varying_n}.
For most of the inputs, the gap is below 3.00E-2, and the quality of the results is stable across significantly different $n$.

We also generated inputs with different number of dimensions $d$, while fixing the number of objects $n$ (and the size of point sets $m_i$ for the convex polytopes and reduced polytopes).
The running time is shown in Figure~\ref{fig:varying_d}. It is clear to see that the running time for points, convex polytopes, AABBs and balls grows approximately linearly with $d$.
The running time for reduced polytopes increases faster for larger $d$. This may be due to the additional work (e.g.\ median search) involved in solving the {\sc Oracle} problems for reduced polytopes.
For ellipsoids, the running time grows approximately quadratically with $d$, which is also consistent with the theoretical analysis.
Table~\ref{table:varying_d} reports the relative duality gap incurred by the final results. 
Interestingly, there is a general trend that the gap decays as $d$ increases.
For large number of dimensions, the gaps for all types of input objects are around 1.00E-2.

\section{Discussion}\label{sec:disc}
The main results of this paper are two approximation algorithms, for the general smallest intersecting ball problem and its soft-margin variant respectively.
The input objects of the problems are only required to be compact and convex, which greatly expands the application domain of the problems and our methods.
We implemented our smallest intersecting ball algorithm for various types of input objects including convex polytopes, reduced polytopes, axis-aligned bounding boxes, etc. Empirical assessment substantiates the practical efficiency of the proposed method.

The advances in smallest intersecting ball algorithms are founded upon our novel algorithmic framework termed symmetric cone games. Future research could explore whether this framework provides advantages for addressing other optimization problems. Regarding the smallest intersecting ball problems, a natural extension would investigate whether our results generalize to broader vector spaces, such as Hilbert and Banach spaces.
With advanced techniques such as mirror-prox and random sampling, it is promising that some of our results could be improved to have faster convergence rate or sublinear running time.
Kernelized versions of and dual methods for the problems would also be valuable topics~to~investigate.

\bibliography{reference}


\begin{thebibliography}{44}
\ifx \bisbn   \undefined \def \bisbn  #1{ISBN #1}\fi
\ifx \binits  \undefined \def \binits#1{#1}\fi
\ifx \bauthor  \undefined \def \bauthor#1{#1}\fi
\ifx \batitle  \undefined \def \batitle#1{#1}\fi
\ifx \bjtitle  \undefined \def \bjtitle#1{#1}\fi
\ifx \bvolume  \undefined \def \bvolume#1{\textbf{#1}}\fi
\ifx \byear  \undefined \def \byear#1{#1}\fi
\ifx \bissue  \undefined \def \bissue#1{#1}\fi
\ifx \bfpage  \undefined \def \bfpage#1{#1}\fi
\ifx \blpage  \undefined \def \blpage #1{#1}\fi
\ifx \burl  \undefined \def \burl#1{\textsf{#1}}\fi
\ifx \doiurl  \undefined \def \doiurl#1{\url{https://doi.org/#1}}\fi
\ifx \betal  \undefined \def \betal{\textit{et al.}}\fi
\ifx \binstitute  \undefined \def \binstitute#1{#1}\fi
\ifx \binstitutionaled  \undefined \def \binstitutionaled#1{#1}\fi
\ifx \bctitle  \undefined \def \bctitle#1{#1}\fi
\ifx \beditor  \undefined \def \beditor#1{#1}\fi
\ifx \bpublisher  \undefined \def \bpublisher#1{#1}\fi
\ifx \bbtitle  \undefined \def \bbtitle#1{#1}\fi
\ifx \bedition  \undefined \def \bedition#1{#1}\fi
\ifx \bseriesno  \undefined \def \bseriesno#1{#1}\fi
\ifx \blocation  \undefined \def \blocation#1{#1}\fi
\ifx \bsertitle  \undefined \def \bsertitle#1{#1}\fi
\ifx \bsnm \undefined \def \bsnm#1{#1}\fi
\ifx \bsuffix \undefined \def \bsuffix#1{#1}\fi
\ifx \bparticle \undefined \def \bparticle#1{#1}\fi
\ifx \barticle \undefined \def \barticle#1{#1}\fi
\bibcommenthead
\ifx \bconfdate \undefined \def \bconfdate #1{#1}\fi
\ifx \botherref \undefined \def \botherref #1{#1}\fi
\ifx \url \undefined \def \url#1{\textsf{#1}}\fi
\ifx \bchapter \undefined \def \bchapter#1{#1}\fi
\ifx \bbook \undefined \def \bbook#1{#1}\fi
\ifx \bcomment \undefined \def \bcomment#1{#1}\fi
\ifx \oauthor \undefined \def \oauthor#1{#1}\fi
\ifx \citeauthoryear \undefined \def \citeauthoryear#1{#1}\fi
\ifx \endbibitem  \undefined \def \endbibitem {}\fi
\ifx \bconflocation  \undefined \def \bconflocation#1{#1}\fi
\ifx \arxivurl  \undefined \def \arxivurl#1{\textsf{#1}}\fi
\csname PreBibitemsHook\endcsname

\bibitem[\protect\citeauthoryear{Chang et~al.}{2013}]{svdd:CLL2013}
\begin{botherref}
\oauthor{\bsnm{Chang}, \binits{W.-C.}},
\oauthor{\bsnm{Lee}, \binits{C.-P.}},
\oauthor{\bsnm{Lin}, \binits{C.-J.}}:
A revisit to support vector data description.
Dept. Comput. Sci., Nat. Taiwan Univ., Taipei, Taiwan, Tech. Rep
(2013)
\end{botherref}
\endbibitem

\bibitem[\protect\citeauthoryear{Tapolcai et~al.}{2020}]{netplan:TRVG2020}
\begin{barticle}
\bauthor{\bsnm{Tapolcai}, \binits{J.}},
\bauthor{\bsnm{Rónyai}, \binits{L.}},
\bauthor{\bsnm{Vass}, \binits{B.}},
\bauthor{\bsnm{Gyimóthi}, \binits{L.}}:
\batitle{{Fast Enumeration of Regional Link Failures Caused by Disasters With Limited Size}}.
\bjtitle{IEEE/ACM Transactions on Networking}
\bvolume{28}(\bissue{6}),
\bfpage{2421}--\blpage{2434}
(\byear{2020})
\end{barticle}
\endbibitem

\bibitem[\protect\citeauthoryear{Löffler and {van Kreveld}}{2010}]{siblowdim:LK2010}
\begin{barticle}
\bauthor{\bsnm{Löffler}, \binits{M.}},
\bauthor{\bsnm{{van Kreveld}}, \binits{M.}}:
\batitle{Largest bounding box, smallest diameter, and related problems on imprecise points}.
\bjtitle{Computational Geometry}
\bvolume{43}(\bissue{4}),
\bfpage{419}--\blpage{433}
(\byear{2010})
\end{barticle}
\endbibitem

\bibitem[\protect\citeauthoryear{Marshall and Olkin}{1960}]{multichebyshev:AI1960}
\begin{botherref}
\oauthor{\bsnm{Marshall}, \binits{A.W.}},
\oauthor{\bsnm{Olkin}, \binits{I.}}:
Multivariate chebyshev inequalities.
The Annals of Mathematical Statistics,
1001--1014
(1960)
\end{botherref}
\endbibitem

\bibitem[\protect\citeauthoryear{Dax}{2006}]{distance:Achiya2006}
\begin{barticle}
\bauthor{\bsnm{Dax}, \binits{A.}}:
\batitle{The distance between two convex sets}.
\bjtitle{Linear Algebra and its Applications}
\bvolume{416}(\bissue{1}),
\bfpage{184}--\blpage{213}
(\byear{2006})
\end{barticle}
\endbibitem

\bibitem[\protect\citeauthoryear{G{\"a}rtner and Jaggi}{2009}]{pd:GJ2009}
\begin{bchapter}
\bauthor{\bsnm{G{\"a}rtner}, \binits{B.}},
\bauthor{\bsnm{Jaggi}, \binits{M.}}:
\bctitle{Coresets for polytope distance}.
In: \bbtitle{Proceedings of the Twenty-fifth Annual Symposium on Computational Geometry},
pp. \bfpage{33}--\blpage{42}
(\byear{2009})
\end{bchapter}
\endbibitem

\bibitem[\protect\citeauthoryear{Bennett and Bredensteiner}{2000}]{svm:BB2000}
\begin{bchapter}
\bauthor{\bsnm{Bennett}, \binits{K.P.}},
\bauthor{\bsnm{Bredensteiner}, \binits{E.J.}}:
\bctitle{Duality and geometry in {SVM} classifiers}.
In: \bbtitle{Proceedings of the Seventeenth International Conference on Machine Learning},
pp. \bfpage{57}--\blpage{64}
(\byear{2000})
\end{bchapter}
\endbibitem

\bibitem[\protect\citeauthoryear{Crisp and Burges}{1999}]{svm:CB1999}
\begin{bchapter}
\bauthor{\bsnm{Crisp}, \binits{D.}},
\bauthor{\bsnm{Burges}, \binits{C.J.C.}}:
\bctitle{A geometric interpretation of $\nu$-{SVM} classifiers}.
In: \bbtitle{Advances in Neural Information Processing Systems},
vol. \bseriesno{12}.
\bpublisher{MIT Press},
\blocation{Cambridge}
(\byear{1999})
\end{bchapter}
\endbibitem

\bibitem[\protect\citeauthoryear{Lanckriet et~al.}{2002}]{minimaxsvm:GLCM2002}
\begin{barticle}
\bauthor{\bsnm{Lanckriet}, \binits{G.R.}},
\bauthor{\bsnm{Ghaoui}, \binits{L.E.}},
\bauthor{\bsnm{Bhattacharyya}, \binits{C.}},
\bauthor{\bsnm{Jordan}, \binits{M.I.}}:
\batitle{A robust minimax approach to classification}.
\bjtitle{Journal of Machine Learning Research}
\bvolume{3}(\bissue{Dec}),
\bfpage{555}--\blpage{582}
(\byear{2002})
\end{barticle}
\endbibitem

\bibitem[\protect\citeauthoryear{Jadhav et~al.}{1996}]{siblowdim:JMB1996}
\begin{barticle}
\bauthor{\bsnm{Jadhav}, \binits{S.}},
\bauthor{\bsnm{Mukhopadhyay}, \binits{A.}},
\bauthor{\bsnm{Bhattacharya}, \binits{B.}}:
\batitle{{An Optimal Algorithm for the Intersection Radius of a Set of Convex Polygons}}.
\bjtitle{Journal of Algorithms}
\bvolume{20}(\bissue{2}),
\bfpage{244}--\blpage{267}
(\byear{1996})
\end{barticle}
\endbibitem

\bibitem[\protect\citeauthoryear{Clarkson et~al.}{2012}]{sublinearopt:CHW2012}
\begin{botherref}
\oauthor{\bsnm{Clarkson}, \binits{K.L.}},
\oauthor{\bsnm{Hazan}, \binits{E.}},
\oauthor{\bsnm{Woodruff}, \binits{D.P.}}:
Sublinear optimization for machine learning.
J. ACM
\textbf{59}(5)
(2012)
\end{botherref}
\endbibitem

\bibitem[\protect\citeauthoryear{Bhattacharya et~al.}{1991}]{siblowdim:BJMR1991}
\begin{bchapter}
\bauthor{\bsnm{Bhattacharya}, \binits{B.K.}},
\bauthor{\bsnm{Jadhav}, \binits{S.}},
\bauthor{\bsnm{Mukhopadhayay}, \binits{A.}},
\bauthor{\bsnm{Robert}, \binits{J.-M.}}:
\bctitle{Optimal algorithms for some smallest intersection radius problems}.
In: \bbtitle{Proceedings of the Seventh Annual Symposium on Computational Geometry}.
\bsertitle{SCG '91},
pp. \bfpage{81}--\blpage{88},
\bconflocation{New York, NY, USA}
(\byear{1991})
\end{bchapter}
\endbibitem

\bibitem[\protect\citeauthoryear{Hazan et~al.}{2011}]{sublinearsvm:HKS2011}
\begin{botherref}
\oauthor{\bsnm{Hazan}, \binits{E.}},
\oauthor{\bsnm{Koren}, \binits{T.}},
\oauthor{\bsnm{Srebro}, \binits{N.}}:
Beating sgd: Learning svms in sublinear time.
Advances in Neural Information Processing Systems
\textbf{24}
(2011)
\end{botherref}
\endbibitem

\bibitem[\protect\citeauthoryear{Son and Afshani}{2015}]{sibhighdim:SA2015}
\begin{bchapter}
\bauthor{\bsnm{Son}, \binits{W.}},
\bauthor{\bsnm{Afshani}, \binits{P.}}:
\bctitle{{Streaming Algorithms for Smallest Intersecting Ball of Disjoint Balls}}.
In: \bbtitle{Theory and Applications of Models of Computation},
pp. \bfpage{189}--\blpage{199}
(\byear{2015})
\end{bchapter}
\endbibitem

\bibitem[\protect\citeauthoryear{Arora et~al.}{2012}]{mwu:AHK2012}
\begin{barticle}
\bauthor{\bsnm{Arora}, \binits{S.}},
\bauthor{\bsnm{Hazan}, \binits{E.}},
\bauthor{\bsnm{Kale}, \binits{S.}}:
\batitle{The multiplicative weights update method: a meta-algorithm and applications}.
\bjtitle{Theory of computing}
\bvolume{8}(\bissue{1}),
\bfpage{121}--\blpage{164}
(\byear{2012})
\end{barticle}
\endbibitem

\bibitem[\protect\citeauthoryear{Canyakmaz et~al.}{2023}]{scmwu:CLPV2023}
\begin{botherref}
\oauthor{\bsnm{Canyakmaz}, \binits{I.}},
\oauthor{\bsnm{Lin}, \binits{W.}},
\oauthor{\bsnm{Piliouras}, \binits{G.}},
\oauthor{\bsnm{Varvitsiotis}, \binits{A.}}:
Multiplicative updates for online convex optimization over symmetric cones.
arXiv preprint arXiv:2307.03136
(2023)
\end{botherref}
\endbibitem

\bibitem[\protect\citeauthoryear{Tao et~al.}{2022}]{gtineq:TWK2022}
\begin{barticle}
\bauthor{\bsnm{Tao}, \binits{J.}},
\bauthor{\bsnm{Wang}, \binits{G.}},
\bauthor{\bsnm{Kong}, \binits{L.}}:
\batitle{{The Araki-Lieb-Thirring inequality and the Golden-Thompson inequality in Euclidean Jordan algebras}}.
\bjtitle{Linear and Multilinear Algebra}
\bvolume{70}(\bissue{19}),
\bfpage{4228}--\blpage{4243}
(\byear{2022})
\end{barticle}
\endbibitem

\bibitem[\protect\citeauthoryear{Zheng}{2024}]{libsib}
\begin{botherref}
\oauthor{\bsnm{Zheng}, \binits{J.}}:
{LIBSIB: A C++ library for computing smallest intersecting balls in arbitrary dimensions}
(2024).
\url{https://github.com/orzzzjq/libsib}
\end{botherref}
\endbibitem

\bibitem[\protect\citeauthoryear{Dyer}{1984}]{Dyer1984}
\begin{barticle}
\bauthor{\bsnm{Dyer}, \binits{M.E.}}:
\batitle{Linear time algorithms for two-and three-variable linear programs}.
\bjtitle{SIAM Journal on Computing}
\bvolume{13}(\bissue{1}),
\bfpage{31}--\blpage{45}
(\byear{1984})
\end{barticle}
\endbibitem

\bibitem[\protect\citeauthoryear{Megiddo}{1983}]{Megiddo1983}
\begin{barticle}
\bauthor{\bsnm{Megiddo}, \binits{N.}}:
\batitle{Linear-time algorithms for linear programming in $\mathbb{R}^{3}$ and related problems}.
\bjtitle{SIAM journal on computing}
\bvolume{12}(\bissue{4}),
\bfpage{759}--\blpage{776}
(\byear{1983})
\end{barticle}
\endbibitem

\bibitem[\protect\citeauthoryear{Nam et~al.}{2012}]{sibtheory:NNS2012}
\begin{barticle}
\bauthor{\bsnm{Nam}, \binits{N.M.}},
\bauthor{\bsnm{An}, \binits{N.T.}},
\bauthor{\bsnm{Salinas}, \binits{J.}}:
\batitle{Applications of convex analysis to the smallest intersecting ball problem}.
\bjtitle{Journal of Convex Analysis}
\bvolume{19},
\bfpage{497}--\blpage{518}
(\byear{2012})
\end{barticle}
\endbibitem

\bibitem[\protect\citeauthoryear{Bădoiu and Clarkson}{2008}]{sebcoreset:BC2008}
\begin{barticle}
\bauthor{\bsnm{Bădoiu}, \binits{M.}},
\bauthor{\bsnm{Clarkson}, \binits{K.L.}}:
\batitle{Optimal core-sets for balls}.
\bjtitle{Computational Geometry}
\bvolume{40}(\bissue{1}),
\bfpage{14}--\blpage{22}
(\byear{2008})
\end{barticle}
\endbibitem

\bibitem[\protect\citeauthoryear{Bădoiu and Clarkson}{2003}]{sescoreset:badoiu2003smaller}
\begin{bchapter}
\bauthor{\bsnm{Bădoiu}, \binits{M.}},
\bauthor{\bsnm{Clarkson}, \binits{K.L.}}:
\bctitle{Smaller core-sets for balls}.
In: \bbtitle{SODA},
vol. \bseriesno{3},
pp. \bfpage{801}--\blpage{802}
(\byear{2003})
\end{bchapter}
\endbibitem

\bibitem[\protect\citeauthoryear{Kumar et~al.}{2003}]{sebcoreset:KMY2003}
\begin{barticle}
\bauthor{\bsnm{Kumar}, \binits{P.}},
\bauthor{\bsnm{Mitchell}, \binits{J.S.}},
\bauthor{\bsnm{Yildirim}, \binits{E.A.}}:
\batitle{Approximate minimum enclosing balls in high dimensions using core-sets}.
\bjtitle{Journal of Experimental Algorithmics (JEA)}
\bvolume{8},
\bfpage{1}--\blpage{1}
(\byear{2003})
\end{barticle}
\endbibitem

\bibitem[\protect\citeauthoryear{Yildirim}{2008}]{seb:yildirim2008}
\begin{barticle}
\bauthor{\bsnm{Yildirim}, \binits{E.A.}}:
\batitle{Two algorithms for the minimum enclosing ball problem}.
\bjtitle{SIAM Journal on Optimization}
\bvolume{19}(\bissue{3}),
\bfpage{1368}--\blpage{1391}
(\byear{2008})
\end{barticle}
\endbibitem

\bibitem[\protect\citeauthoryear{Clarkson}{2010}]{sebcoreset:Clarkson2010}
\begin{barticle}
\bauthor{\bsnm{Clarkson}, \binits{K.L.}}:
\batitle{Coresets, sparse greedy approximation, and the frank-wolfe algorithm}.
\bjtitle{ACM Transactions on Algorithms (TALG)}
\bvolume{6}(\bissue{4}),
\bfpage{1}--\blpage{30}
(\byear{2010})
\end{barticle}
\endbibitem

\bibitem[\protect\citeauthoryear{G{\"a}rtner and Sch{\"o}nherr}{2000}]{sebopt:BS2000}
\begin{bchapter}
\bauthor{\bsnm{G{\"a}rtner}, \binits{B.}},
\bauthor{\bsnm{Sch{\"o}nherr}, \binits{S.}}:
\bctitle{An efficient, exact, and generic quadratic programming solver for geometric optimization}.
In: \bbtitle{Proceedings of the Sixteenth Annual Symposium on Computational Geometry},
pp. \bfpage{110}--\blpage{118}
(\byear{2000})
\end{bchapter}
\endbibitem

\bibitem[\protect\citeauthoryear{Zhou et~al.}{2005}]{sebball:ZTS2005}
\begin{barticle}
\bauthor{\bsnm{Zhou}, \binits{G.}},
\bauthor{\bsnm{Toh}, \binits{K.-C.}},
\bauthor{\bsnm{Sun}, \binits{J.}}:
\batitle{Efficient algorithms for the smallest enclosing ball problem}.
\bjtitle{Computational Optimization and Applications}
\bvolume{30},
\bfpage{147}--\blpage{160}
(\byear{2005})
\end{barticle}
\endbibitem

\bibitem[\protect\citeauthoryear{Allen-Zhu et~al.}{2016}]{sebopt:ALY2016}
\begin{bchapter}
\bauthor{\bsnm{Allen-Zhu}, \binits{Z.}},
\bauthor{\bsnm{Liao}, \binits{Z.}},
\bauthor{\bsnm{Yuan}, \binits{Y.}}:
\bctitle{{Optimization Algorithms for Faster Computational Geometry}}.
In: \bbtitle{43rd International Colloquium on Automata, Languages, and Programming (ICALP 2016)},
vol. \bseriesno{55},
pp. \bfpage{53}--\blpage{1536}
(\byear{2016})
\end{bchapter}
\endbibitem

\bibitem[\protect\citeauthoryear{Li et~al.}{2021}]{l1lpgame:LWCW2021}
\begin{bchapter}
\bauthor{\bsnm{Li}, \binits{T.}},
\bauthor{\bsnm{Wang}, \binits{C.}},
\bauthor{\bsnm{Chakrabarti}, \binits{S.}},
\bauthor{\bsnm{Wu}, \binits{X.}}:
\bctitle{Sublinear classical and quantum algorithms for general matrix games}.
In: \bbtitle{Proceedings of the AAAI Conference on Artificial Intelligence},
vol. \bseriesno{35},
pp. \bfpage{8465}--\blpage{8473}
(\byear{2021})
\end{bchapter}
\endbibitem

\bibitem[\protect\citeauthoryear{Zheng et~al.}{2024}]{pdscp:ZVTL2024}
\begin{botherref}
\oauthor{\bsnm{Zheng}, \binits{J.}},
\oauthor{\bsnm{Varvitsiotis}, \binits{A.}},
\oauthor{\bsnm{Tan}, \binits{T.-S.}},
\oauthor{\bsnm{Lin}, \binits{W.}}:
A primal-dual framework for symmetric cone programming.
arXiv preprint arXiv:2405.09157
(2024)
\end{botherref}
\endbibitem

\bibitem[\protect\citeauthoryear{Faraut and Kor{\'a}nyi}{1994}]{symmetriccone:FK1994}
\begin{bbook}
\bauthor{\bsnm{Faraut}, \binits{J.}},
\bauthor{\bsnm{Kor{\'a}nyi}, \binits{A.}}:
\bbtitle{Analysis on Symmetric Cones}.
\bsertitle{Oxford mathematical monographs}.
\bpublisher{Clarendon Press},
\blocation{Oxford}
(\byear{1994})
\end{bbook}
\endbibitem

\bibitem[\protect\citeauthoryear{Orlitzky}{2022}]{eja:Orlitzky2022}
\begin{bbook}
\bauthor{\bsnm{Orlitzky}, \binits{M.}}:
\bbtitle{{Euclidean Jordan Algebras}},
(\byear{2022})
\end{bbook}
\endbibitem

\bibitem[\protect\citeauthoryear{Vandenberghe}{2016}]{symmetriccone:Vandenberghe2016}
\begin{botherref}
\oauthor{\bsnm{Vandenberghe}, \binits{L.}}:
Symmetric Cones.
Lecture notes, Electrical and Computer Engineering Department, UCLA, Spring
(2016)
\end{botherref}
\endbibitem

\bibitem[\protect\citeauthoryear{Sion}{1958}]{minimax:Sion1958}
\begin{barticle}
\bauthor{\bsnm{Sion}, \binits{M.}}:
\batitle{On general minimax theorems.}
\bjtitle{Pacific J. Math.}
\bvolume{8},
\bfpage{171}--\blpage{176}
(\byear{1958})
\end{barticle}
\endbibitem

\bibitem[\protect\citeauthoryear{Beck}{2014}]{projtheory:Beck2014}
\begin{bbook}
\bauthor{\bsnm{Beck}, \binits{A.}}:
\bbtitle{Introduction to Nonlinear Optimization}.
\bpublisher{Society for Industrial and Applied Mathematics},
\blocation{Philadelphia, PA}
(\byear{2014})
\end{bbook}
\endbibitem

\bibitem[\protect\citeauthoryear{Cormen et~al.}{2022}]{intro2algo:CLRS2022}
\begin{bbook}
\bauthor{\bsnm{Cormen}, \binits{T.H.}},
\bauthor{\bsnm{Leiserson}, \binits{C.E.}},
\bauthor{\bsnm{Rivest}, \binits{R.L.}},
\bauthor{\bsnm{Stein}, \binits{C.}}:
\bbtitle{Introduction to Algorithms, Fourth Edition}.
\bpublisher{MIT Press},
\blocation{Cambridge}
(\byear{2022})
\end{bbook}
\endbibitem

\bibitem[\protect\citeauthoryear{John}{1948}]{ellipsoid:fritz1948}
\begin{bchapter}
\bauthor{\bsnm{John}, \binits{F.}}:
\bctitle{Extremum problems with inequalities as subsidiary conditions}.
In: \bbtitle{Studies and {E}ssays {P}resented to {R}. {C}ourant on His 60th {B}irthday, {J}anuary 8, 1948},
pp. \bfpage{187}--\blpage{204}.
\bpublisher{Interscience Publishers},
\blocation{New York}
(\byear{1948})
\end{bchapter}
\endbibitem

\bibitem[\protect\citeauthoryear{Vandenberghe and Boyd}{1996}]{sdp:LS1996}
\begin{barticle}
\bauthor{\bsnm{Vandenberghe}, \binits{L.}},
\bauthor{\bsnm{Boyd}, \binits{S.}}:
\batitle{Semidefinite programming}.
\bjtitle{SIAM review}
\bvolume{38}(\bissue{1}),
\bfpage{49}--\blpage{95}
(\byear{1996})
\end{barticle}
\endbibitem

\bibitem[\protect\citeauthoryear{Shivaswamy et~al.}{2006}]{ellipsoidsvm:PCA2006}
\begin{botherref}
\oauthor{\bsnm{Shivaswamy}, \binits{P.K.}},
\oauthor{\bsnm{Bhattacharyya}, \binits{C.}},
\oauthor{\bsnm{Smola}, \binits{A.J.}}:
Second order cone programming approaches for handling missing and uncertain data.
Journal of Machine Learning Research,
1283--1314
(2006)
\end{botherref}
\endbibitem

\bibitem[\protect\citeauthoryear{Trafalis and Alwazzi}{2010}]{ballsvm:TS2010}
\begin{barticle}
\bauthor{\bsnm{Trafalis}, \binits{T.B.}},
\bauthor{\bsnm{Alwazzi}, \binits{S.A.}}:
\batitle{Support vector machine classification with noisy data: a second order cone programming approach}.
\bjtitle{International Journal of General Systems}
\bvolume{39}(\bissue{7}),
\bfpage{757}--\blpage{781}
(\byear{2010})
\end{barticle}
\endbibitem

\bibitem[\protect\citeauthoryear{Shenmaier}{2015}]{kSEBhardness:Vladimir2015}
\begin{barticle}
\bauthor{\bsnm{Shenmaier}, \binits{V.}}:
\batitle{Complexity and approximation of the smallest k-enclosing ball problem}.
\bjtitle{European Journal of Combinatorics}
\bvolume{48},
\bfpage{81}--\blpage{87}
(\byear{2015})
\end{barticle}
\endbibitem

\bibitem[\protect\citeauthoryear{Cohen et~al.}{2016}]{geometricmedian:CLMPS2016}
\begin{bchapter}
\bauthor{\bsnm{Cohen}, \binits{M.B.}},
\bauthor{\bsnm{Lee}, \binits{Y.T.}},
\bauthor{\bsnm{Miller}, \binits{G.}},
\bauthor{\bsnm{Pachocki}, \binits{J.}},
\bauthor{\bsnm{Sidford}, \binits{A.}}:
\bctitle{Geometric median in nearly linear time}.
In: \bbtitle{Proceedings of the Forty-eighth Annual ACM Symposium on Theory of Computing},
pp. \bfpage{9}--\blpage{21}
(\byear{2016})
\end{bchapter}
\endbibitem

\bibitem[\protect\citeauthoryear{Arashloo}{2022}]{lpsvdd:Arashloo2022}
\begin{barticle}
\bauthor{\bsnm{Arashloo}, \binits{S.R.}}:
\batitle{$\ell_p$-norm support vector data description}.
\bjtitle{Pattern Recognition}
\bvolume{132},
\bfpage{108930}
(\byear{2022})
\end{barticle}
\endbibitem

\end{thebibliography}

\end{document}